\documentclass[a4paper,fleqn,usenatbib]{mnras}
\usepackage[dvipdfmx]{graphicx}
\usepackage{newtxtext,newtxmath}
\usepackage[T1]{fontenc}
\usepackage{ae,aecompl}

\usepackage{color}
\usepackage[dvipsnames]{xcolor}
\usepackage{ulem}

\usepackage{graphicx}	% Including figure files
\usepackage{amsmath}	% Advanced maths commands
\usepackage{amssymb}	% Extra maths symbols

\graphicspath{{./figure/}}

\newcommand{\stol}{\mathcal{D}_\mathrm{S} / \mathcal{D}_\mathrm{L}}
\title[]{Evolution of dust extinction curves in galaxy simulation}

\author[K. C. Hou et al.]
{Kuan-Chou Hou$^{1,2}$\thanks{E-mail: kchou@asiaa.sinica.edu.tw},
Hiroyuki Hirashita$^{2}$,
Kentaro Nagamine$^{3,4}$,
Shohei Aoyama$^{3,5}$,
\newauthor and 
Ikkoh Shimizu$^{3,5}$\\
$^{1}$Department of Physics, Institute of Astrophysics, National Taiwan University, Taipei 10617, Taiwan \\
$^{2}$Institute of Astronomy and Astrophysics, Academia Sinica, PO Box 23-141, Taipei 10617, Taiwan\\
$^{3}$Theoretical Astrophysics, Department of Earth \& Space Science, Osaka University, 1-1 Machikaneyama, Toyonaka, Osaka 560-0043, Japan\\
$^{4}$Department of Physics \& Astronomy, University of Nevada Las Vegas, 4505 S. Maryland Pkwy, Las Vegas, NV 89154-4002, USA \\
$^{5}$College of General Education, Osaka Sangyo University, 3-1-1, Nakagaito, Daito, Osaka, 574-8530, Japan\\
}

\date{Accepted XXX. Received YYY; in original form ZZZ}

\pubyear{2016}

\begin{document}
\label{firstpage}
\pagerange{\pageref{firstpage}--\pageref{lastpage}}
\maketitle

\begin{abstract}
To understand the evolution of extinction curve,
we calculate the dust evolution in a galaxy 
using smoothed particle hydrodynamics simulations
incorporating stellar dust production, dust destruction in supernova
shocks, grain growth by accretion and coagulation, and grain
disruption by shattering. 
The dust species are separated into 
carbonaceous dust and silicate.
The evolution of grain size distribution 
is considered by dividing grain population 
into large and small gains, 
which allows us to estimate extinction curves. 
We examine the dependence of extinction 
curves on the position, gas density, and metallicity in the galaxy,
and find that extinction curves are flat at $t \lesssim 0.3$\,Gyr
because stellar dust production dominates 
the total dust abundance. 
The 2175\,\AA\ bump and far-ultraviolet (FUV) rise become
prominent after dust growth by accretion. 
At $t \gtrsim 3$\,Gyr, shattering works efficiently 
in the outer disc and low density regions,
so extinction curves show a very strong 2175\,\AA\ bump and steep FUV rise.
The extinction curves at $t\gtrsim 3$\,Gyr are consistent with the Milky Way
extinction curve, which implies that we successfully included the necessary dust processes in the model. 
The outer disc component caused by stellar feedback has 
an extinction curves with a weaker 2175\,\AA\ bump and flatter FUV slope.
The strong contribution of carbonaceous dust tends to underproduce the
FUV rise in the Small Magellanic Cloud extinction curve, which
supports selective loss of small carbonaceous dust in the galaxy.
The snapshot at young ages also explain the extinction curves in high-redshift quasars.
\end{abstract}


%%%%%%%%%%%%%%%%%%%%%%%%%%%%%%%%%%%%%%%%%%%%%%%%%%

%%%%%%%%%%%%%%%%% BODY OF PAPER %%%%%%%%%%%%%%%%%%

\section{Introduction}

Studying dust is necessary for the understanding of galaxy evolution.
Dust surfaces are the main sites for 
the efficient production of molecular hydrogen,
which is the main constituent of molecular clouds 
and an important coolant in 
low-metallicity environments \citep[e.g.][]{Hirashita:2002aa,Cazaux:2004aa}.
Moreover, dust cooling determines the mass of final fragments
of star-forming clouds \citep{Omukai:2005aa,Schneider:2006aa}.
Therefore, star formation in galaxies is strongly affected by dust grains.
In terms of the observational properties of galaxies, 
dust shapes the spectral energy distribution (SED)
by absorbing stellar light and re-emitting it in the far infrared
\citep[e.g.][for recent modelling]{Yajima:2014aa,Schaerer:2015aa};
thus, correcting for dust extinction is required to derive the intrinsic stellar SED.

Extinction curves, which represent the wavelength dependence 
of dust extinction (absorption + scattering), 
provide useful information on the grain size distribution and 
chemical composition \citep[e.g.][]{Weingartner:2001aa},
both of which are important in determining the aforementioned processes,
i.e.\ grain surface reactions and dust cooling \citep{Yamasawa:2011aa}.
The Milky Way extinction curve has been observed in detail
\citep{Pei:1992aa,Fitzpatrick:2007aa}, and 
by fitting that, \citet{Mathis:1977aa} derived a
grain size distribution of a power law $\propto a^{-3.5}$ 
in the range of $a\simeq 0.005$--0.25 $\micron$ ($a$ is the grain radius)
with a mixture of silicate and graphite. 
Furthermore, \citet{Weingartner:2001aa}
suggested detailed functional forms of grain size
distribution for carbonaceous dust and silicate 
to explain the extinction curves of the Milky Way and Large/Small 
Magellanic Clouds (LMC/SMC).
Although these models are successful in explaining 
the extinction curves, 
it is still necessary to clarify what processes 
govern the evolution of grain size distribution in galaxies.

There have been a lot of efforts not only 
to model the evolution of dust abundance
in the solar neighbourhood
\citep{Dwek:1980aa,Zhukovska:2008aa}, and in both local galaxies
\citep{Dwek:1998aa,Lisenfeld:1998aa,Hirashita:1999aa,Inoue:2003aa}
and high redshift galaxies 
\citep{Kuo:2012aa,de-Bennassuti:2014aa,Mancini:2016aa,Popping:2016aa,Wang:2017aa},
but also to model the evolution of
grain size distribution in galaxies 
\citep{Liffman:1989aa,ODonnell:1997aa,Yamasawa:2011aa}.
\citet{Asano:2013aa}, treating a galaxy as a single zone,
constructed a 
framework for calculating the evolution of grain size distribution
in a consistent manner with galaxy evolution
over the entire galaxy history.
Furthermore, \citet{Asano:2014aa} used their model to calculate
the evolution of extinction curve and found the following
evolutionary features. The extinction curve is flat
at the earliest evolutionary stage because the dust content is
dominated by stellar sources. The subsequent contribution from
shattering and accretion increases the abundance of small grains
drastically; as a consequence, the extinction curve becomes steep,
and eventually even steeper than the Milky Way extinction curve.
Using the same framework, \citet{Nozawa:2015aa}
successfully reproduced the Milky Way extinction curve
by newly considering dense clouds that increases the 
efficiency of coagulation.
The above studies basically treat a galaxy as a single-zone object, 
and this treatment inevitably and implicitly adopt some 
strong assumptions such as instantaneous mixing and homogeneity
(i.e., the spatial structure in a galaxy is neglected). 
\citet{Fitzpatrick:2007aa} showed the variation of 
the Milky Way extinction curves toward different lines of sight, 
which indicates the inhomogeneity in dust properties 
\citep[see also][]{Nozawa:2013aa}.
This inhomogeneity might cause uncertainty in interpreting 
the extinction properties of unresolved extragalactic objects.

To calculate the evolution of dust 
in realistic conditions and in a spatially resolved manner,
we use a hydrodynamical simulation of a galaxy.
Hydrodynamic simulations have been recognized as
a useful tool in studying galaxy formation and evolution.
There have been some hydrodynamic simulations 
incorporating the effects of dust in recent years.
\citet{Yajima:2015aa} estimated the dust abundance in 
high-resolution zoom-in cosmological 
simulations of galaxies with a fixed dust-to-metal ratio.
The ultraviolet (UV) and infrared luminosities of individual galaxies 
are estimated by post-processing using a radiative transfer code.
\citet{Bekki:2015aa} performed a smoothed particle 
hydrodynamics (SPH) simulation 
with a dust evolution model, which includes
dust formation in stellar ejecta, dust growth by accretion,
and dust destruction in SN shocks.
In particular, they treated dust as new particles
besides dark matter, gas and stellar particles.
\citet{McKinnon:2016ab} treated dust as an attached property of gas.
They implemented dust formation and destruction in a moving-mesh code 
and performed cosmological zoom-in simulations.
Their result showed that dust growth by accretion is important 
and that an accurate treatment of stellar and 
active galactic nuclei feedback is needed to reproduce the observation.
Moreover, they also ran cosmological simulations
\citep{McKinnon:2016aa} to investigate the dust mass function 
and radial profile of dust in the circum-galactic environment.
The simulated dust mass function is consistent with
observations in the local Universe but has a tendency of underestimating
the number of dust-rich galaxies at high redshift.
\citet{Zhukovska:2016aa} focused on the dominant processes,
dust growth and destruction in the interstellar medium (ISM), to examine the
temperature-dependent sticking coefficient in the accretion of gas-phase metals onto dust.
However, the grain size distribution is not considered in 
the above simulations.

Our previous work \citep[][hereafter Paper~I]{Aoyama:2017aa}
simulated dust abundance in an isolated galaxy 
using SPH simulations with the dust enrichment model
developed by \citet{Hirashita:2015aa}.
The model includes the dust production in stellar ejecta, 
destruction in supernova (SN) shocks, dust growth by accretion, 
grain growth by coagulation, and grain disruption by shattering.
Shattering and coagulation, which are caused by grain--grain collisions, 
are important in determining the grain size distribution.
To treat the evolution of grain size distribution within 
a reasonable computational cost, Paper~I adopted the 
two-size approximation proposed by \citet{Hirashita:2015aa} 
in which the whole range of grain size is represented by two sizes 
divided at around $0.03 \mu$m. In Paper I,
we found the following evolutionary features of dust.
Stellar dust production dominates the dust content
at young ages ($t \lesssim$ 0.2\,Gyr). After that,
dust processing in the ISM is much more important than stellar
dust production. In particular, dust growth by the accretion of
gas-phase metals is triggered by small grains produced as
a result of shattering, and drastically increases the dust-to-gas ratio
and the dust-to-metal ratio.
Coagulation becomes efficient at later times of $t \gtrsim$ 1\,Gyr.
We also computed the radial profile of dust-to-gas mass ratio and 
dust-to-metal mass ratio (i.e.\ depletion), both of which are broadly
consistent with the 
observational data of a sample of spatially resolved nearby galaxies
in \citet{Mattsson:2012aa}.

In this paper, we further compare our simulation results with observed
extinction curves in the nearby Universe to obtain more constraints
on the grain size distribution.
To calculate extinction curves, we separate the dust species into 
carbonaceous dust and silicate.
For the first time, extinction curves can be predicted by hydrodynamical
galaxy evolution simulations
with a physically plausible model 
including dust production and destruction.
Taking advantage of the simulations, 
we examine how the evolution of extinction curve depends on 
the physical conditions such as the galactic radius,
gas density and metallicity.
We compare the calculated extinction curves with 
observational extinction curves in the Milky Way, 
LMC and SMC, 
in which a precise determination of extinction curves 
are possible because individual bright stars can be spatially resolved
\citep[e.g.][]{Pei:1992aa,Gordon:2003aa}. 
We also discuss the evolution of extinction curves and its implication for the extinction curves at high redshifts.

This paper is organized as follows. 
In Section 2, we summarise the simulation,
the dust evolution model and the extinction curve calculation. 
The evolution of extinction curves
and the dependence on the radial distance, 
gas density, and metallicity are shown in Section 3.
In Section 4, we make a comparison with observational data and 
discuss the results. 
Finally, we provide the conclusions in Section 5.

\section{Model}

\subsection{Galaxy simulation with dust enrichment}

We briefly summarize the simulation with the dust enrichment model.
Our model is based on Paper I, and the
main difference in this work is that we separate the dust species into carbonaceous 
dust and silicate. 
Separate treatment of grain species is fundamental in calculating
extinction curves \citep[e.g.][]{Draine:1984aa}.
We focus on the difference from Paper I, but also provide a brief summary of
Paper I. We refer the interested reader to Paper I for details.

We adopt the modified version of {\sf GADGET-3} 
$N$-body/SPH code for this study \citep{Springel:2005aa}, and
the Grackle\footnote{https://grackle.readthedocs.org/} 
chemistry and cooling library \citep{Bryan:2014aa,Kim:2014aa,Smith:2017aa} 
to solve non-equilibrium primordial chemistry network.
Initial condition is the low-resolution model of 
{\sf AGORA} simulations \citep{Kim:2014aa} with
the minimum gravitational softening length of $\epsilon_{\rm grav}=80$\,pc.
The star formation is assumed to occur on a local free-fall time with an
efficiency of 0.01. Stellar feedback from SNe and stellar winds is also
taken into account. All the feedback and metal enrichment from stars
are assumed to occur in 4 Myr after the star formation; thus, in our framework,
the injection of metals and dust occurs almost instantaneously after the
star formation, and it is difficult to include delayed dust and metal input
from asymptotic giant branch (AGB) stars. Although we discuss the effect of dust production by AGB stars later
(Section \ref{subsubsec:stellar}),
we leave the treatment of
delayed dust production by AGB stars to the future work.

For dust enrichment,
we adopt the two-size dust enrichment model developed by \citet{Hirashita:2015aa}.
In this model, we divide the whole grain size range into two at
around $a\sim 0.03~\micron$ ($a$ is the grain radius) and
consider the evolution of large and small grains.
This approximate treatment is introduced because solving the full
grain size distribution in individual SPH particles is computationally
too expensive. \citet{Hirashita:2015aa} has already 
clarified that the two-size approach
reproduces the evolution of extinction curves calculated by
a full grain size treatment in \citet{Asano:2014aa}
if a functional form of grain size distribution is properly chosen
(see Section \ref{subsec:extinc_curve}).

For dust evolution processes, we consider stellar dust production, SN destruction,
grain disruption by shattering in the diffuse ISM, and grain growth by 
coagulation and accretion in the dense ISM.
On each gas (SPH) particle, 
we consider the mass of small and large grains, $m_{i,\mathrm{X}(j)}$, where
subscripts $i$, X, and $j$ denote the grain size ($i=$ L and S
for large and small grains, respectively), grain species (X=Si and C
for silicate and carbonaceous dust, respectively), and the label of the gas particle,
respectively. We formulate the dust evolution using the dust-to-gas mass ratio:
\begin{eqnarray} 
\mathcal{D}_{i,\mathrm{X}(j)}\equiv m_{i,\mathrm{X}(j)}/m_{\mathrm{gas}(j)},
\label{eq:dust-to-gas}
\end{eqnarray} 
where $m_{\mathrm{gas}(j)}$ is the gas mass of the $j$-th particle.
We calculate the evolution of $\mathcal{D}_{i,\mathrm{X}(j)}$ for each gas
particle.
We follow Paper I for the basic equations of dust evolution except that
we consider two species.
The time evolution of the large and small grain abundances in the $j$-th particle 
from time $t$ to the next time step $t + \Delta t$ is formulated as
(for brevity, we omit the subscript $j$, and we always solve the equations for
each gas particle):
\begin{align} 
\mathcal{D}_{{\rm L,X}}(t+\Delta t)
&= \mathcal{D}_{{\rm L,X}}(t) - \Delta \mathcal{D}^{\rm SN}_{{\rm L,X}}(t) \notag \\ 
&- \left( \dfrac{\mathcal{D}_{{\rm L,X}}(t)}{\tau_{\rm sh}}
- \dfrac{\mathcal{D}_{{\rm S,X}}(t)}{\tau_{\rm co}}\right)\Delta t \notag \\
&+ f_{\rm in,X }\mathcal{Y}^{\prime }_{\rm X}\dfrac{\Delta M_{{\rm return}}}{m_{\rm gas}}
( 1 - \delta )~,
\label{timeL}\\ 
\mathcal{D}_{{\rm S,X}}(t+\Delta t)
&=\mathcal{D}_{{\rm S,X}}(t) - \Delta \mathcal{D}^{\rm SN}_{{\rm S,X}}(t) \notag \\ 
&+ \left(\dfrac{\mathcal{D}_{{\rm L,X}}(t)}{\tau_{\rm sh}}
- \dfrac{\mathcal{D}_{{\rm S,X}}(t)}{\tau_{\rm co}} 
+\dfrac{\mathcal{D}_{{\rm S,X}}(t)}{\tau_{\rm acc}}\right)\Delta t~,\label{timeS}
\end{align}
where $\mathcal{D}_{{\rm L,X}}$ ($\mathcal{D}_{{\rm S,X}}$) is
large (small) grain dust-to-gas mass ratio defined in Eq. (\ref{eq:dust-to-gas}), 
$\Delta \mathcal{D}^{\rm SN}_{i,\mathrm{X}}(t)$
is the abundance of the pre-existing dust 
destroyed by SN shock,
$f_{\rm in,X }$ is dust condensation efficiency in stellar ejecta,
$\Delta M_{{\rm return}}$ is returned gas mass, 
$\delta$ is the fraction of newly formed dust that is destroyed by SNe,
and $\mathcal{Y}^{\prime }_{\rm X}$ is the metal yield, and
$\tau_{\rm sh}$, $\tau_{\rm co}$ and $\tau_{\rm acc}$ are the time-scales of
shattering, coagulation and accretion respectively.
For the metal yield, we adopt $\mathcal{Y}^{\prime }_{\rm Z} = 0.02896$
(Paper I) and multiply $Z_\mathrm{X\odot}/Z_\odot$
(the mass ratio of the element X to the total metal content in the
solar abundance)
for the Si and C yields by assuming the
solar metallicity pattern \citep{Lodders:2003aa}, 
where $Z_\mathrm{X\odot}$ is the solar abundance (X = Si and C).
Following Paper I, the solar metallicity $Z_\odot = 0.02$ is adopted.
We notice that the precise solar metallicity is less than 0.02 
\citep{Lodders:2003aa,Asplund:2009aa}. However, in this work, 
the solar metallicity is just for normalizing metallicity in the figures,  
and the results are not affected by which value we adopted.
We assume that Si occupies a mass fraction of 
0.166 in silicate while
C is the only constituent of carbonaceous dust. 
The parameters adopted in this work are summarised 
in Table \ref{table:Adopted_quantities}. 
For simplicity, we assume that the two dust species evolve independently
in this work.

We briefly explain each dust evolution process in what follows.

\subsubsection{Stellar dust formation}\label{subsubsec:stellar}
Stars produce dust at their final stage of evolution. The formed dust 
is distributed to the surrounding SPH particles in the same 
way as metals by assuming the dust condensation efficiency
$f_\mathrm{in,X}$.
The dust condensation efficiencies in stellar ejecta
for carbonaceous dust and silicate (X = C and Si) are
calculated based on \citet{Hou:2016aa} except that we
adopt a \citet{Chabrier:2003ab} initial mass function
with a stellar mass range of 0.1--100 $M_\odot$. The values are
listed in Table \ref{table:Adopted_quantities}.
In our simulation, metals are assumed to be produced only by SNe.
Thus, to derive the dust condensation efficiency in stellar ejecta,
we integrate the metal and dust yields of SNe for 
the progenitors in the mass range
$m=8$--40 $M_\odot$. 
The metal and dust yields taken
from \citet{Kobayashi:2006aa} and \citet[][with an ambient hydrogen number density of
1 cm$^{-3}$ and the unmixed helium core]{Nozawa:2007aa}, respectively.
We assume that neither metals nor dust is ejected from stars whose mass is
above 40 $M_\odot$ \citep{Heger:2003aa}.

In this study, dust production by AGB stars is not included for the
following two reasons.
The first reason is that the stellar dust production is more important in the early stage
($\lesssim 0.3$\,Gyr) than in the later stage, at least in our simulation:
at $\lesssim 0.3$\,Gyr, the progenitors of AGB stars have no time to evolve
to a dust production phase \citep{Valiante:2009aa,Valiante:2011aa}.
The second reason is related to a technical aspect:
our simulation assumes almost instantaneous metal production 
after star formation. Thus, the metal and dust production by
AGB stars, whose progenitor lifetimes have a large variety, cannot be included
in the current simulation framework. We could effectively include the dust
production by AGB stars by adopting the dust condensation efficiency
($f_\mathrm{in,X}$)
that takes into account the contribution from AGB stars. Paper I already
clarified the effects of changing
$f_\mathrm{in,X}$. Extinction curves are governed by the small-to-large
grain abundance ratio ($\stol$) in our model, but the value of
$\stol$ is affected not by stellar dust production but by
interstellar dust processing,
especially accretion, at the epoch when AGB stars contribute to
the dust production ($t\gtrsim 1$\,Gyr). At this stage, the dust evolution
is insensitive to
stellar dust production or $f_\mathrm{in,X}$.
Moreover, as shown by \citet{Asano:2013ab}, the epoch at
which interstellar processing starts to dominate the dust
evolution over stellar dust production is determined by the metallicity
and is insensitive to the
choice of $f_\mathrm{in,X}$.
Therefore, we do not attempt a detailed modelling of
$f_\mathrm{in,X}$ by including the contribution from AGB stars,
although we plan to modify the framework to include the delayed
dust input from AGB stars in the future.
The parameter $f_\mathrm{in,X}$ is important for determining the 
$\stol$ and C/Si ratio at early evolutionary stage, 
which will be discussed in Sections \ref{subsection:c_to_si} and 
\ref{subsec:species_ratio}.

\subsubsection{SN destruction}
Each SN destroys the dust in its sweeping radius.
Because the simulations do not resolve individual SNe, a sub-grid model
is necessary to treat dust destruction by SNe.
We separately treat the destruction of the pre-existing dust and
the newly formed dust by the same SNe. The terms for the
destruction of the pre-existing dust in equations (\ref{timeL}) and (\ref{timeS}) can be written as
\begin{eqnarray} 
{\Delta \mathcal{D}^{{\rm SN}}_{{\rm L\slash S,X}} (t)}=
\left[ 1-( 1 - \eta )^{N}\right] {\mathcal{D}^{{\rm SN}}_{{\rm L\slash S,X}}(t)},
\end{eqnarray}
where $\eta \equiv\min [\epsilon_\mathrm{SN}(m_\mathrm{sw}/m_\mathrm{gas}),\,\epsilon_\mathrm{SN}]$ 
($m_\mathrm{sw}$ is the gas mass swept by a single SN; see Paper I for its estimation,
and $\epsilon_\mathrm{SN}$ is the efficiency of dust destruction in a single SN blast),
and $N$ is the number of SN explosions in the gas particle. 
We adopt $\epsilon_\mathrm{SN}=0.4$ based on \citet{Nozawa:2006aa}. 
The SN destruction of newly formed dust is absorbed in the stellar ejecta term by giving
a destruction fraction $\delta$ in Eq. (\ref{timeL}).
The derivation of this quantity was shown in Paper I.

\subsubsection{Shattering}
Shattering is a process in which large grains collide with each
other and shattered or fragmented into small grains.
Shattering occurs in the diffuse ISM \citep{Hirashita:2009aa}
because it needs high relative velocity between grains;
therefore, we allow for shattering only in gas particles with 
$n_\mathrm{gas} < 1 ~\mathrm{cm}^{-3}$.

The shattering time-scale is estimated as
\begin{eqnarray} 
\tau_\mathrm{sh} = 5.408 \times 10^{8}\ \mathrm{yr}\ \left(\frac{\rho_\mathrm{X}}{\mathrm{3\ g\ cm^{-3} }}\right)\left(\frac{n_{\rm gas}}{1~\mathrm{cm}^{-3}}\right)^{-1} \left( \dfrac{\mathcal{D}_{\rm L,X}}{0.01} \right)^{-1},
\label{tau_sh}
\end{eqnarray} 
where $\rho_\mathrm{X}$ is the material density of the dust species listed in Table \ref{table:Adopted_quantities}.
For the particles with $n_\mathrm{gas} \ge 1 ~\mathrm{cm}^{-3}$, 
we turn off shattering, i.e. $\tau_\mathrm{sh} = \infty$.

\subsubsection{Coagulation and accretion}
Coagulation is a process in which small grains collide 
with each other and turn into large grains. 
It occurs under the condition that grains have low relative velocity. 
Accretion is a process in which 
small grains gain their mass by accreting gas-phase metals.
Coagulation and accretion are expected to occur in dense clouds,
where relative velocities of grains and gas are low
but collision still happens because of high number density; 
however, our simulations are not capable of resolving them since 
the highest density is determined by the minimum gravitational 
softening length of $\epsilon_{\rm grav}=80$\ pc, 
which is larger than the typical size of a dense cloud. 
We apply a sub-grid model for gas particles with 
$n_\mathrm{gas} \ge 10 ~\mathrm{cm}^{-3}$
and $T_\mathrm{gas} < 10^3$ K by assuming that a fraction, $f_{\rm dense}$, 
of gas content is in the dense cloud phase with fixed density and temperature
as
$n_\mathrm{gas} = 10^3 ~\mathrm{cm}^{-3}$ and $T_\mathrm{gas} = 50$ K. 
The time-scale of coagulation is formulated as
\begin{eqnarray} 
\tau_\mathrm{co} = 2.704 \times 10^{5}\ \mathrm{yr}\ \left(\frac{\rho_\mathrm{X}}{\mathrm{3\ g\ cm^{-3} }}\right) \left( \dfrac{\mathcal{D}_{\rm S,X}}{0.01} \right)^{-1}\slash f_{\rm dense},
\label{tau_co}
\end{eqnarray} 
and the time-scales of accretion for carbonaceous dust and silicate are
formulated as
\begin{eqnarray}
\tau_\mathrm{acc,C} = 0.95 \times 10^{6}\ \mathrm{yr} \left(\dfrac{Z_\mathrm{C}}{Z_\mathrm{C\odot}} \right)^{-1}\left(1-\dfrac{\mathcal{D}_\mathrm{tot,C}}{Z_\mathrm{C}}\right)^{-1}\slash f_{\rm dense}, 
\label{eq:tau_acc_C}\\
\tau_\mathrm{acc,Si} = 1.05 \times 10^{6}\ \mathrm{yr} \left(\dfrac{Z_\mathrm{Si}}{Z_\mathrm{Si\odot}} \right)^{-1}\left(1-\dfrac{\mathcal{D}_\mathrm{tot,Si}}{Z_\mathrm{Si}}\right)^{-1}\slash f_{\rm dense},
\label{eq:tau_acc_Si}
\end{eqnarray} 
where 
$\mathcal{D}_\mathrm{tot,X} = \mathcal{D}_\mathrm{S,X} + \mathcal{D}_\mathrm{L,X}$
(X = C and Si).
We assume $f_{\rm dense} = 0.5$ as a fiducial value. 
We set $\tau_\mathrm{acc,X} = \infty$ (X = C and Si)
and $\tau_\mathrm{co} = \infty$ if
the particles do not satisfy the criteria 
($n_\mathrm{gas} \ge 10 ~\mathrm{cm}^{-3}$
and $T_\mathrm{gas} < 10^3$ K).

\subsection{Extinction curve}\label{subsec:extinc_curve}

\begin{table}%table
\caption{
Parameters adopted in equations (\ref{timeL})--(\ref{eq:tau_acc_Si}).
In this work, we set the solar elemental abundance pattern 
($Z_\mathrm{X\odot}$) for C and Si, 
and assume that Si occupies a \textbf{mass} fraction of 
0.166 in silicate.
$\rho_\mathrm{X}$ is the material density, and
$f_\mathrm{in,X}$ is condensation efficiency 
in stellar ejecta (see Section \ref{subsubsec:stellar})
}
\label{table:Adopted_quantities}
\begin{center}
\begin{minipage}{130mm}
\begin{tabular}{lccccccc}
\hline
Species & X & $Z_\mathrm{X\odot}$ & $\rho_\mathrm{X}~(\mathrm{g~cm^{-3}})$ & $f_\mathrm{in,X}$ \\
\hline
Carbonaceous dust & C & $2.47 \times 10^{-3}$ & 2.24 & 0.34  \\
Silicate & Si &  $8.17 \times 10^{-4}$ & 3.5 & 0.13 \\
\hline
\end{tabular}
\end{minipage}
\end{center}
\end{table}%

We calculate the extinction curve for each gas particle in the following way.
To calculate the extinction curve, grain size distribution 
is required; however, the whole range of grain size is represented by
the large and small grain populations in our simulation.
Thus, we need to assume a specific functional form for the grain
size distribution of each population.
Following \citet{Hirashita:2015aa}, we adopt a modified-lognormal function for the
grain size distribution. 
In the $j$th particle, the grain size distribution is 
(we omit the subscript $j$):
\begin{equation}
  n_{i,\mathrm{X}}(a)=\frac { C_{i,\mathrm{X}} }{ a^{ 4 } } \exp\left\{ - \frac { \left[ \ln(a/a_{ 0,i }) \right] ^{ 2 } }{ 2\sigma^2 }  \right\},
\end{equation}
where subscript $i$ indicates the small ($i = \mathrm{S}$) or large 
($i = \mathrm{L}$) grain component, subscript X represents dust species
[carbonaceous dust (X = C) or silicate (X = Si)], 
$C_{i,\mathrm{X}}$ is the normalization 
constant for the $j$th particle and 
$a_{0,i}$, and $\sigma$ are the central grain radius 
and the standard deviation of the lognormal part, respectively.
We adopt $a_\mathrm{0,s} = 0.005~\micron$, 
$a_\mathrm{0,l} = 0.1~\micron$ and $\sigma = 0.75$
since these values reproduce the Milky Way extinction curve
when the small-to-large grain abundance ratio is the 
same as the \citet[][MRN]{Mathis:1977aa} size distribution \citep{Hirashita:2015aa}.
The normalization $C_{i,\mathrm{X}}$ is determined by
\begin{equation}
  {\mu}m_\mathrm{H}\mathcal{D}_{i,\mathrm{X}}=\int_{0}^{\infty}\frac{4}{3}{\pi}a^3\rho_\mathrm{X}n_{i,\mathrm{X}}(a)\,\mathrm{d}a,
\end{equation}
where $\mu = 1.4$ is the gas mass per hydrogen nucleus, and
$m_\mathrm{H}$ is the mass of hydrogen atom.

The extinction $A_{\lambda,\mathrm{X}}$ (in units of magnitude) normalized to the column density of hydrogen nuclei ($N_\mathrm{H}$) is written as
\begin{equation}
  \frac{A_{\lambda,\mathrm{X}}}{N_\mathrm{H}}=2.5\,\log_{10} \mathrm{e}\, \sum_{i}\int_{0}^{\infty}n_{i,\mathrm{X}}(a){\pi}a^2Q_\mathrm{ext}(a,\lambda,\mathrm{X})\,\mathrm{d}a,
\end{equation}
where %%e is the mathematical constant and 
$Q_\mathrm{ext}(a,\lambda,\mathrm{X})$ is the extinction coefficient 
(extinction cross section normalized to the geometric cross section) as a 
function of grain size, wavelength and dust species. 
$Q_\mathrm{ext}(a,\lambda,\mathrm{X})$ is calculated by using the Mie theory 
\citep{Bohren:1983aa} based on the same optical constants for silicate and 
graphite as in \citet{Weingartner:2001aa} unless otherwise stated, 
on the assumption that all the grains 
are spherical with a uniform composition.

\section{Results}

The new features calculated by our models are the time evolution
and the spatially resolved
information of extinction curves. In this section, we describe
how the calculated extinction curves depend on
various quantities that characterize the environmental or
evolutionary properties.
We focus on the silicate--graphite modelling as a standard
\citep{Draine:1984aa,Pei:1992aa},
but the qualitative behaviour of overall steepness
(other than the carbon bump) does not change even if we
adopt other dust species (we mention other possible dust species
in Section \ref{subsec:MCs}).

Following Paper I, we adopt the ages 0.3, 1, 3, and 10\,Gyr
to represent four galaxy evolution stages that cover a wide range of
specific star formation rate (sSFR), 3--0.1\,Gyr$^{-1}$.
To discuss the extinction curves, we mainly focus on two features, 
2175 \AA\ bump and far-UV (FUV) slope.
Since the extinction curves reflect the variation
of grain size distribution, the small-to-large grain abundance ratio ($\stol$)
determines the extinction curve in our model.

\subsection{Radial dependence}\label{result:radial}

\begin{figure*}%figure
	\begin{center}
	\includegraphics[width=0.33\textwidth]{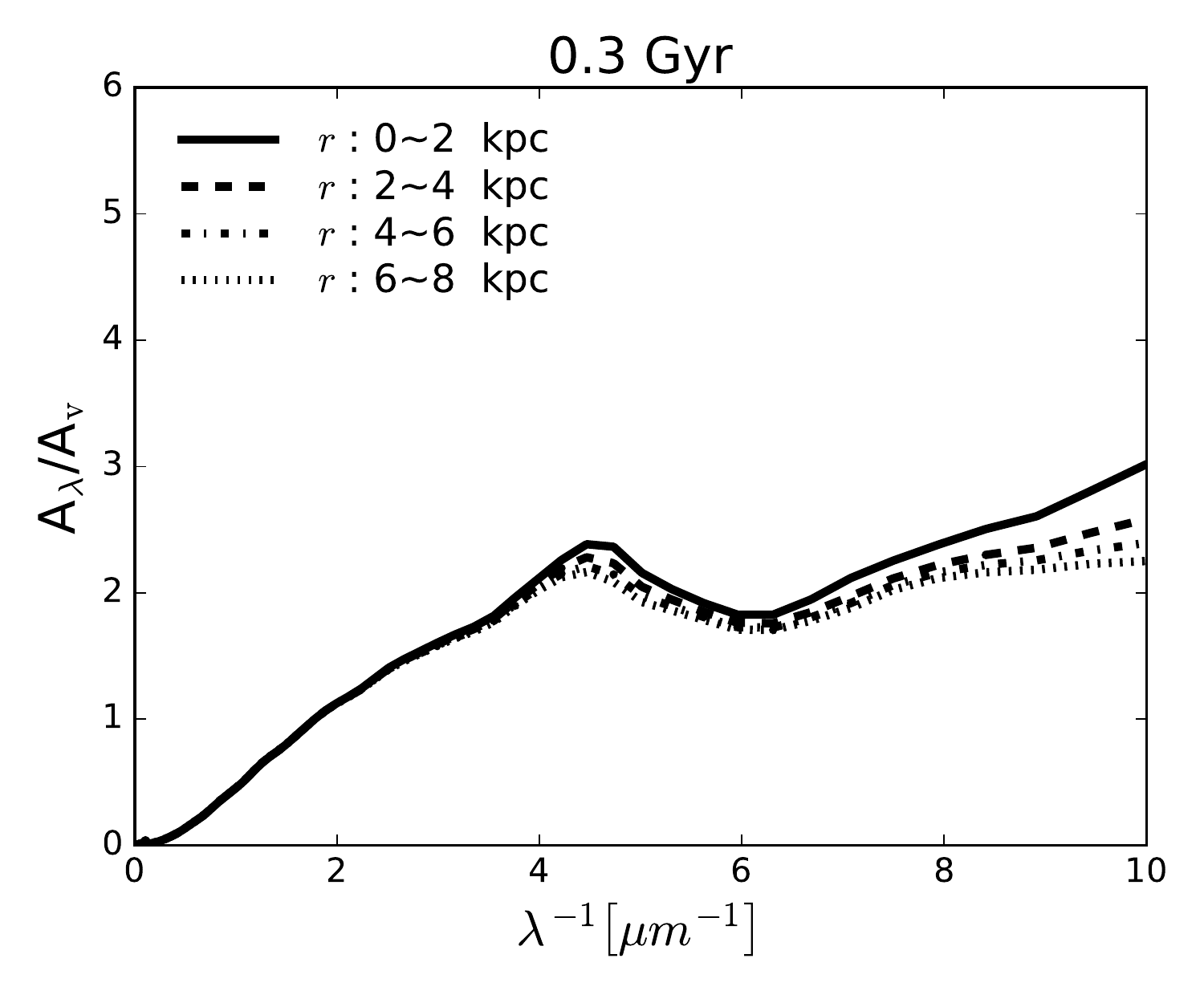}
	\includegraphics[width=0.285\textwidth]{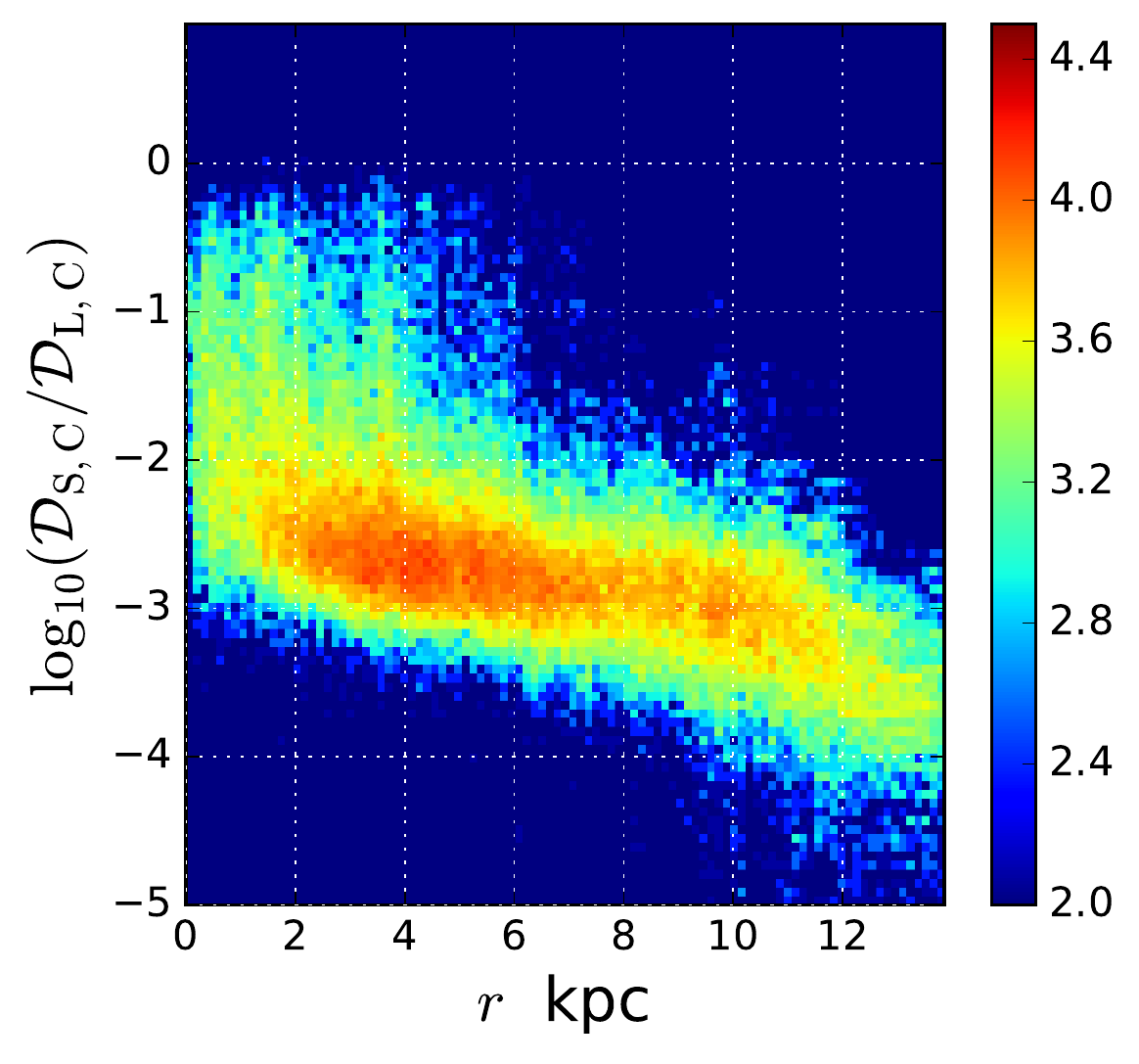}
	\includegraphics[width=0.285\textwidth]{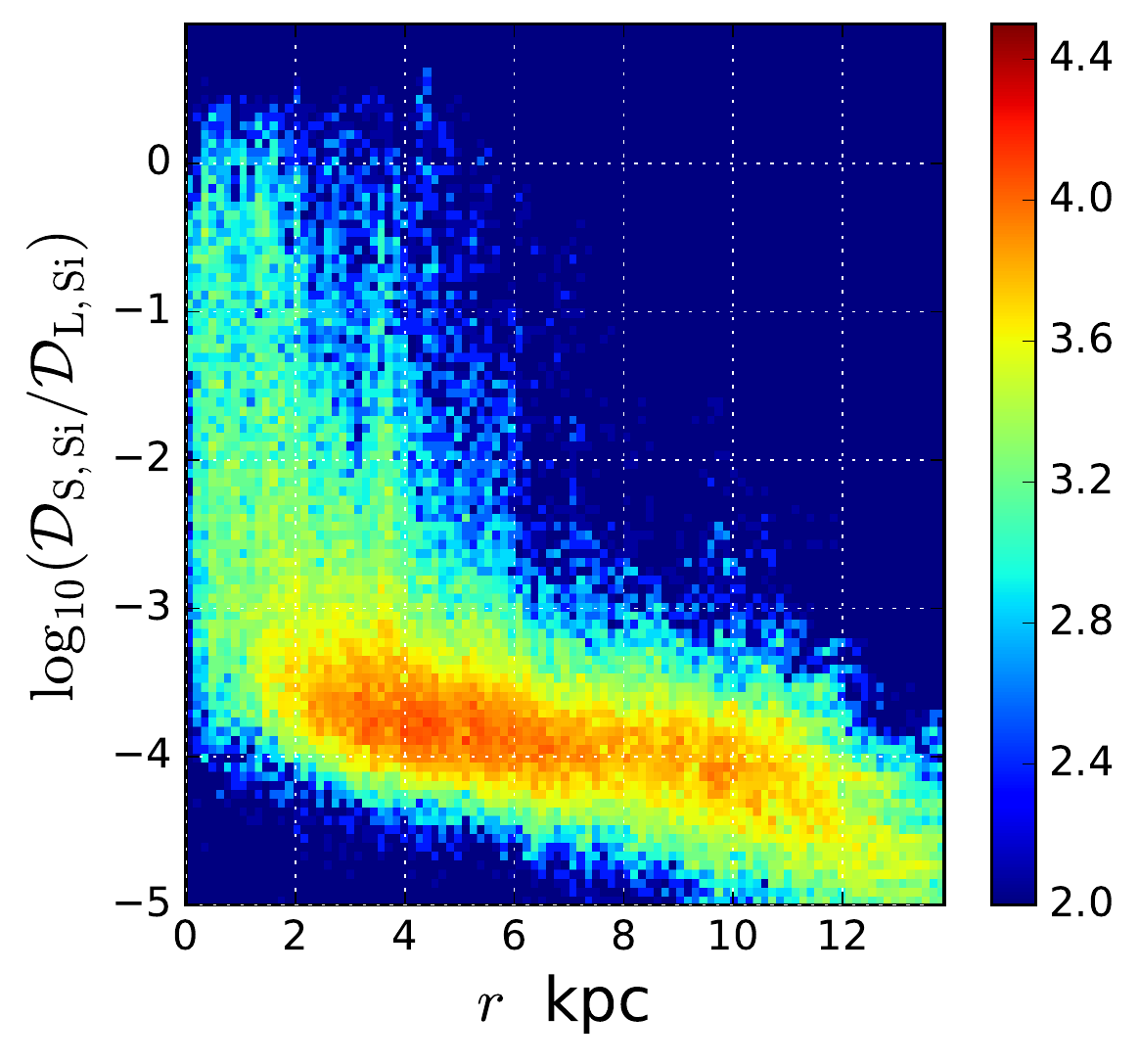}

	\includegraphics[width=0.33\textwidth]{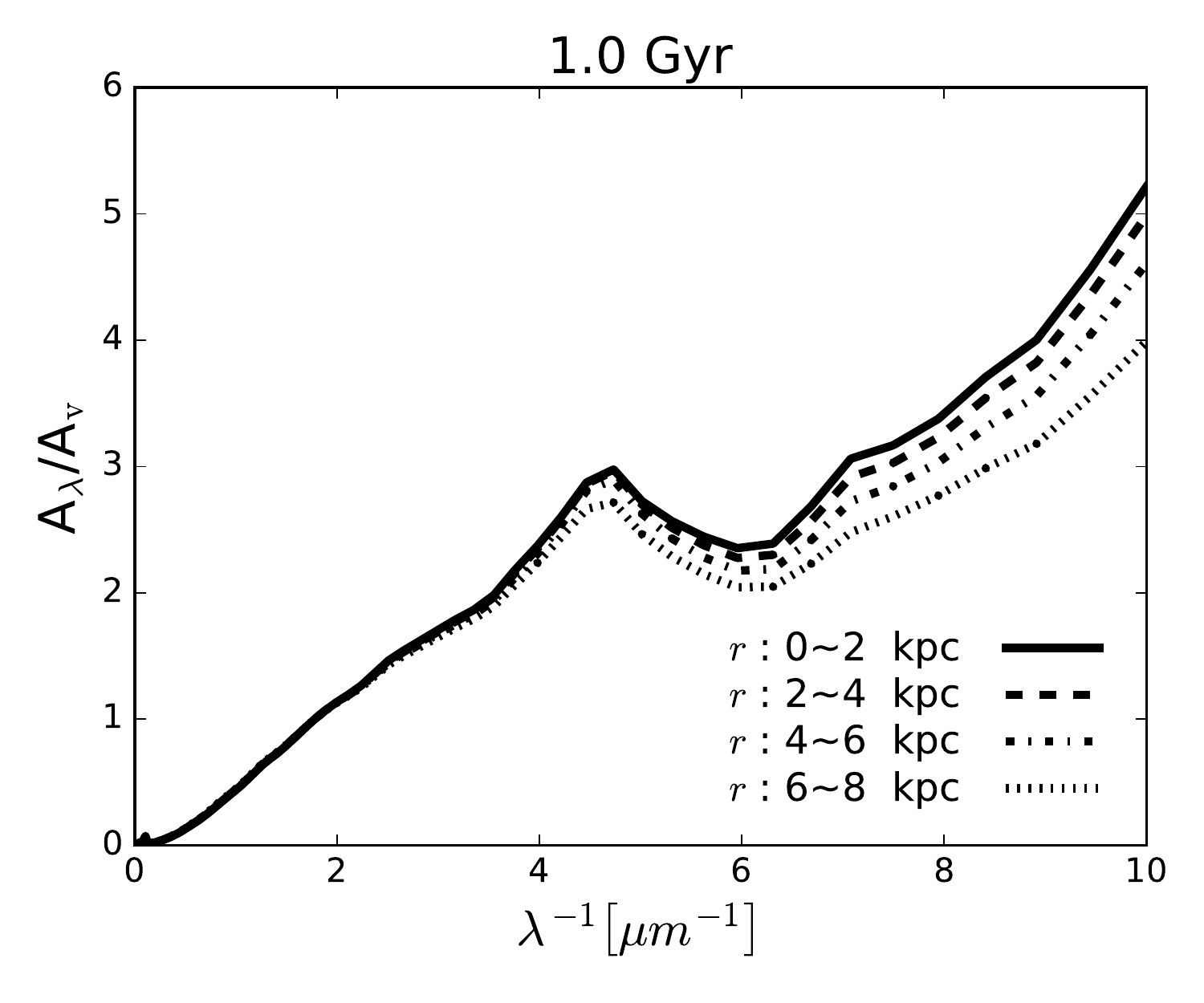}
	\includegraphics[width=0.285\textwidth]{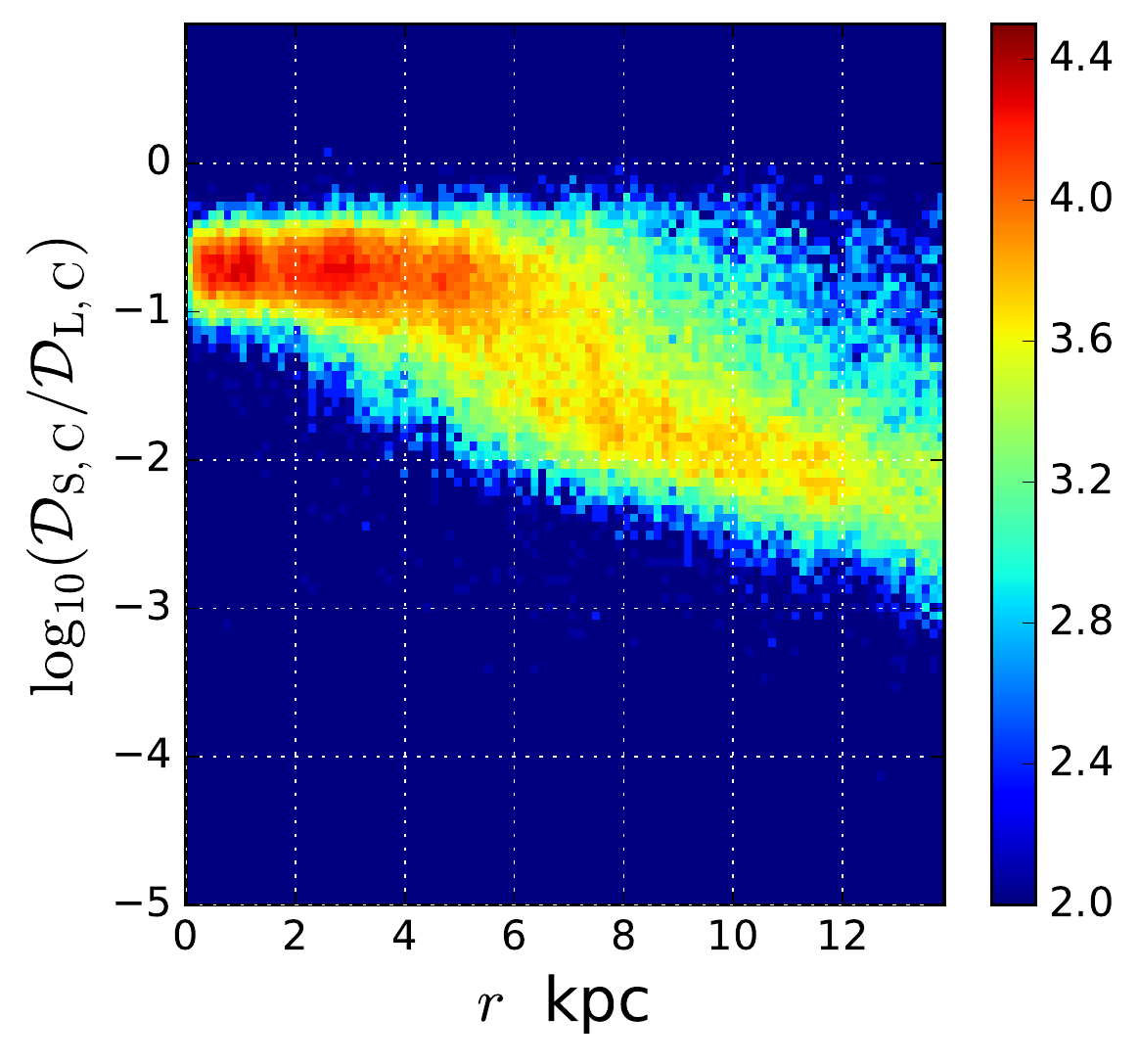}
	\includegraphics[width=0.285\textwidth]{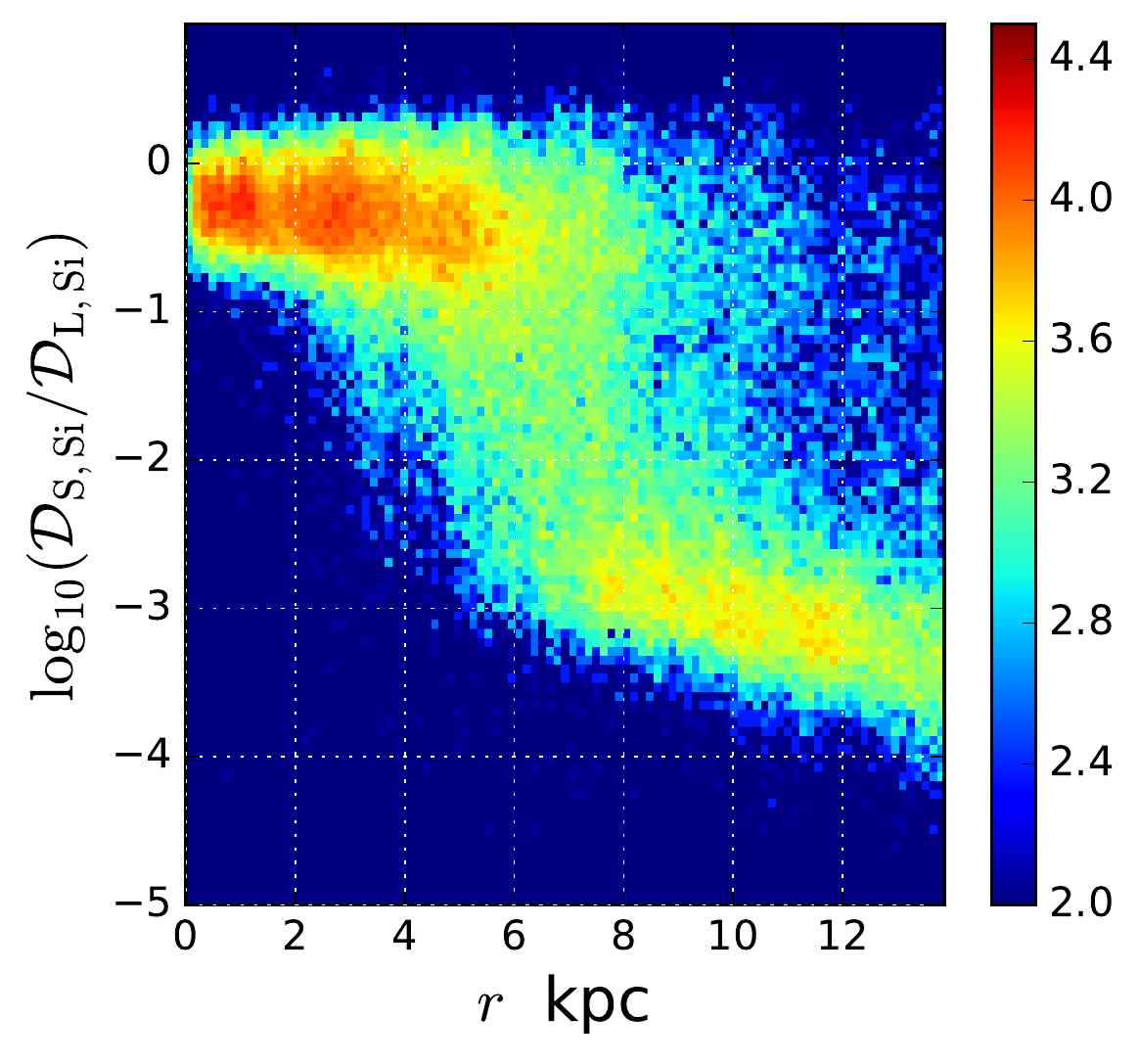}

	\includegraphics[width=0.33\textwidth]{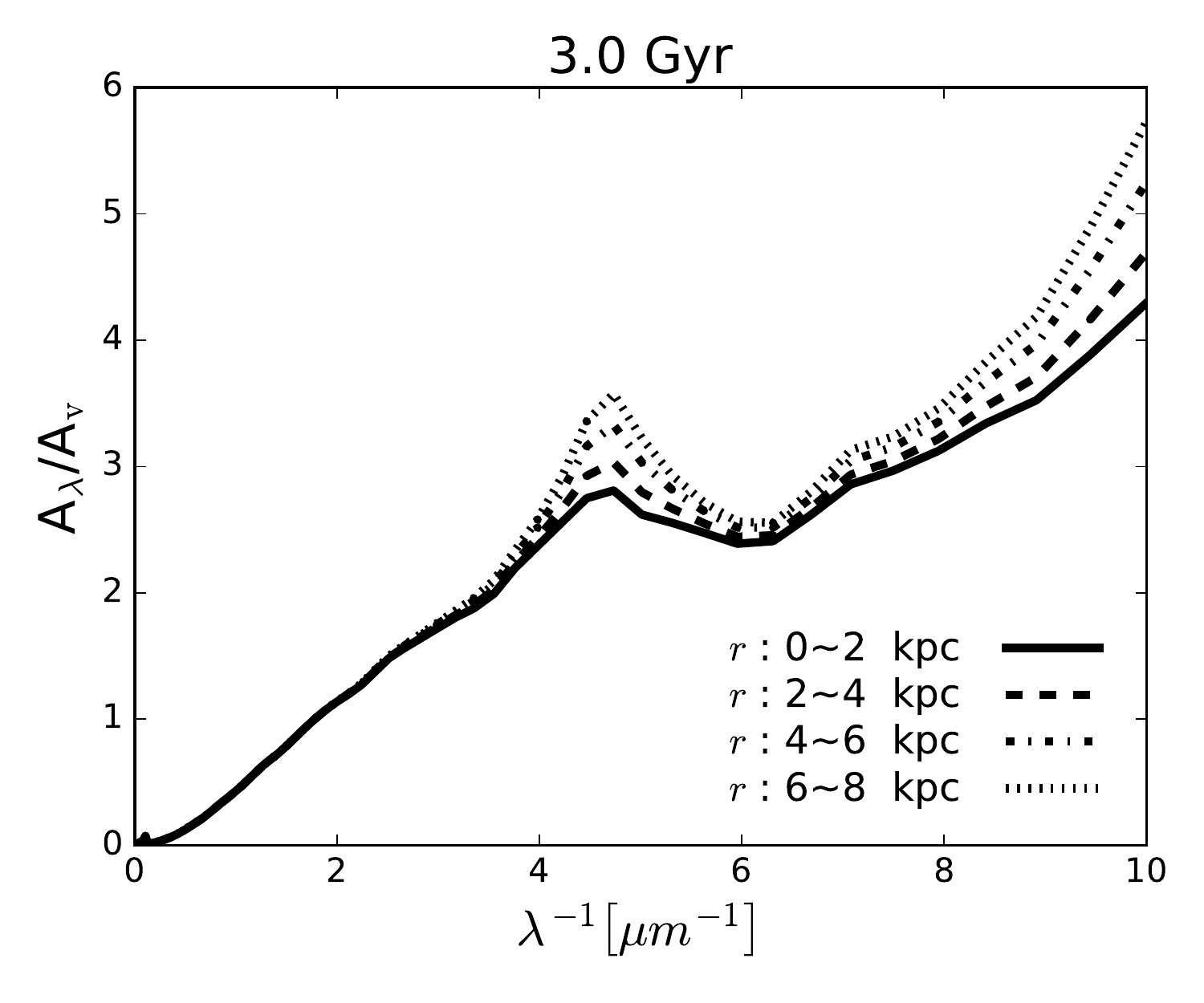}
	\includegraphics[width=0.285\textwidth]{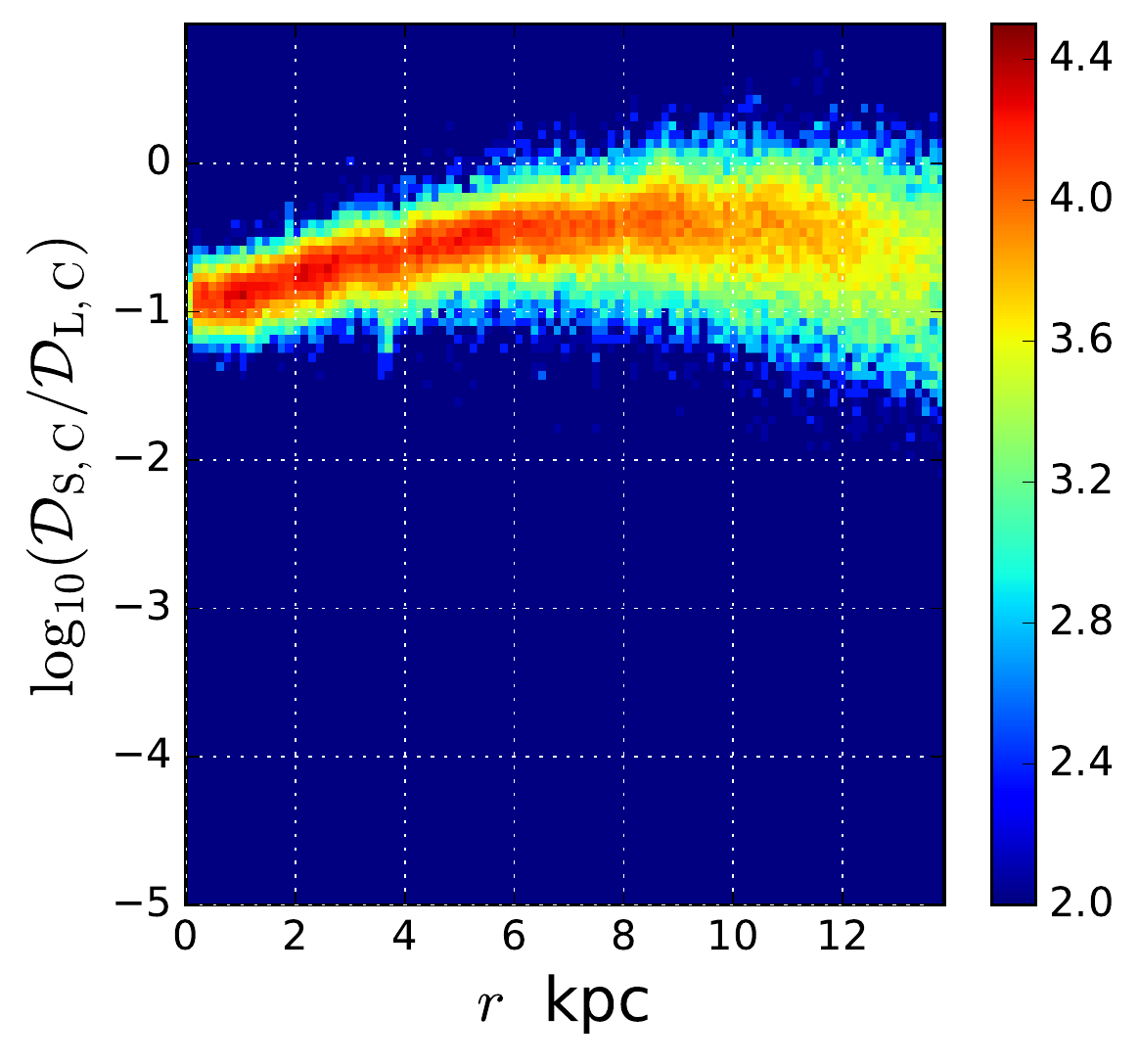}
	\includegraphics[width=0.285\textwidth]{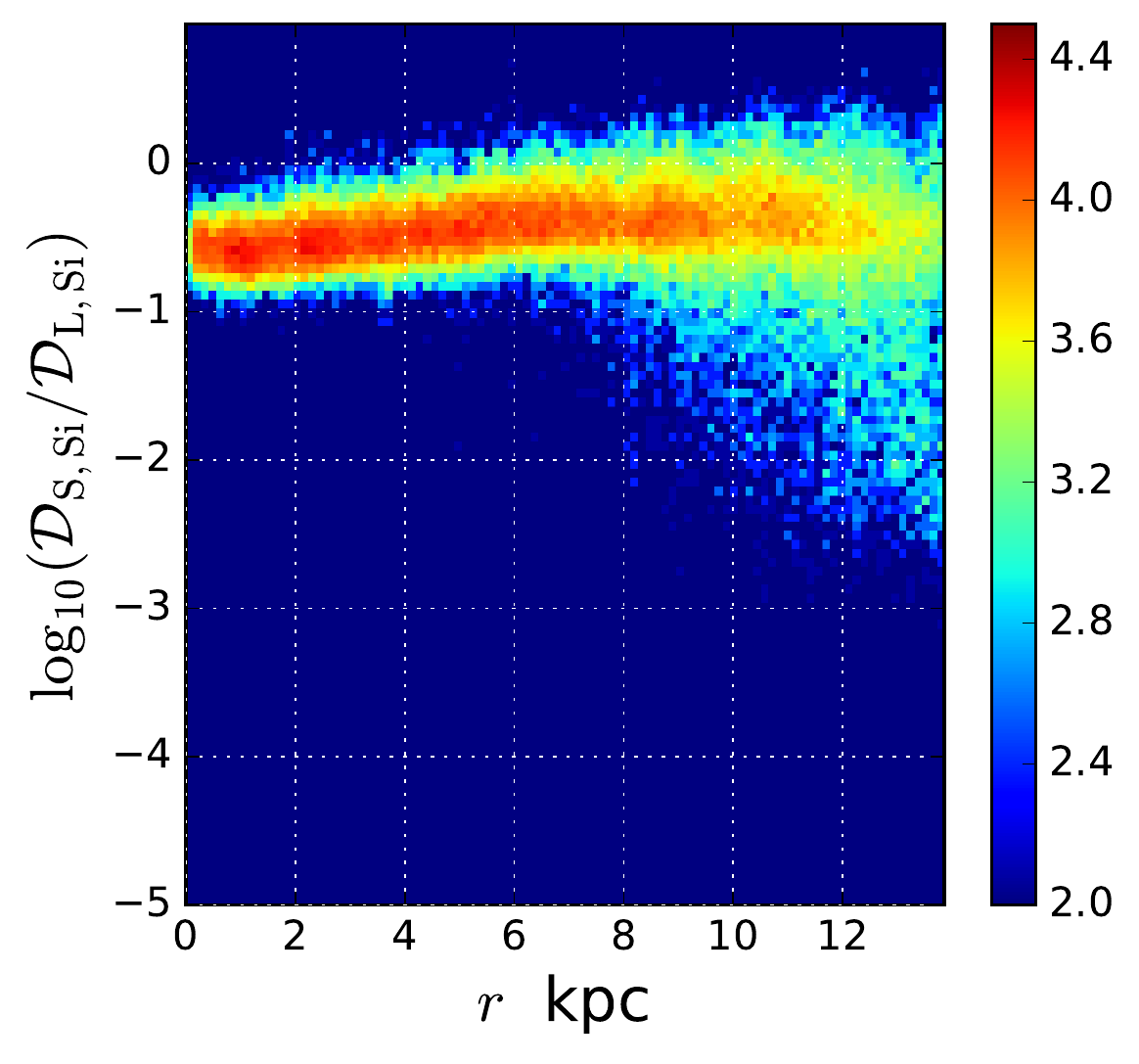}

	\includegraphics[width=0.33\textwidth]{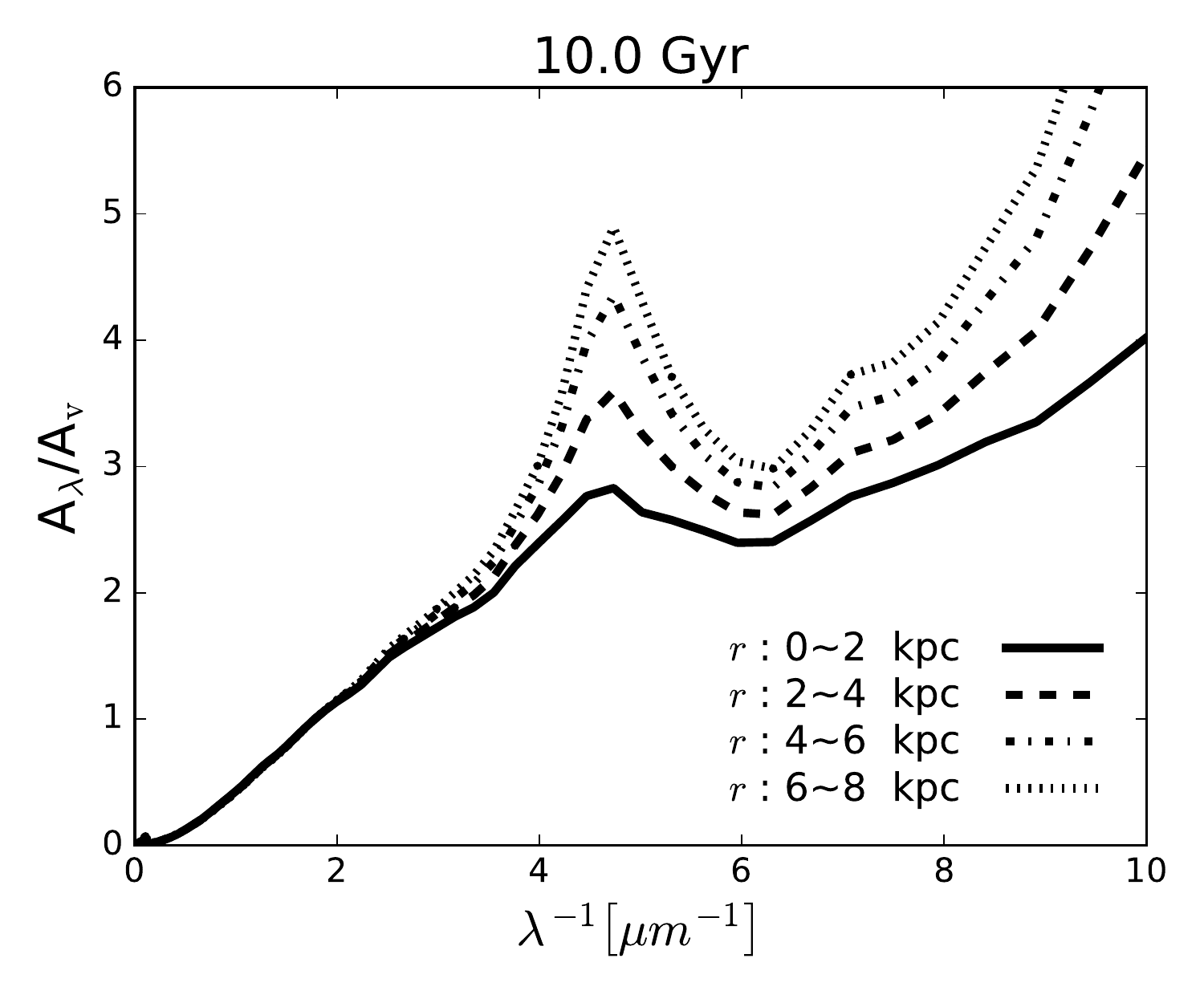}
	\includegraphics[width=0.285\textwidth]{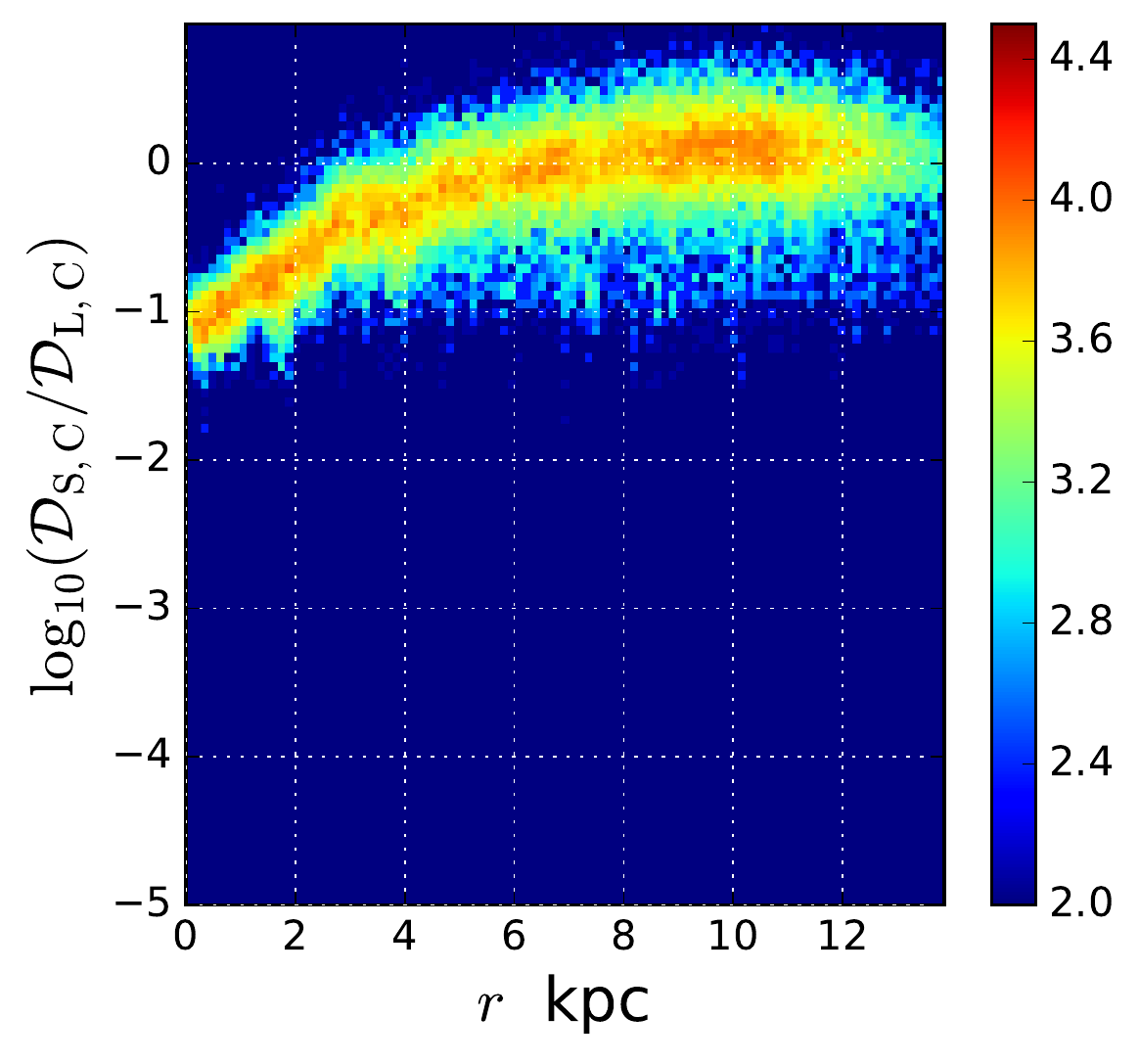}
	\includegraphics[width=0.285\textwidth]{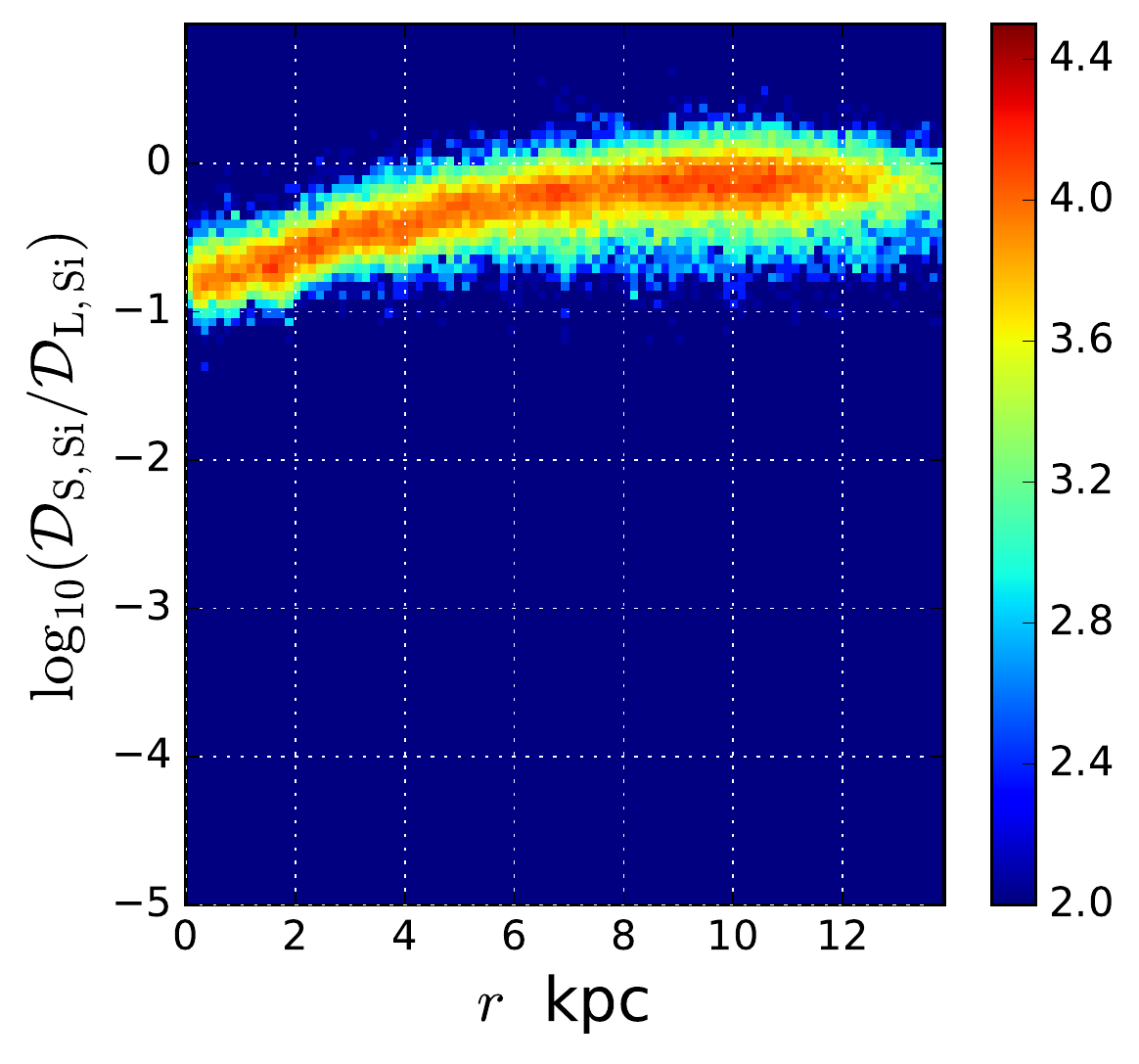}
	\caption{Extinction curves and small-to-large grain abundance ratio 
($\stol$) as a function of the radius.
Columns from left to right show the extinction curves, 
the distribution of $\stol$ of carbonaceous dust, 
and the distribution of $\stol$ of silicate.
In the left panels, 
the solid, dashed, dash-dotted, and dotted lines represent 
the extinction curves in radial ranges, 0--2, 2--4, 4--6 and 6--8 kpc, respectively.
The colour indicates the logarithmic surface density of gas particles in relative units
as indicated by the colour bar.
From top to bottom show the snapshots at 0.3, 1, 3 and 10\,Gyr. }
	\label{Fig:extinction_radii}
	\end{center}
\end{figure*}

We investigate the radial dependence here to represent the spatial
variation. For extinction curves, we usually obtain an average over
a line of sight, which is often through a galactic disc for a disc
galaxy. Thus, for the radial dependence, we consider a scale
larger than the disc scale height (but still
smaller than the disc scale length) by adopting
a radius bin width of 2 kpc and take the mass-weighted 
average for all gas particles in each radius bin.
We analyse extinction curves up to $r$ = 8 kpc 
($r$ is the radius in cylindrical coordinate),
which is close to $R_{25}$ of our simulated galaxy.
The centre of the galaxy is determined by the centre of mass for 
the stellar particles.
Fig.\ \ref{Fig:extinction_radii} shows the extinction curves in 
the four radial bins in the four evolution epochs.

We also show the small-to-large grain abundance ratio as a
function of radial distance for silicate and 
carbonaceous dust (note that we plot the distribution of all gas
particles on this diagram), since this ratio determines the
shape of the extinction curve. In general, a higher
small-to-large grain abundance ratio ($\stol$) shows an
extinction curve with a more prominent 2175~\AA\ bump and
a steeper UV slope. The evolution of $\stol$ has already been
investigated in Paper I, but we repeat some discussions on $\stol$, since
this is the central quantity in interpreting the calculated extinction curves.
We refer the interested reader to Paper I for detailed
discussions on $\stol$.

We observe in Fig.\ \ref{Fig:extinction_radii} that
the features of extinction curves (the 2175\,\AA\ bump
and the UV slope) become stronger with
age, which is caused by the increase of $\stol$.
The small-to-large grain abundance ratio 
increases in the centre first since dust growth by accretion 
becomes the most efficient in the most metal-enriched region.
The regions with high small-to-large grain abundance ratio extend
to the outer radii with age.

At 0.3\,Gyr, only the extinction curve of the central region shows 
a slight FUV rise, and the other regions have flat extinction curves.
A portion of gas particles have high small-to-large grain abundance ratio
in the central region, where some particles are dense and
already metal-rich so that accretion already enhances the $\stol$.
The small-to-large grain abundance ratios at outer radii 
are smaller than $10^{-2}$ and $10^{-3}$ for carbonaceous dust and 
silicate, respectively. The low $\stol$ is due to the dominance of
stellar dust production, and the existence of small amounts of small
grains is explained by shattering.

At 1\,Gyr, the extinction curves at all radii show a prominent 2175 \AA~
bump and FUV rise, and 
these extinction curve features are the strongest in the 
central part of the galaxy.
The small-to-large grain abundance ratio is greater than $10^{-1}$ at
$r \lesssim 6$ kpc. There is a transition of the small-to-large grain
abundance ratio at $6 \lesssim r \lesssim 8$ kpc, and silicate shows
a stronger transition than carbonaceous dust
(see Section \ref{subsection:c_to_si} for further discussions).
This clear rise of $\stol$ at $r \sim$ 6--8 kpc
is caused by a sensitive dependence of accretion on
metallicity. Therefore, the strong radial dependence of extinction
curves is driven by accretion in this stage.

The radial variation of extinction curves at 3\,Gyr shows 
the opposite trend  to that at 1\,Gyr, with outer regions 
having steeper extinction curves.
The small-to-large grain abundance ratio shows a decrease toward
the centre. Because accretion dominates the dust evolution in the entire
galaxy regions, the
small-to-large grain abundance ratio is as high as $\stol$ $\gtrsim 0.1$.
The decrease of $\stol$ in the central region is caused by coagulation
since the ISM is relatively dense in the central region.

At 10\,Gyr, the features in the extinction curves become more
prominent than at 3\,Gyr except in the centre.
At this stage, the small grain abundance is almost comparable to
the large grain abundance ($\stol$ $\sim 1$) in the outer disc, 
and $\stol$ is smaller in the centre than at 3\,Gyr.
The highest value of $\stol$ gradually increases
at $t \gtrsim $ 3\,Gyr. We notice that shattering continuously converts
large grains to small grains, while coagulation becomes
less efficient because of the general trend that the gas becomes less dense in the late
evolutionary stage mainly as a result of gas consumption by star
formation.

\subsection{Density dependence}\label{result:density}

\begin{figure*}%figure
	\begin{center}
	\includegraphics[width=0.33\textwidth]{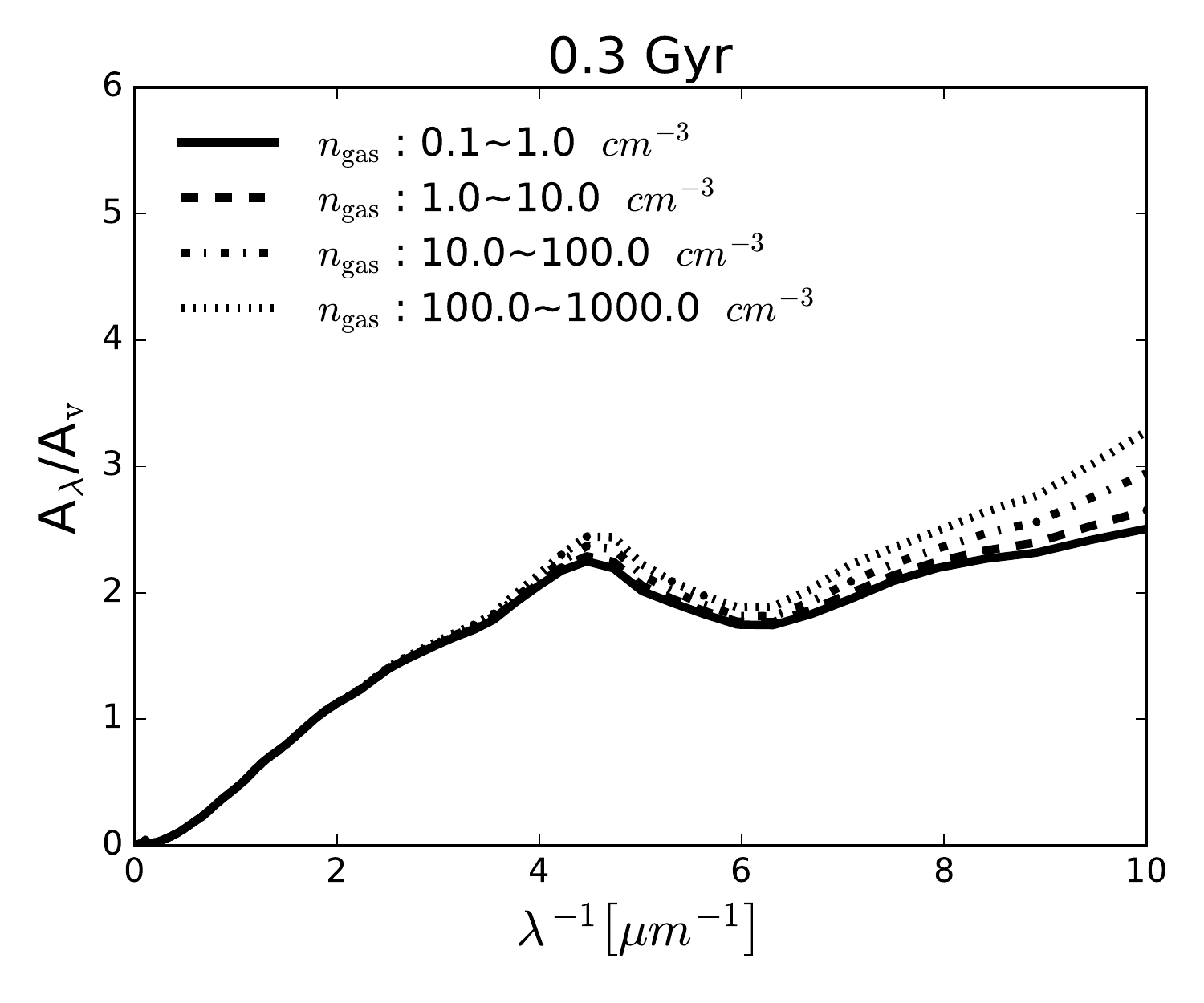}
	\includegraphics[width=0.285\textwidth]{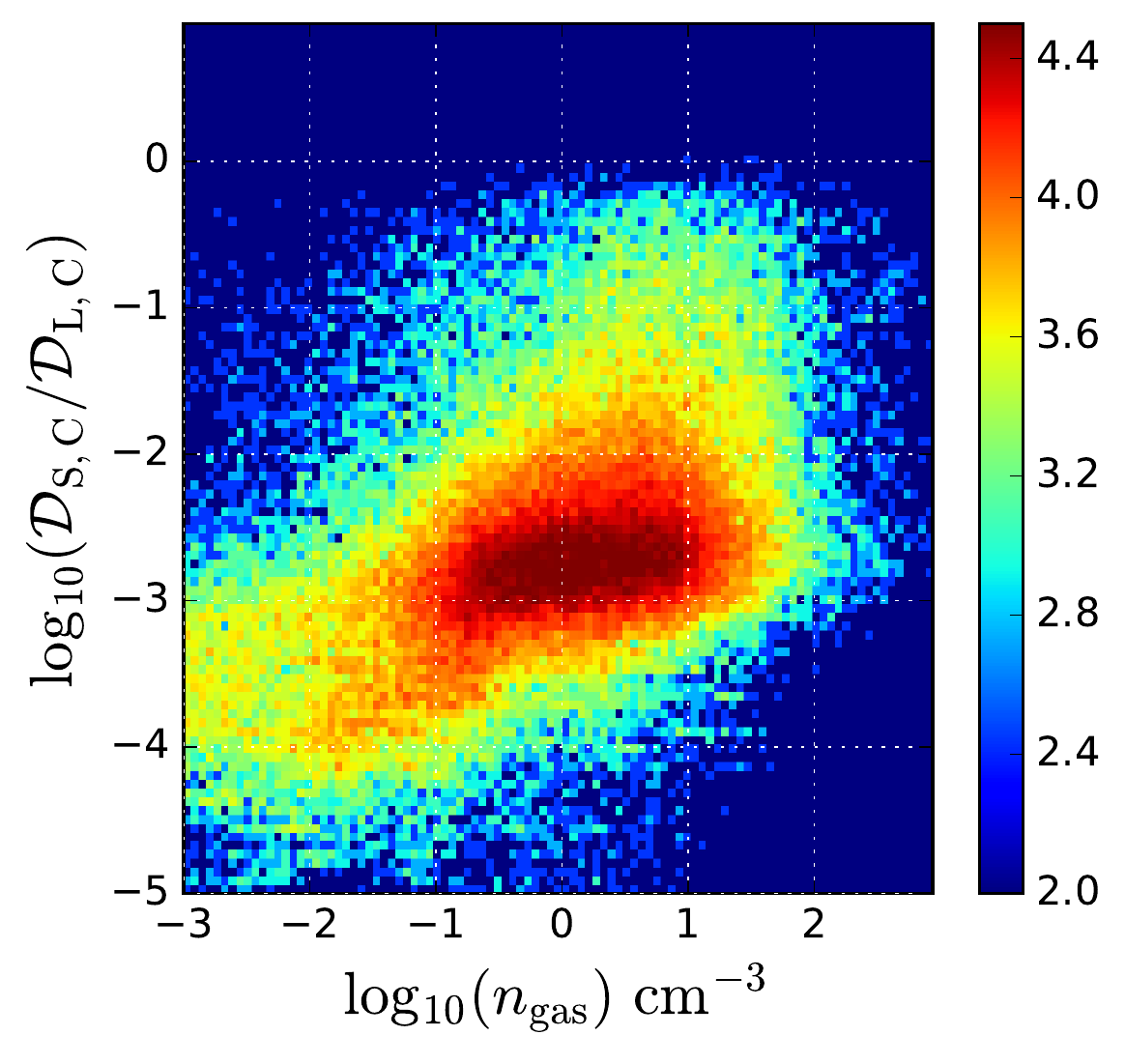}
	\includegraphics[width=0.285\textwidth]{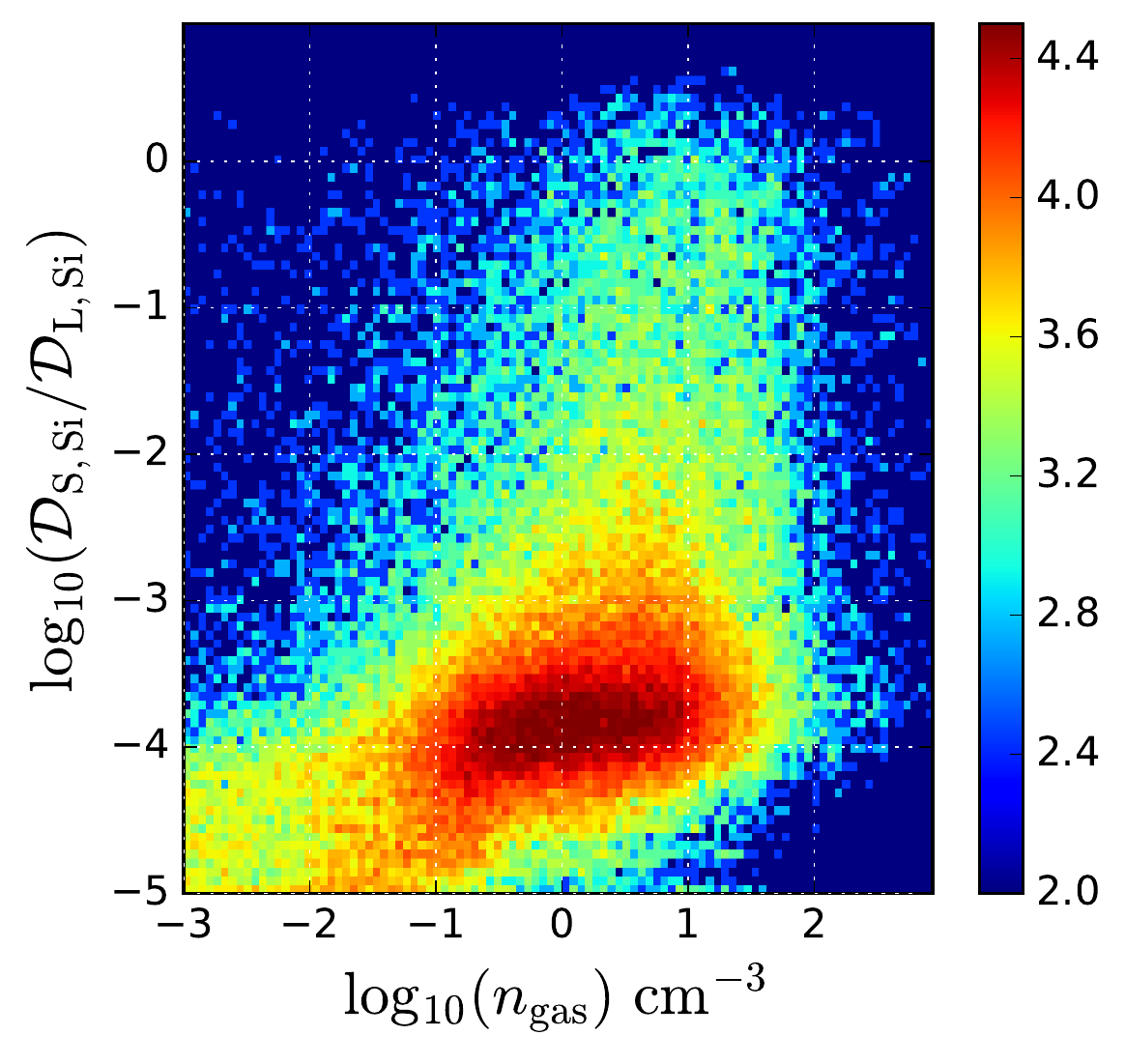}

	\includegraphics[width=0.33\textwidth]{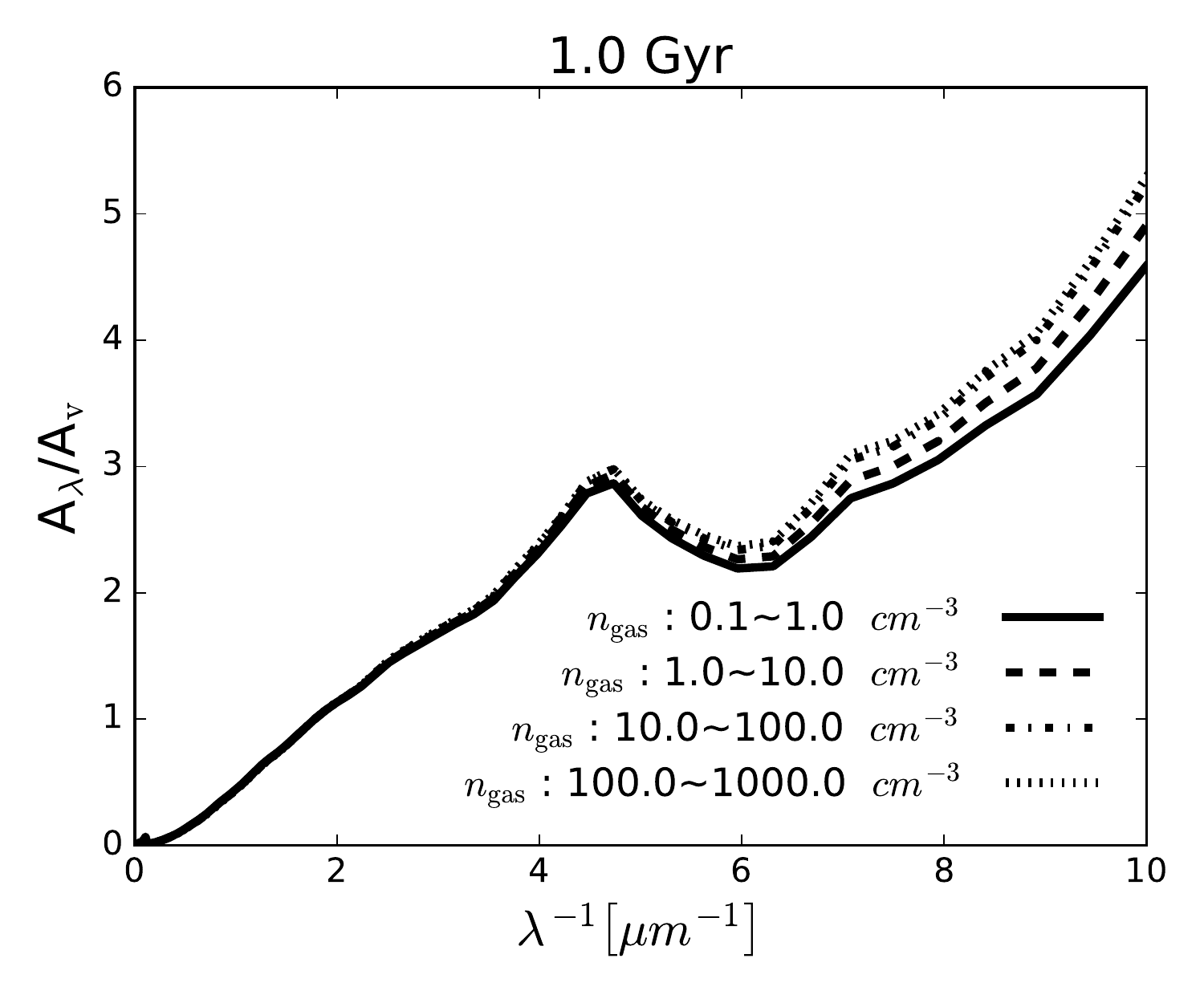}
	\includegraphics[width=0.285\textwidth]{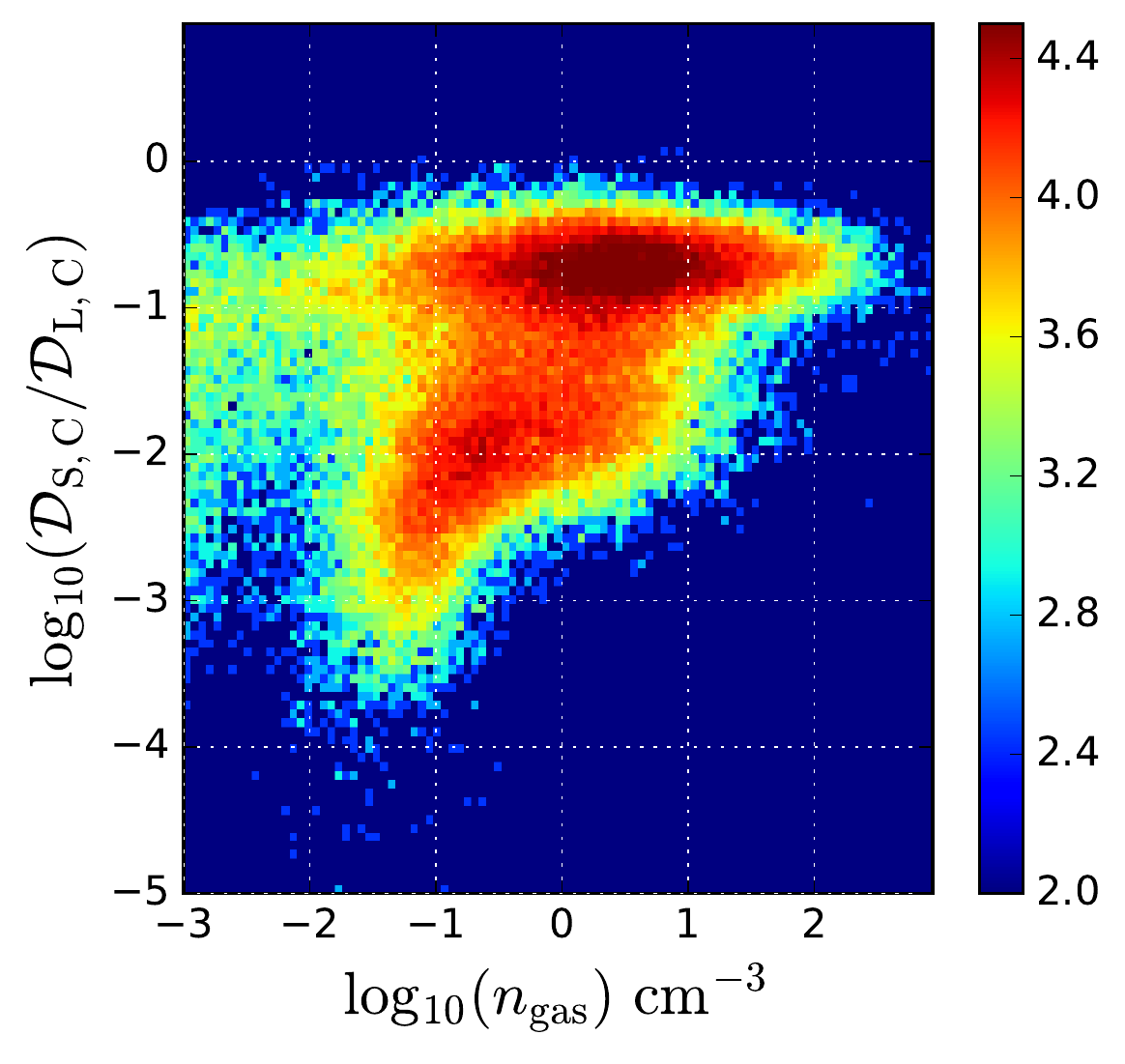}
	\includegraphics[width=0.285\textwidth]{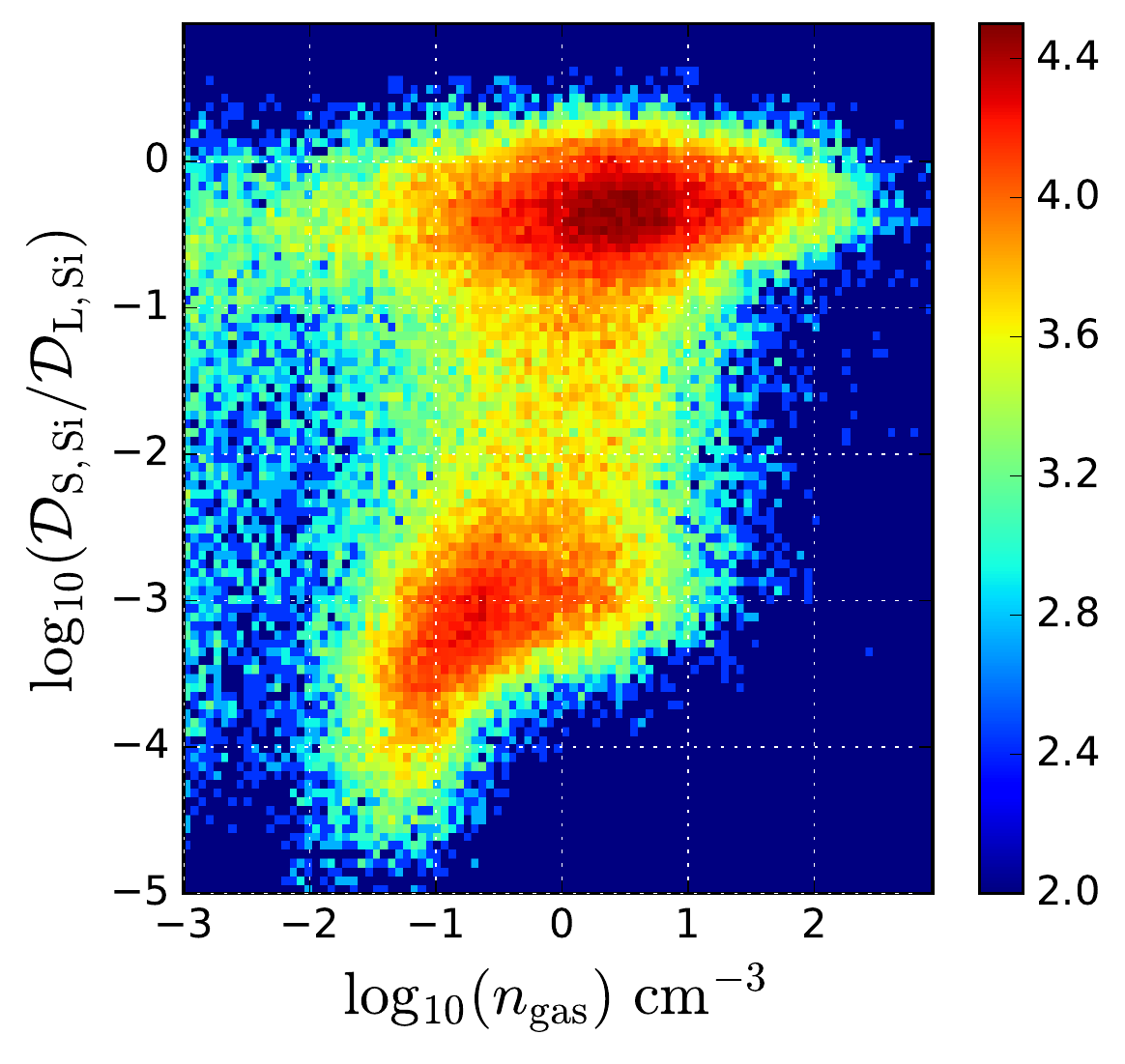}

	\includegraphics[width=0.33\textwidth]{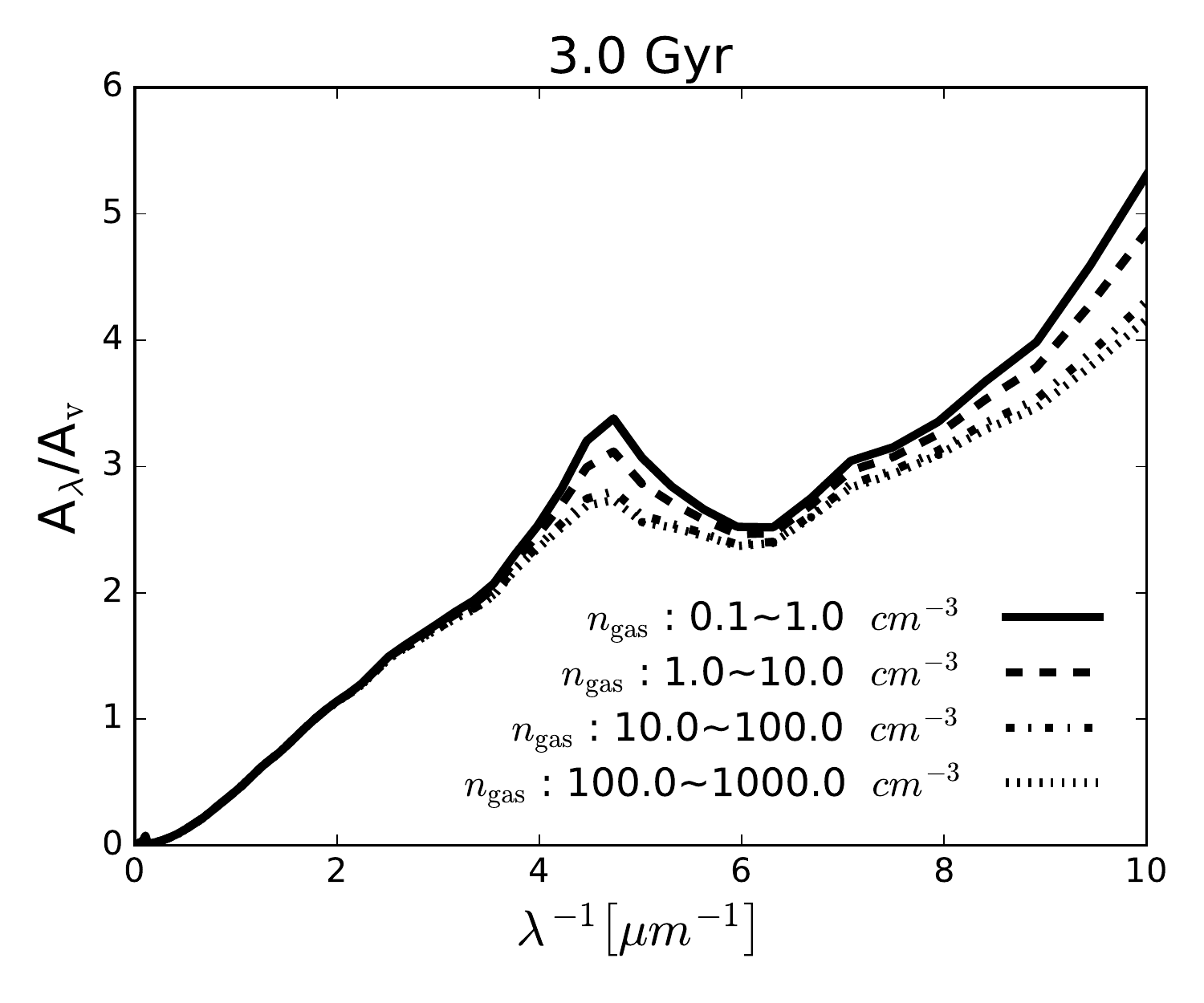}
	\includegraphics[width=0.285\textwidth]{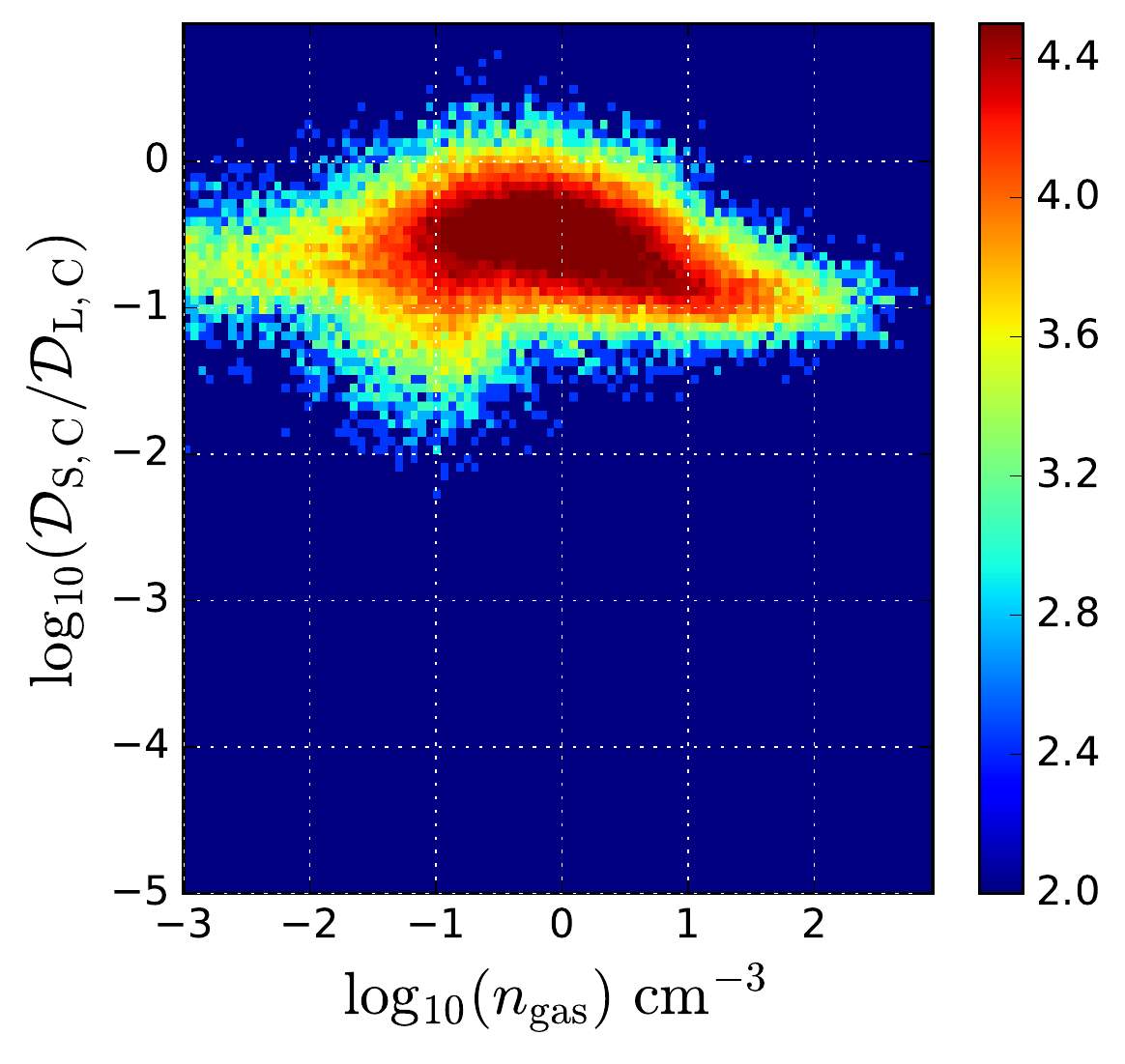}
	\includegraphics[width=0.285\textwidth]{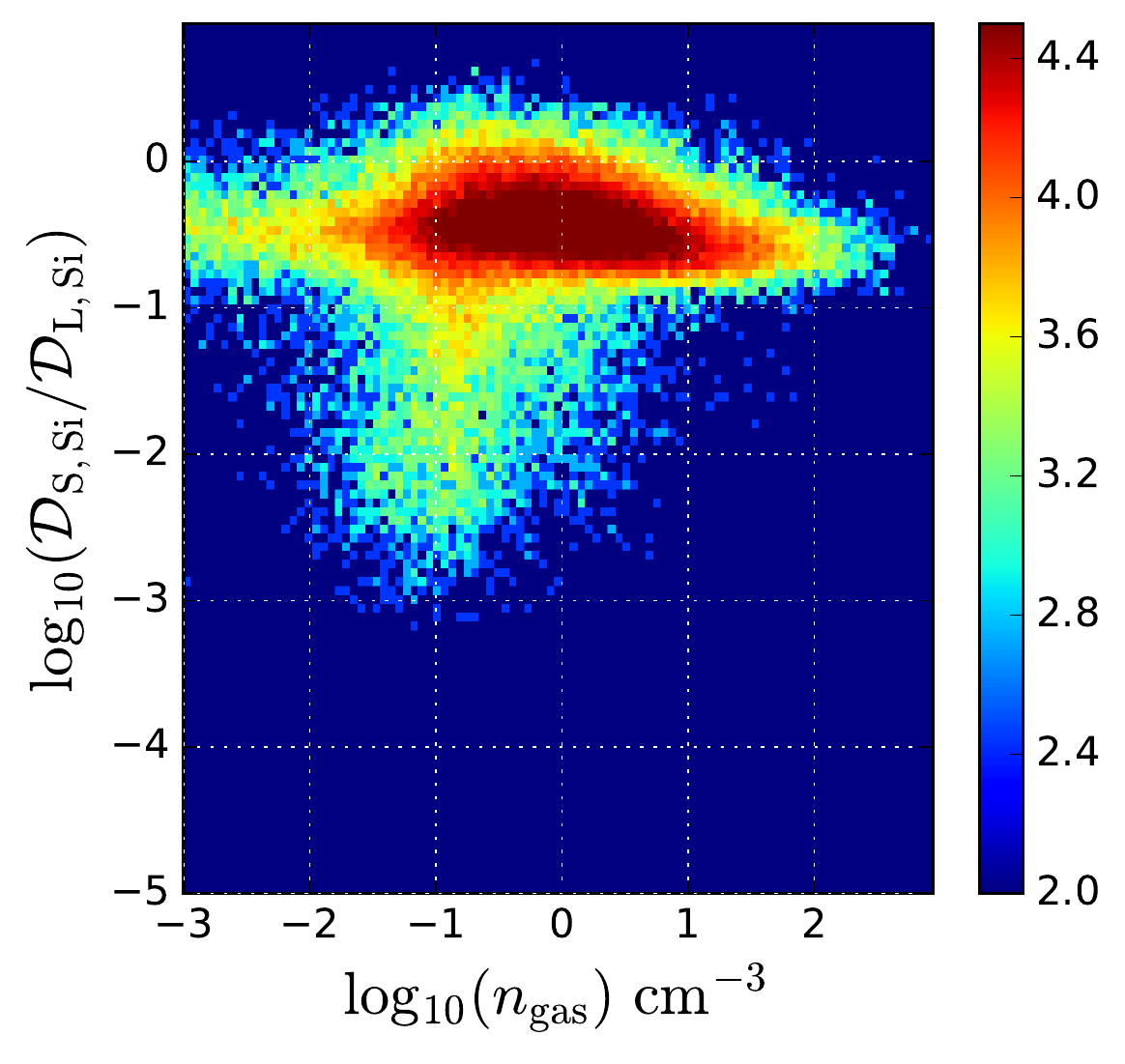}	
	
	\includegraphics[width=0.33\textwidth]{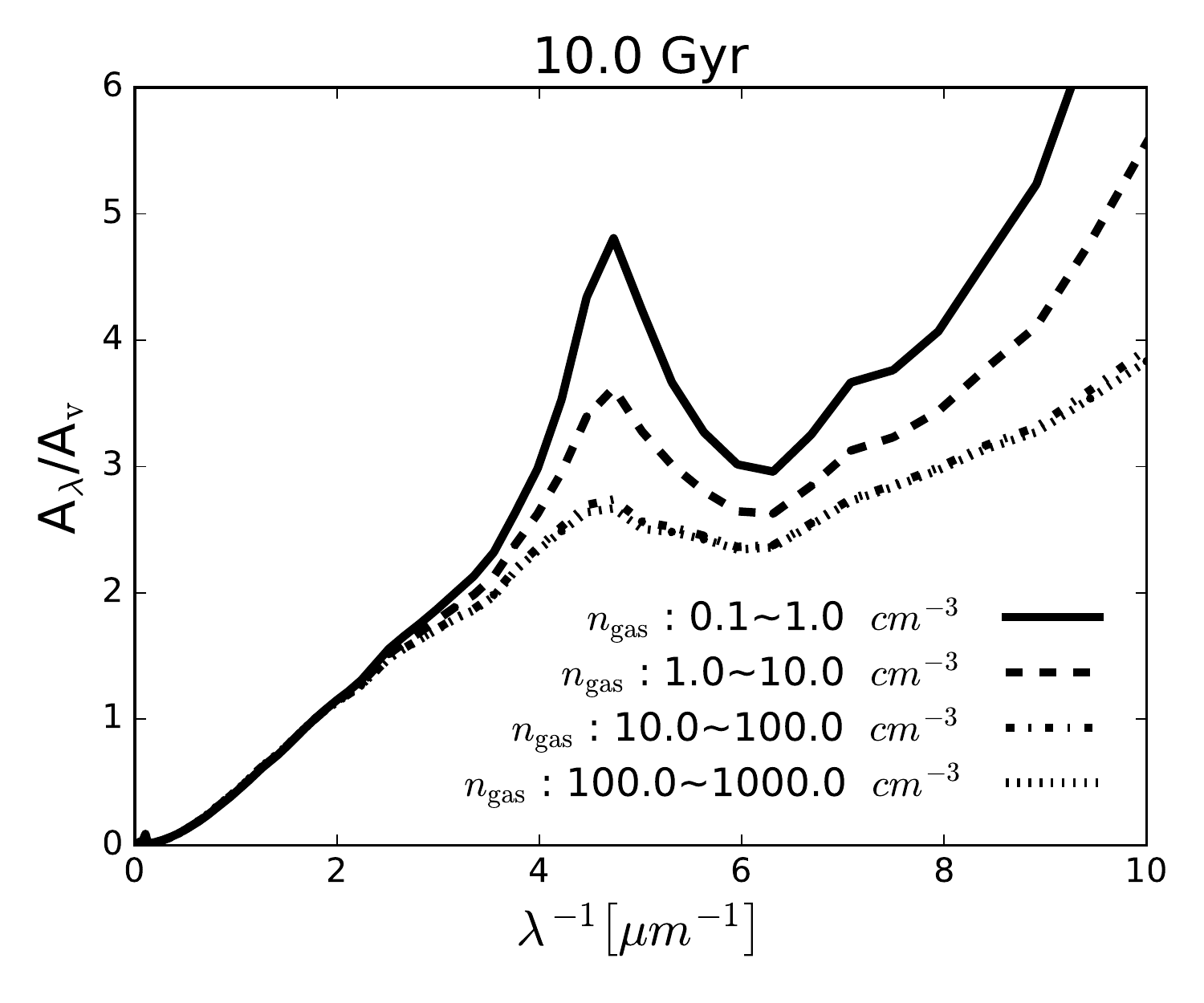}
	\includegraphics[width=0.285\textwidth]{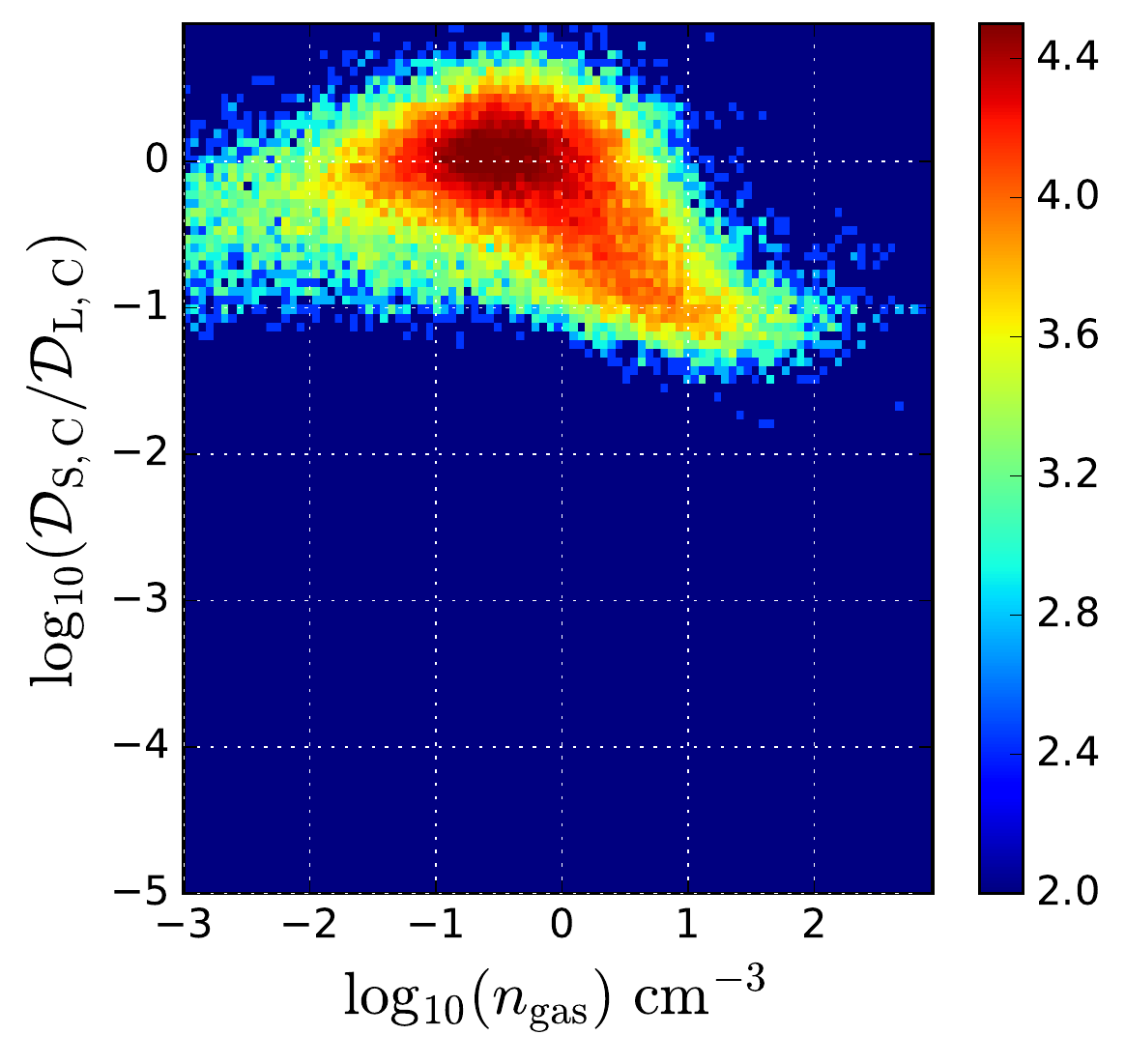}
	\includegraphics[width=0.285\textwidth]{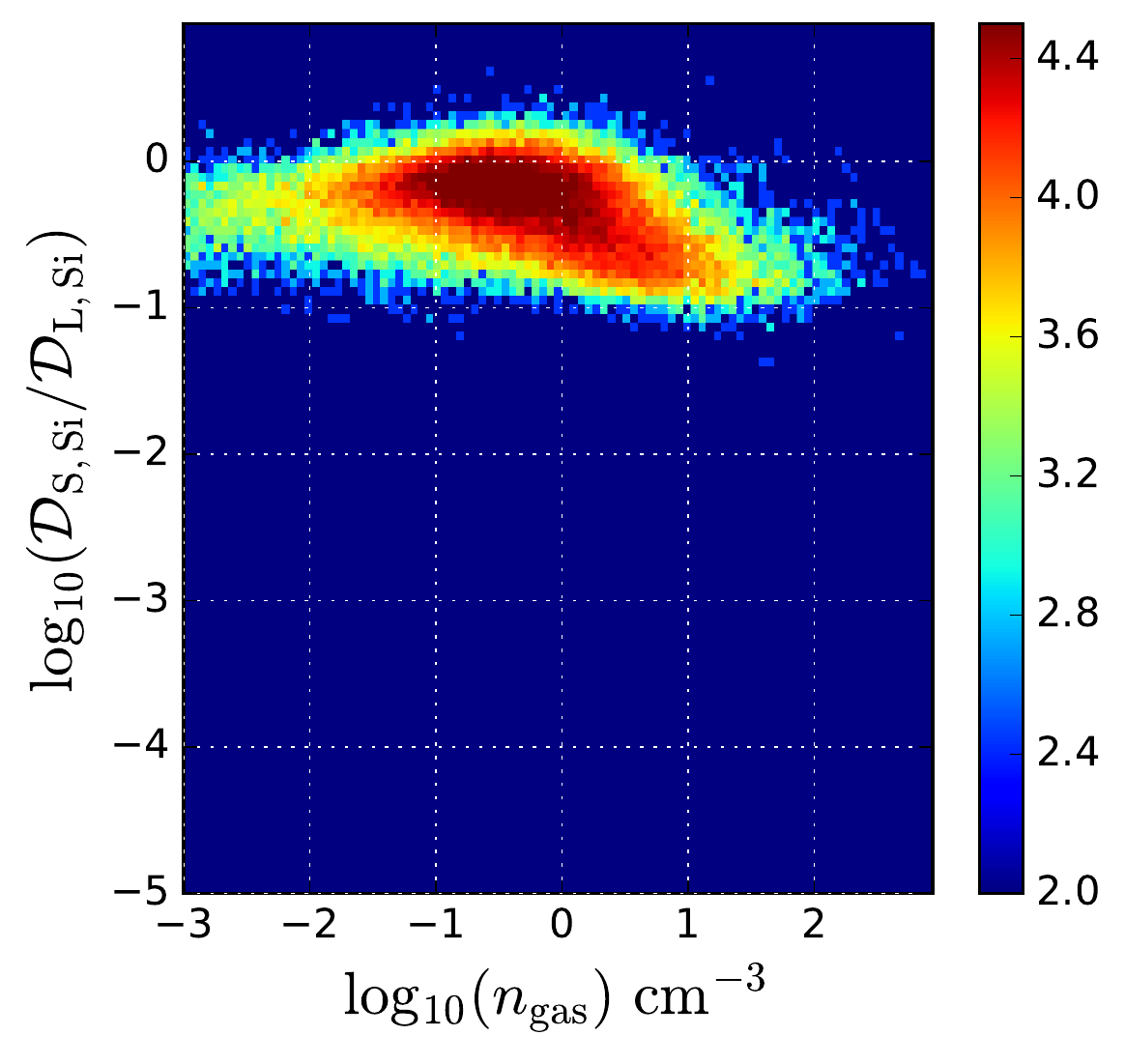}

	\caption{Same as Fig. \ref{Fig:extinction_radii} but gas density dependence.
	In the left panels, the solid, dashed, dash-dotted, 
	and dotted lines represent the extinction curves in gas 
	density ranges, 0.1--1, 1--10, 10--$10^2$ and $10^2$--$10^3$ cm$^{-3}$, respectively.}
	\label{Fig:extinction_density}
	\end{center}
\end{figure*}

In our model, the efficiencies of shattering, coagulation and accretion have 
gas density dependence, so it is worth investigating how the extinction curve
varies in different density environments. To this aim, we separate the
gas particles according to the logarithmic density in
$n_\mathrm{gas} = 0.1$ to $10^3$ cm$^{-3}$ and derived
the mass-weighted average in each density bin. The
resulting extinction curves are shown in
Fig.\ \ref{Fig:extinction_density}.

In the early evolutionary stage ($t \sim 0.3$\,Gyr), all the density ranges show 
a weak 2175\,\AA\ bump and a flat FUV curve. There is a slight trend
of higher densities showing relatively steep 
FUV slopes. 
Although $\stol$ increases with density, 
most of the particles have low $\stol$,
which is determined by stellar dust production.
The effect of shattering and accretion can be seen in a small fraction of particles with
high density, causing the upward spread of $\stol$
at $n_\mathrm{gas} \gtrsim 1$ cm$^{-3}$.
Since the steepening of FUV slope is seen in the dense medium,
it is caused by accretion rather than shattering, although
shattering is also necessary to produce the `seed' small grains for
accretion.

At 1\,Gyr, all the density ranges show a prominent 2175\,\AA\ bump
and a steep FUV rise. These features become clearer as the
density becomes higher. Therefore, at this stage, the extinction curves
are steeper in the dense ISM than in the diffuse ISM. This is explained
by the $\stol$ ratio as follows.
The small-to-large grain abundance ratio rapidly 
increases in $n_\mathrm{gas} \gtrsim 0.1$ $\mathrm{cm}^{-3}$
from 0.3 to 1\,Gyr.
Gas particles with $n_\mathrm{gas} \gtrsim 10$ $\mathrm{cm}^{-3}$ 
all reach $\stol \gtrsim 0.1$ by active accretion.
When such dense regions become diffuse by stellar feedback, etc.,
the small grains formed in the dense regions are dispersed into the
diffuse ISM. Thus, the high-$\stol$ regions extend down to
low densities. In the diffuse regions with 
$n_\mathrm{gas}<10$~cm$^{-3}$, there is another concentration of $\stol$
around $\sim 10^{-2}$ in carbonaceous dust 
($\stol \sim 10^{-3}$ in silicate), 
which is the branch of inefficient accretion
(i.e.\ the main source of small grains is shattering).
Thus, there is a large dispersion in $\stol$ at $n_\mathrm{gas}<10$ cm$^{-3}$. 
We expect that, although the extinction curves in the diffuse ISM
are on average flatter than in the dense ISM, 
the dispersion in the curves is larger than the difference
between the mean extinction curves in the dense and diffuse
ISM because of the large dispersion in $\stol$.

At 3\,Gyr, the trend of extinction curves for the density is
reversed: the diffuse regions have extinction curves with
a high 2175\,\AA\ bump and a steep FUV rise.
There are few particles that have 
$\stol <10^{-2}$ over all the
density range, because the increase of small grains by accretion
has been prevalent in the entire galaxy after the increase of
metallicity.
We observe a slight trend of decreasing
$\stol$ at $n_\mathrm{gas} \gtrsim 10$ $\mathrm{cm}^{-3}$ 
because of coagulation. This effect of coagulation
makes the 2175 \AA\ bump and FUV slope less prominent. 
In contrast, shattering efficiently occurs
in $0.1 \lesssim n_\mathrm{gas} \lesssim 1$ $\mathrm{cm}^{-3}$,
creating an enhanced small grain abundance
in this density range. This as well as the enhanced accretion
efficiency is the reason for a prominent 2175\,\AA\ bump and
a steep FUV rise in the extinction curve in the lowest density
range.

At 10\,Gyr, the trend seen at 3\,Gyr is enhanced.
The extinction curve features are very strong in 
$0.1 \lesssim n_\mathrm{gas} \lesssim 1$ $\mathrm{cm}^{-3}$ and 
less prominent in denser regions.
This is due to the small grain formation by shattering as seen
in the high values of
$\stol$ between $n_\mathrm{gas}\sim 0.1$ and 1 $\mathrm{cm}^{-3}$.
At the highest densities, $n_\mathrm{gas} \gtrsim 10$ $\mathrm{cm}^{-3}$, 
the extinction curve is as flat as that seen at 0.3\,Gyr, 
but this flatness at 10\,Gyr is due to coagulation,
not the predominant stellar dust production.
$\stol$ decreases at $n_\mathrm{gas} \gtrsim 1$ $\mathrm{cm}^{-3}$:
accretion is saturated because a large fraction of gas-phase metals
are locked into dust, while coagulation is efficient.

\subsection{Metallicity dependence}\label{result:metal}

\begin{figure*}%figure
	\begin{center}
	\includegraphics[width=0.33\textwidth]{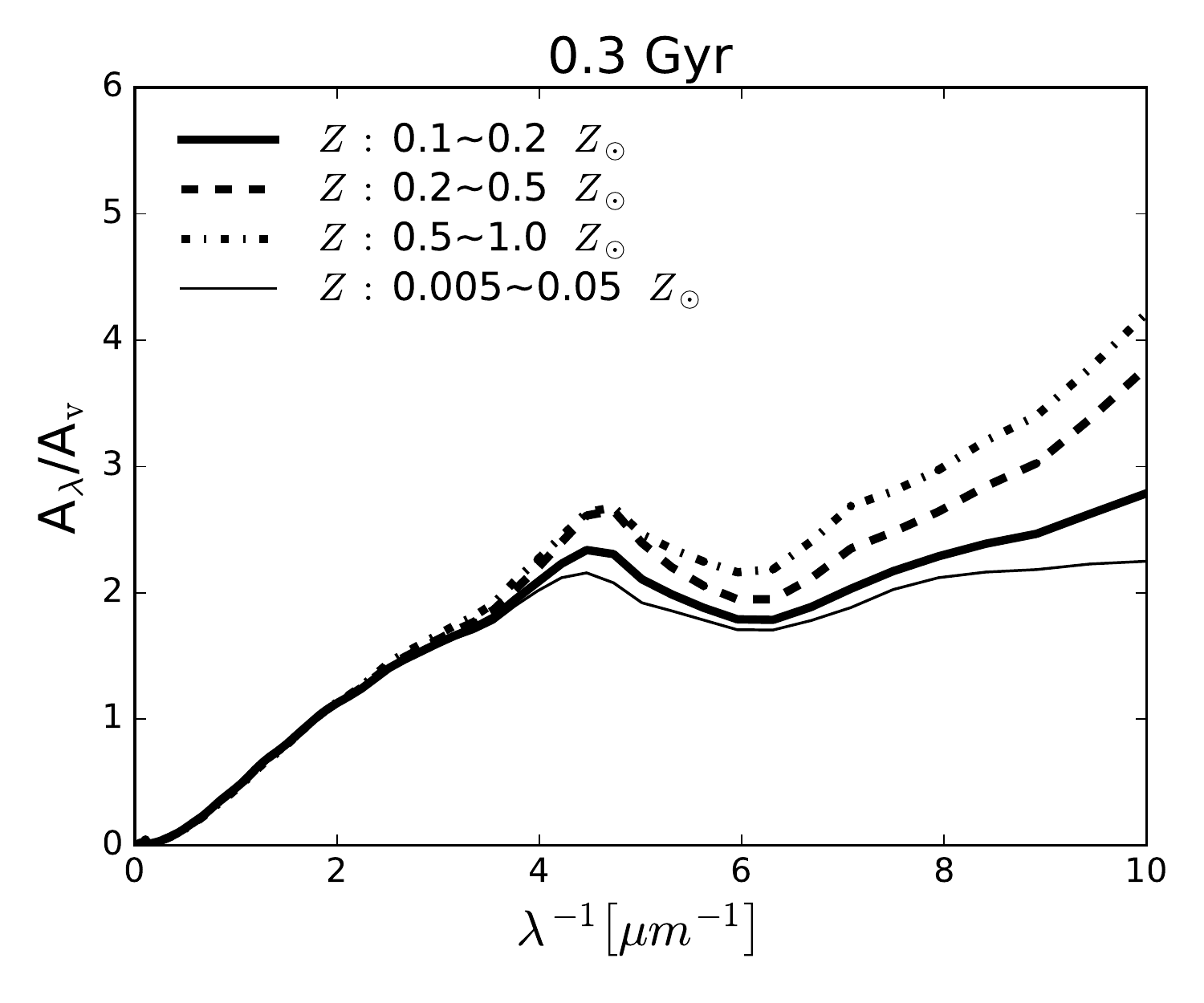}
	\includegraphics[width=0.285\textwidth]{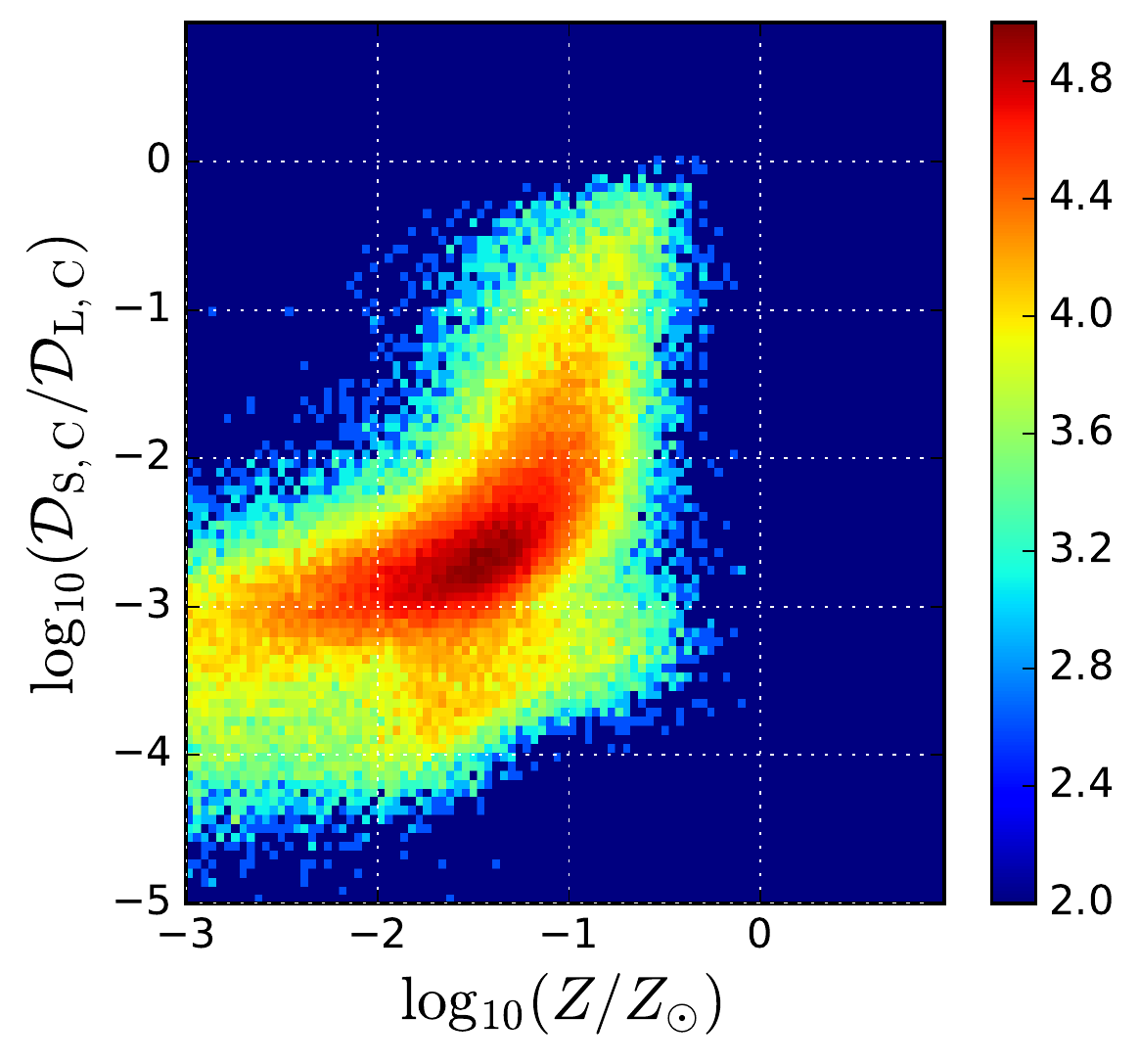}
	\includegraphics[width=0.285\textwidth]{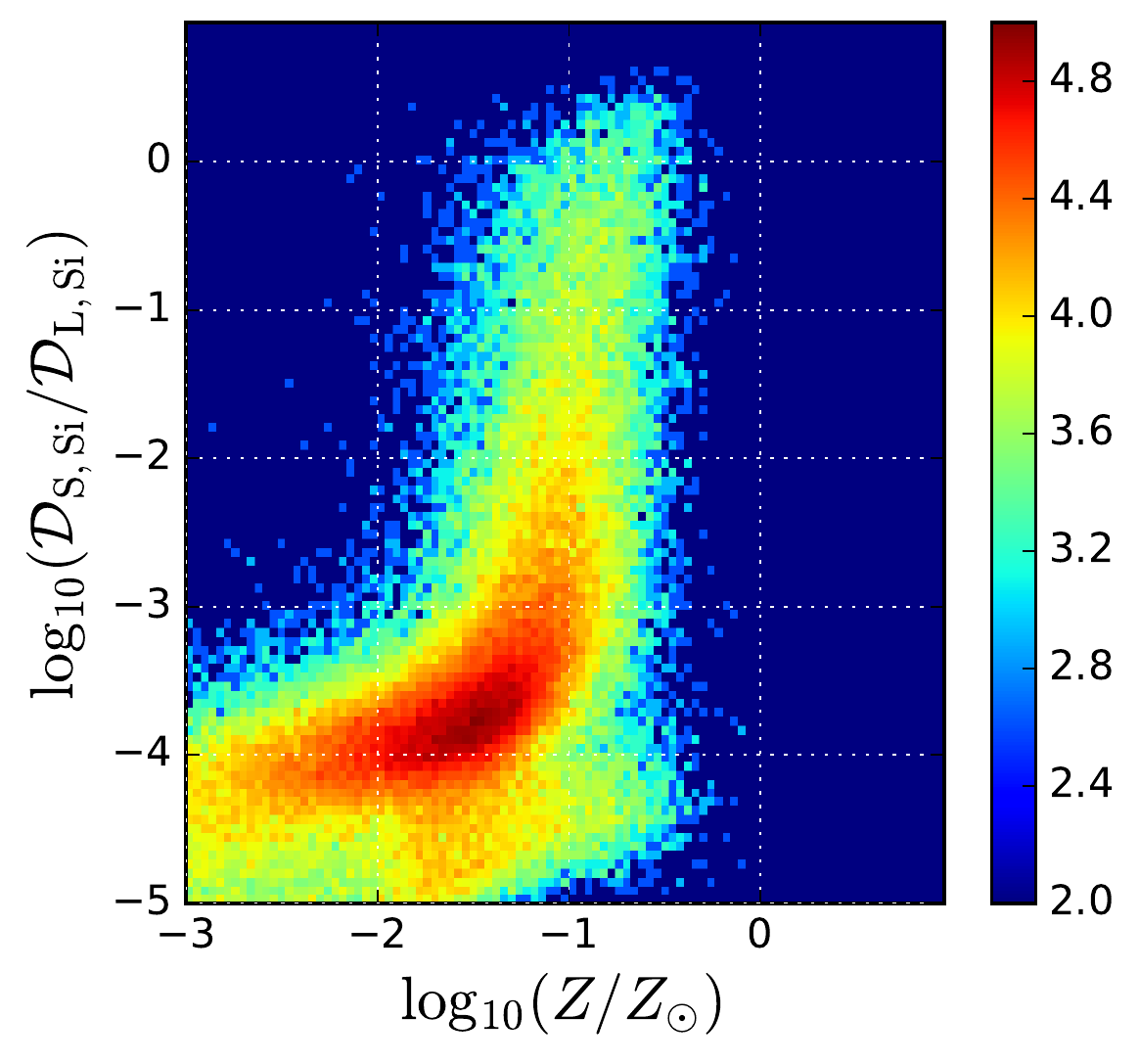}

	\includegraphics[width=0.33\textwidth]{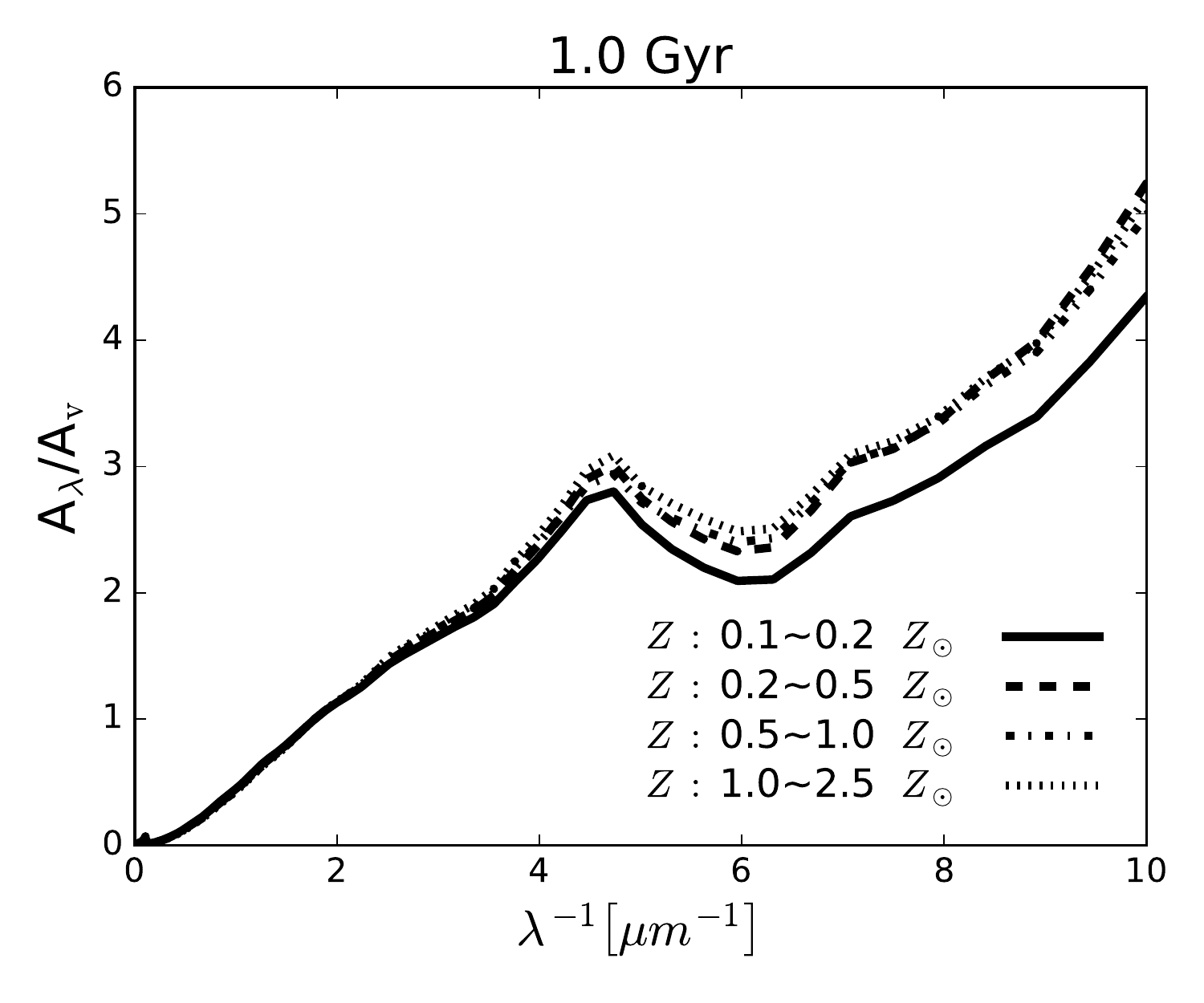}
	\includegraphics[width=0.285\textwidth]{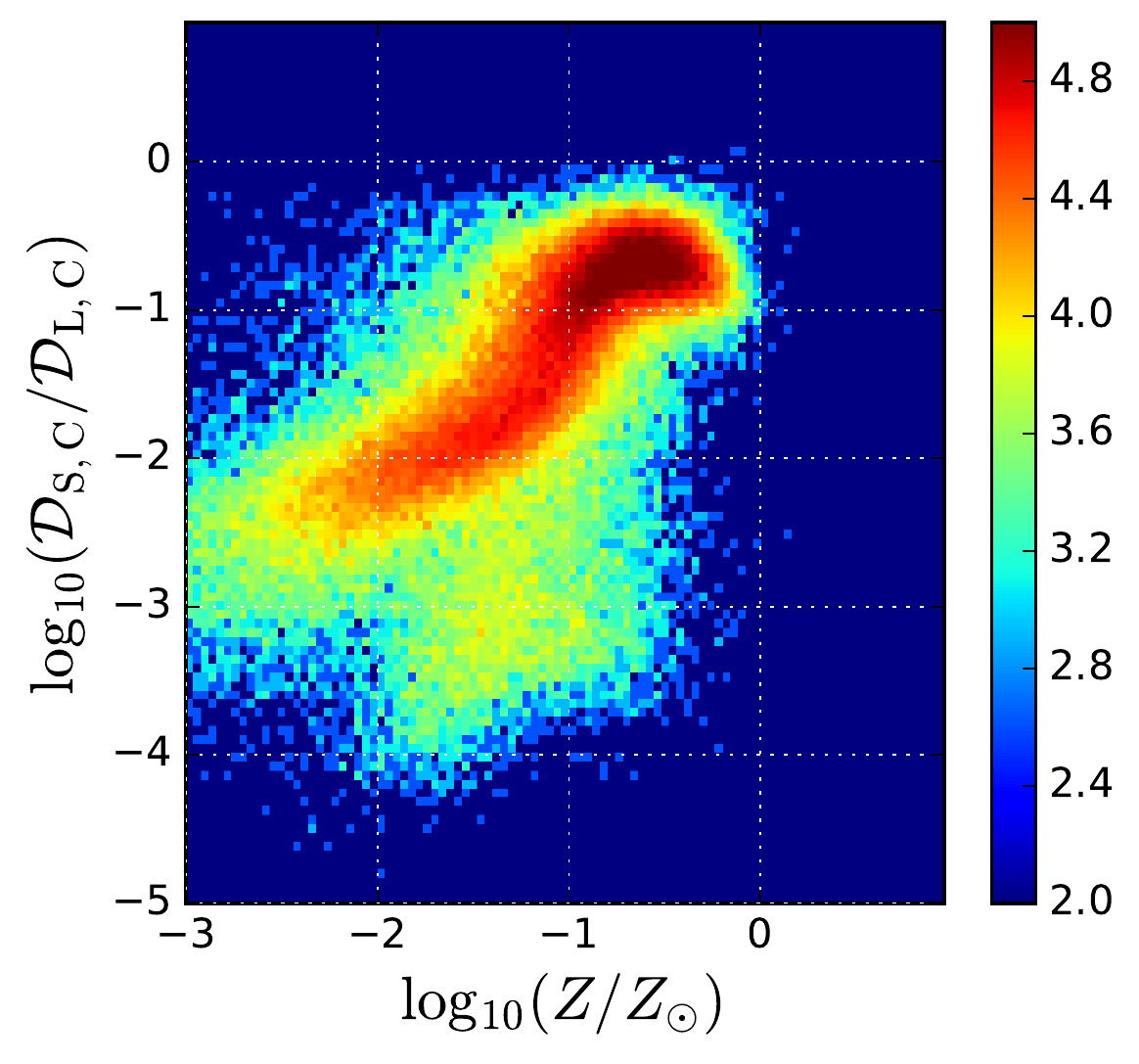}
	\includegraphics[width=0.285\textwidth]{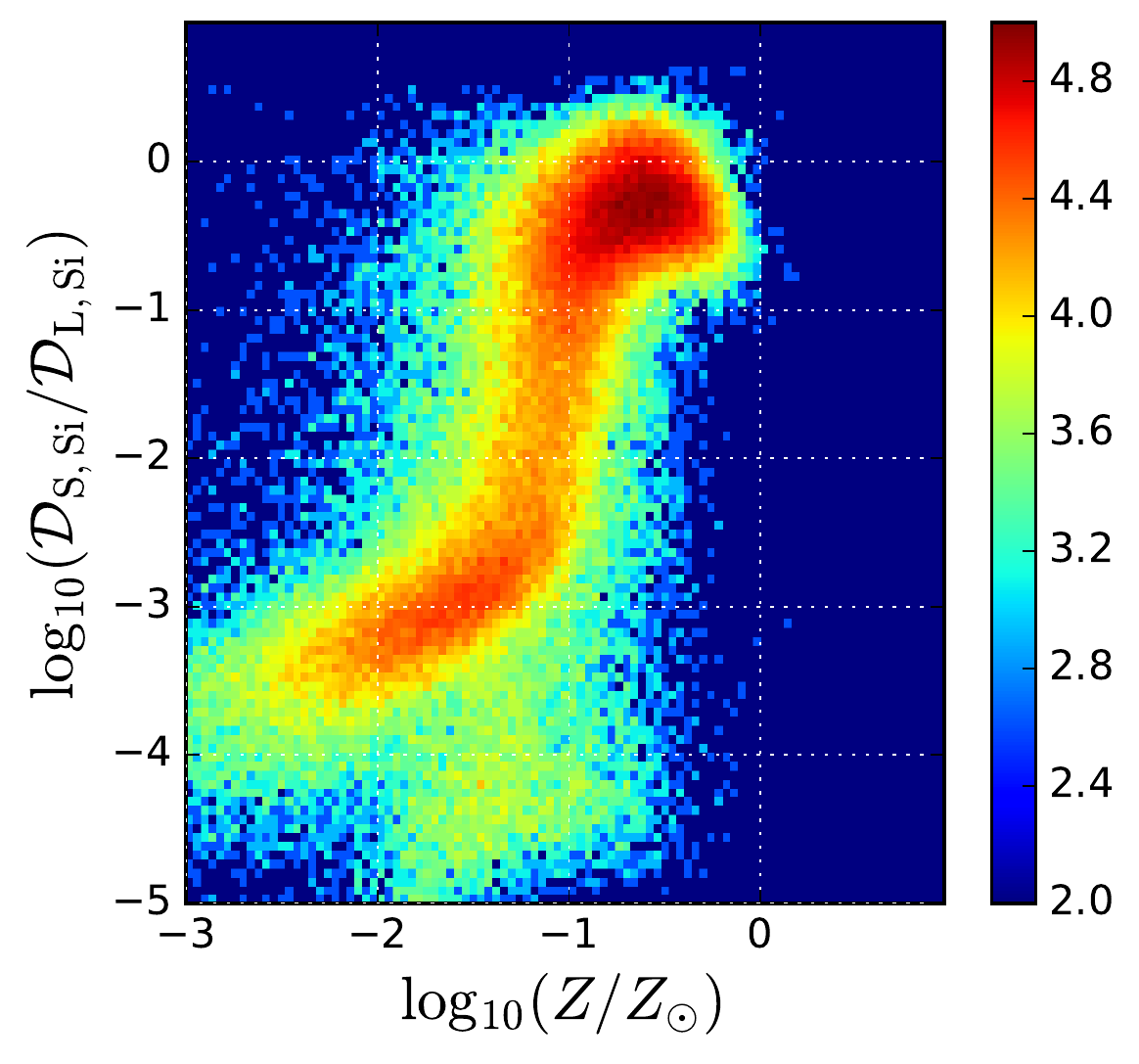}

	\includegraphics[width=0.33\textwidth]{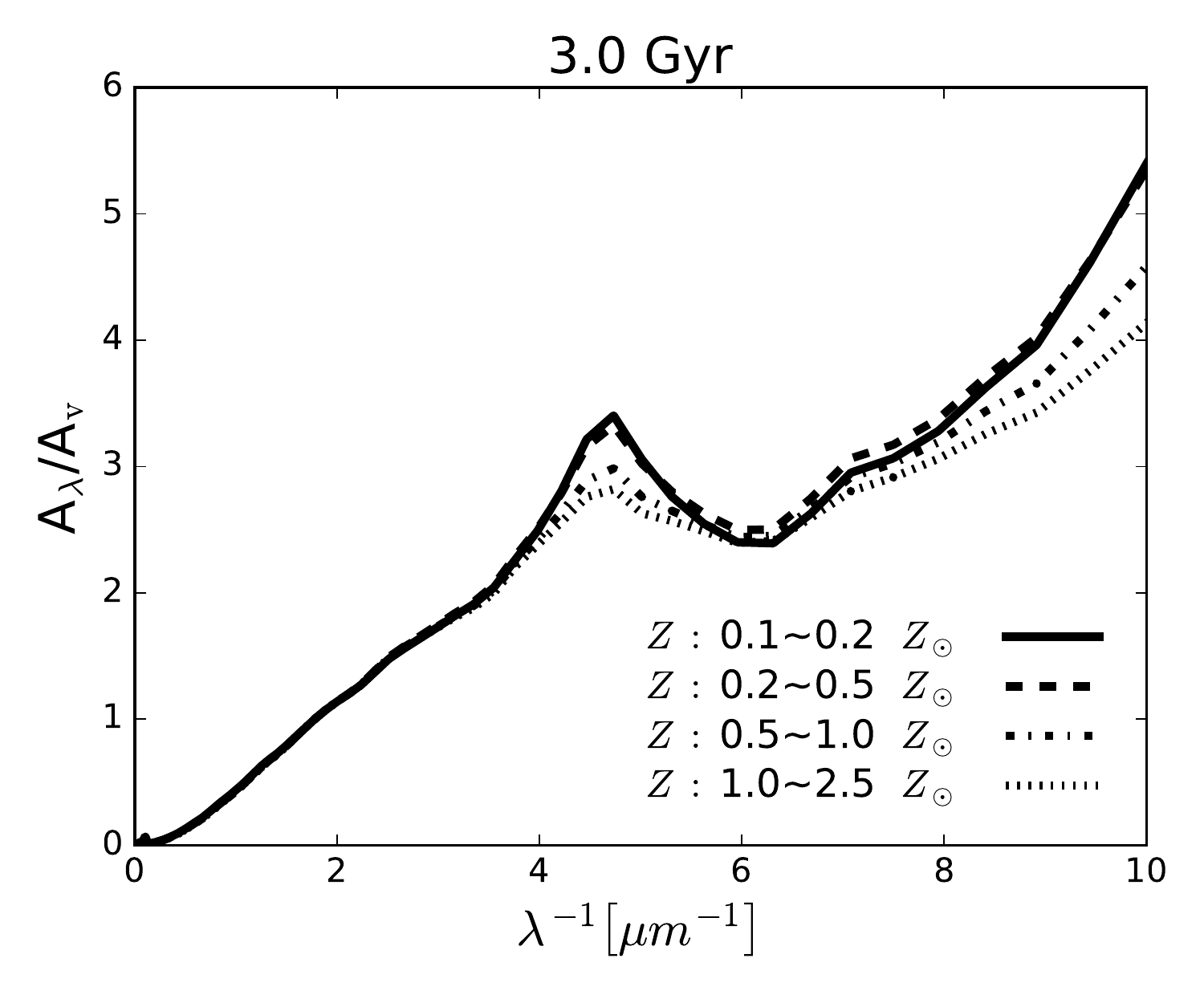}
	\includegraphics[width=0.285\textwidth]{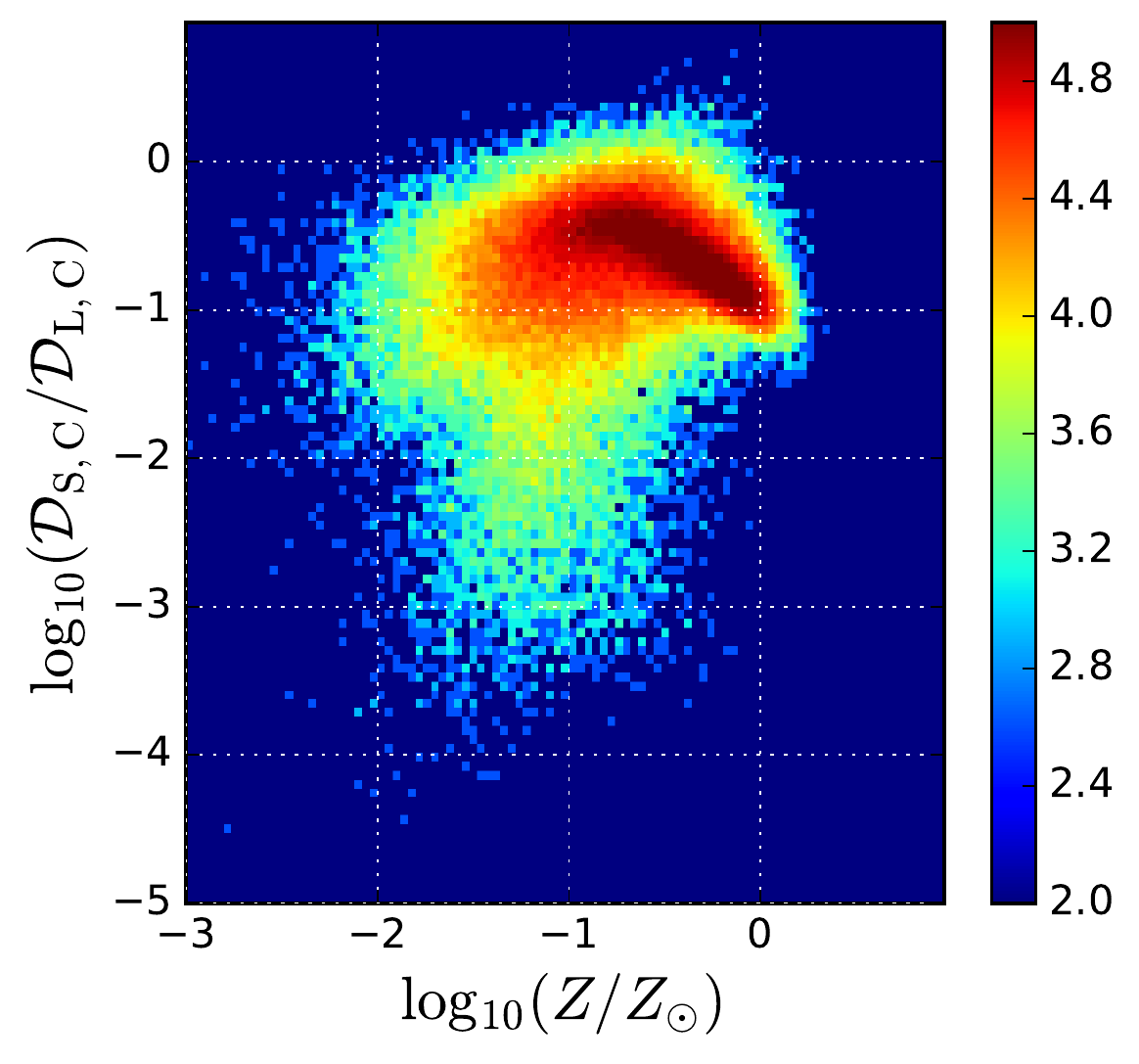}
	\includegraphics[width=0.285\textwidth]{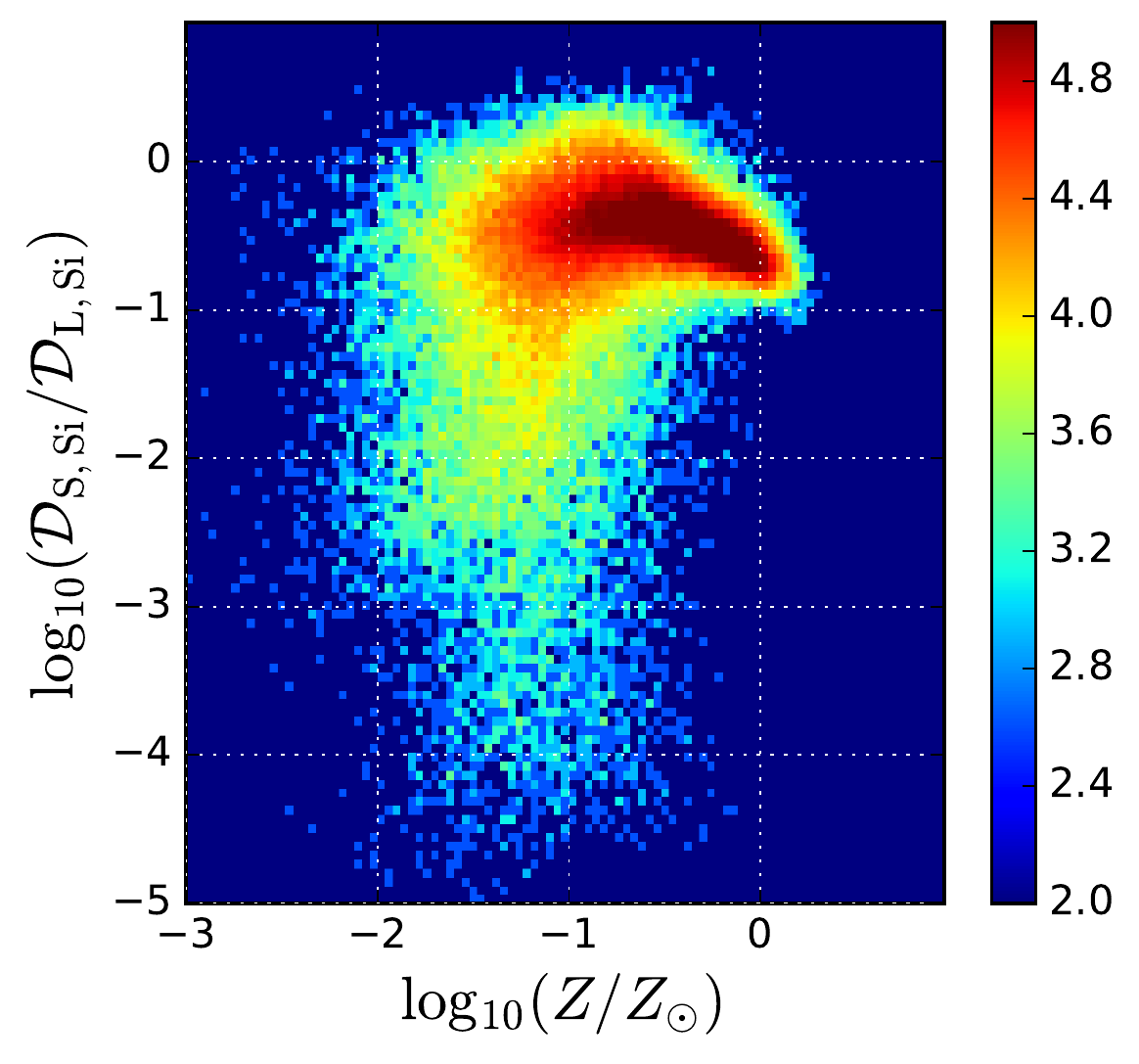}

	\includegraphics[width=0.33\textwidth]{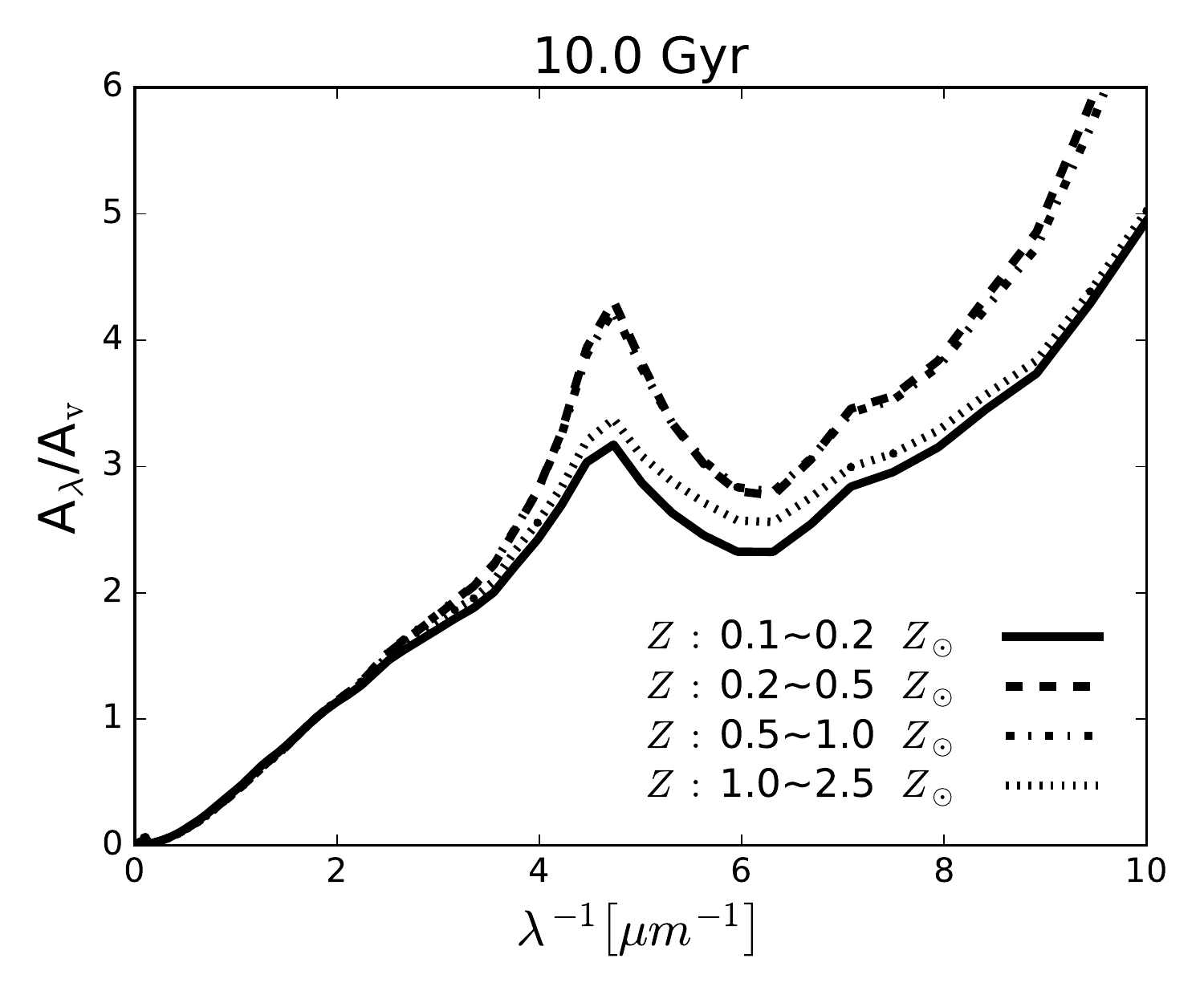}
	\includegraphics[width=0.285\textwidth]{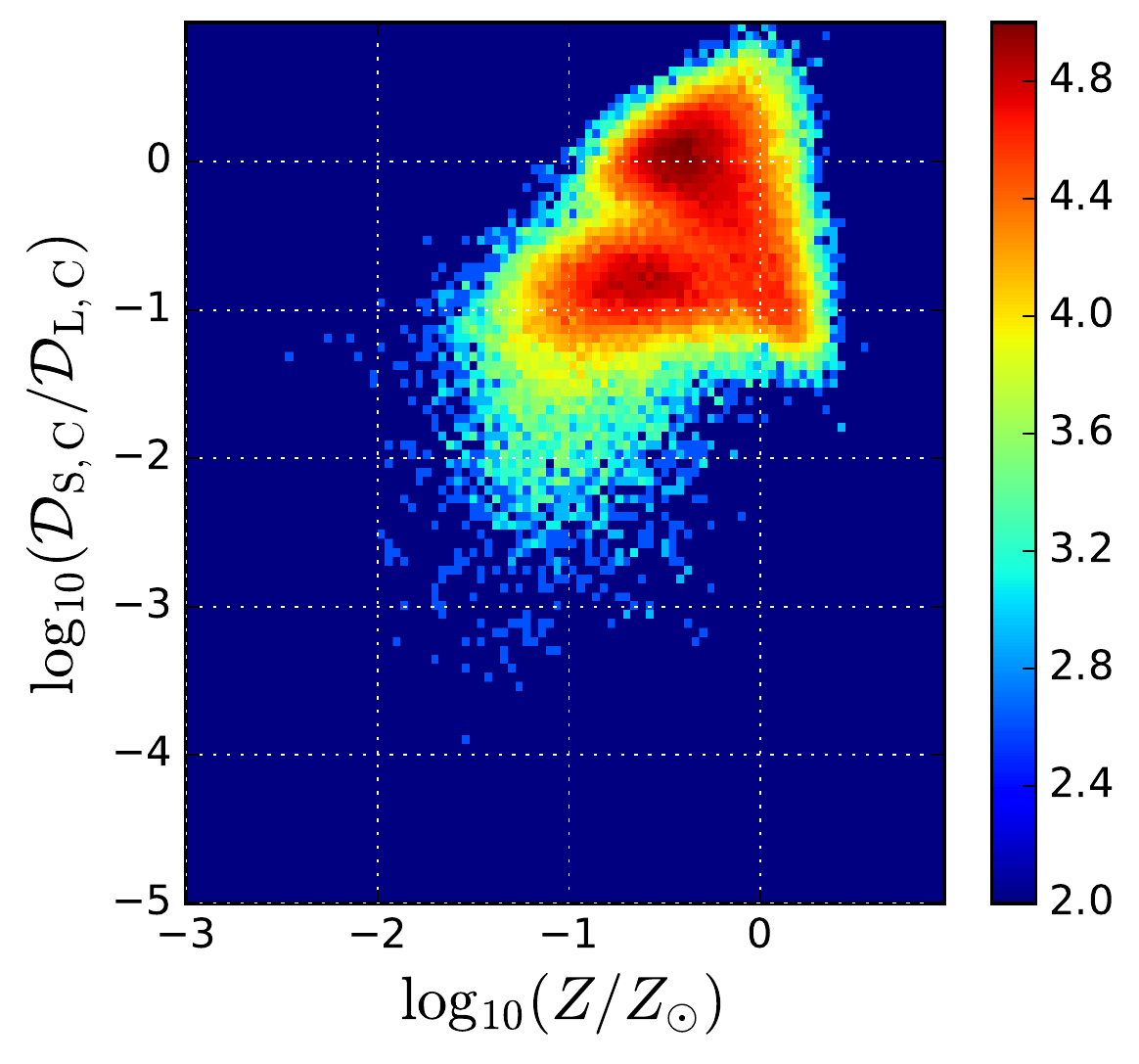}
	\includegraphics[width=0.285\textwidth]{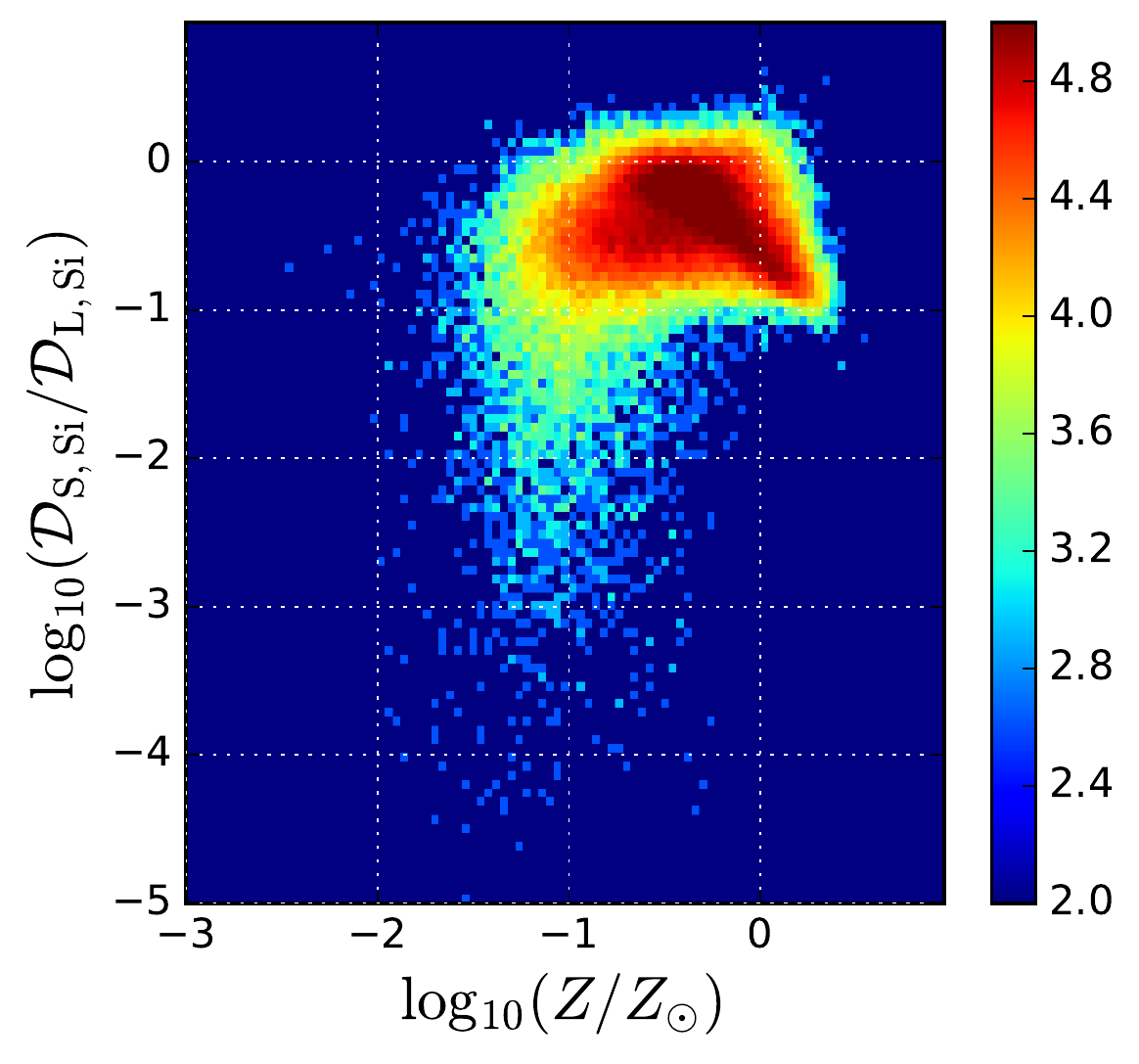}
	\caption{Same as Fig \ref{Fig:extinction_radii} but for
	metallicity dependence.
	In the left panels, the (thick) solid, dashed, dash-dotted, 
	and dotted lines represent the extinction curve in metallicity
	ranges, 0.1--0.2, 0.2--0.5, 0.5--1 and 1--2.5 $Z_\odot$, respectively.
	In the panel of extinction curve at 0.3 Gyr, the 
	additional thin solid line shows the extinction curve in
	extremely low metallicity range 0.005--0.05 $Z_\odot$. }
	\label{Fig:extinction_metal}
	\end{center}
\end{figure*}

Metallicity could also be an important driver of dust evolution
because of the coevolution of dust and metals 
\citep[e.g.][]{Remy-Ruyer:2014aa}. 
To investigate the
effect of metal enrichment, we show the metallicity dependence of
extinction curves and small-to-large grain abundance ratios at
various ages in
Fig. \ref{Fig:extinction_metal}. We derived mass-weighted
extinction curves for gas particles that have the corresponding 
metallicity in each metallicity bin. The whole metallicity range
considered is from $0.1 Z_\odot$ to $2.5 Z_\odot$.

At 0.3\,Gyr, there are few particles with $Z$ > 1 $Z_\odot$ 
and the intensity of 2175\,\AA\ bump 
and the steepness of FUV rise increase with metallicity, 
although both of these features are weak and the extinction
curves are overall flat.
The $\stol$ value has a rapid increase at $Z \gtrsim 0.1 Z_\odot$; thus the 
features of extinction curves become prominent.
The value of $\stol$ at $Z \lesssim 0.01 Z_\odot$ is determined 
by stellar dust production and shattering.
Although there is no observational data at such
low metallicity, we calculate the extinction curve for 
a metallicity range of
$0.005 \lesssim Z \lesssim 0.05 Z_\odot$ to show prediction for
an extremely metal-poor phase, which could be important for
future high-redshift observations.
The extinction curve in the lowest metallicity range
is much flatter than the other metallicity ranges because 
of the lowest $\stol$.
%%\sout{
%%although we did not
%%calculate the extinction curves for this metallicity range
%%because there is no observational data with such low metallicity.}
At $Z \gtrsim 0.01 Z_\odot$, the efficiencies of shattering and accretion increase with
metallicity. As a result, the small grain abundance increases drastically, 
especially at $Z \gtrsim 0.1 Z_\odot$.

At 1\,Gyr, the extinction curves of all the metallicity range
show a clear 2175\,\AA\ bump and FUV rise because of
the overall increase of $\stol$ at all metallicities.
The FUV slope is relatively flat at the lowest metallicity
range, $0.1 \lesssim Z \lesssim 0.2 Z_\odot$.
The $\stol$ value does not increase any further when particles reach $Z \sim 0.5 Z_\odot$,
because the coagulation efficiency is comparable to the accretion efficiency
at such high metallicities.

At 3\,Gyr, the features of extinction curves are stronger
in $0.1 \lesssim Z \lesssim 0.5 Z_\odot$ than
in $0.5 \lesssim Z \lesssim 2.5 Z_\odot$. 
$\stol$ shows a clear decline at $Z \gtrsim 0.5 Z_\odot$ since
coagulation is efficient enough to convert small grains
to large grains.
At $Z \sim 0.1 Z_\odot$, $\stol$ rises to the 
highest value in the entire metallicity range, because shattering
dominates the grain size distribution around that 
metallicity. 
Thus, low-metallicity ($Z\sim 0.1$ $Z_{\sun}$) gas has
steeper extinction curves than solar-metallicity gas. This trend
at 3\,Gyr is opposite to that seen above at 1\,Gyr.

At 10\,Gyr, particles with $0.2 \lesssim Z \lesssim 1 Z_\odot$ show
a very strong 2175 \AA~ bump and steep FUV slope. 
Although the extinction curves of $Z \lesssim 0.2~\mathrm{Z}_\odot$ 
have relatively weak features, 
the 2175 \AA~ bump and FUV rise are still prominent. 
In this stage, shattering completely dominates the grain size
distribution at 
$Z \lesssim 1 Z_\odot$, raising $\stol$ to the 
highest value seen at $Z \sim 1 Z_\odot$.
%%Is it important to say the following? I have just deleted it.
%%, although the highest $\stol$ by taking mass average is shown 
%%at $Z \sim 0.5 Z_\odot$.
We note that $\stol$ has a bimodal distribution especially 
for carbonaceous dust. As mentioned in Paper I, 
the higher-$\stol$ component, 
which makes the extinction curve features prominent, 
is associated with the diffuse medium in the galactic disc, 
while the lower component is
associated with low density ($n_\mathrm{gas} < 10^{-3}$ cm$^{-3}$) 
and hot ($T_\mathrm{gas} > 10^4$ K) gas more extended in the halo. 
This clear separation is due to stellar feedback and 
is further discussed in Section \ref{feedback}.
Therefore, shattering in the diffuse ISM in the disc is
important in the steepening of extinction curves in the latest
($\sim 10$\,Gyr) evolutionary phase of the galactic disc.

\section{discussion}

\subsection{Comparison with the Milky Way extinction curve}

\begin{figure*}%figure
	\begin{center}
	\includegraphics[width=0.45\textwidth]{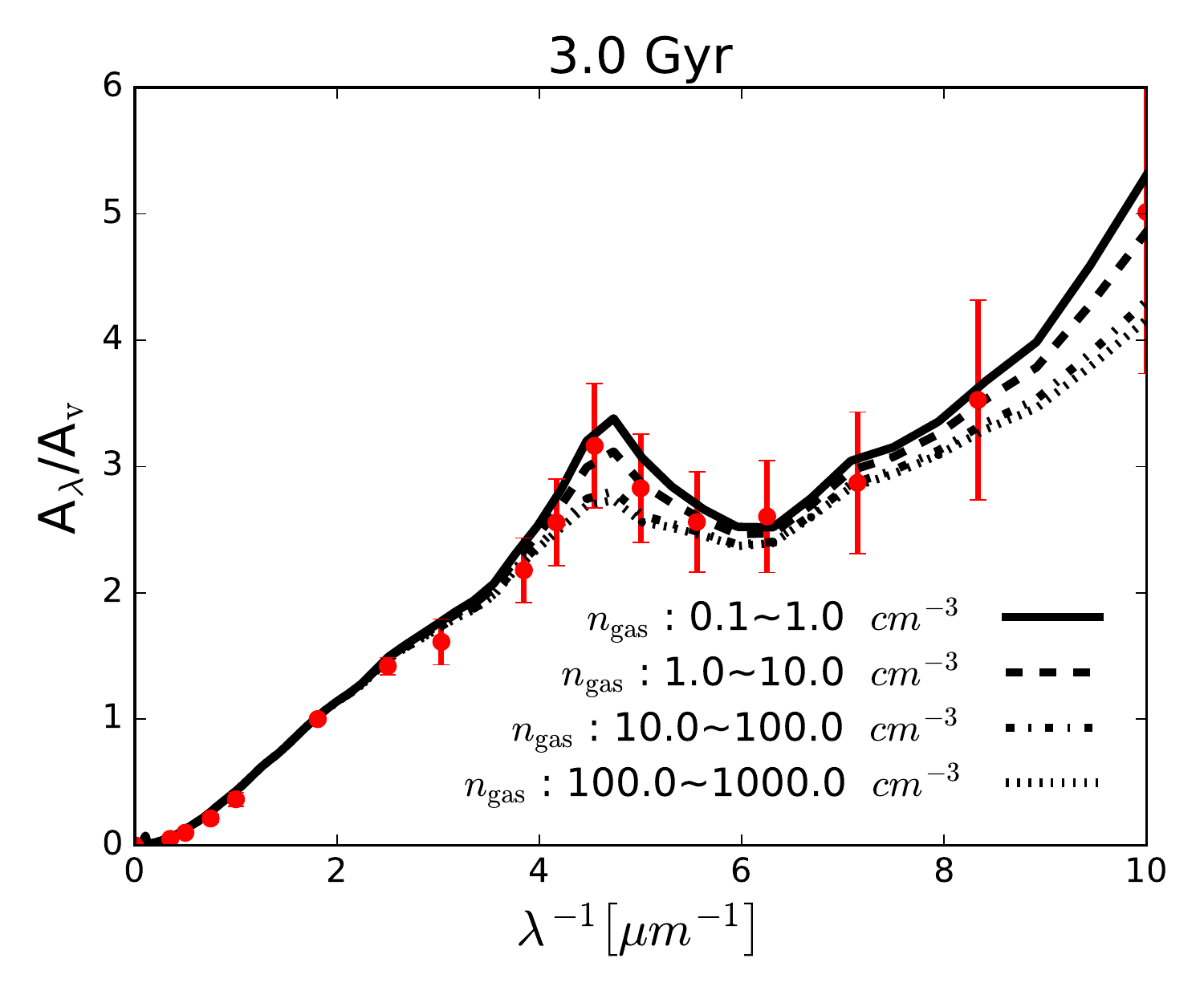}
	\includegraphics[width=0.45\textwidth]{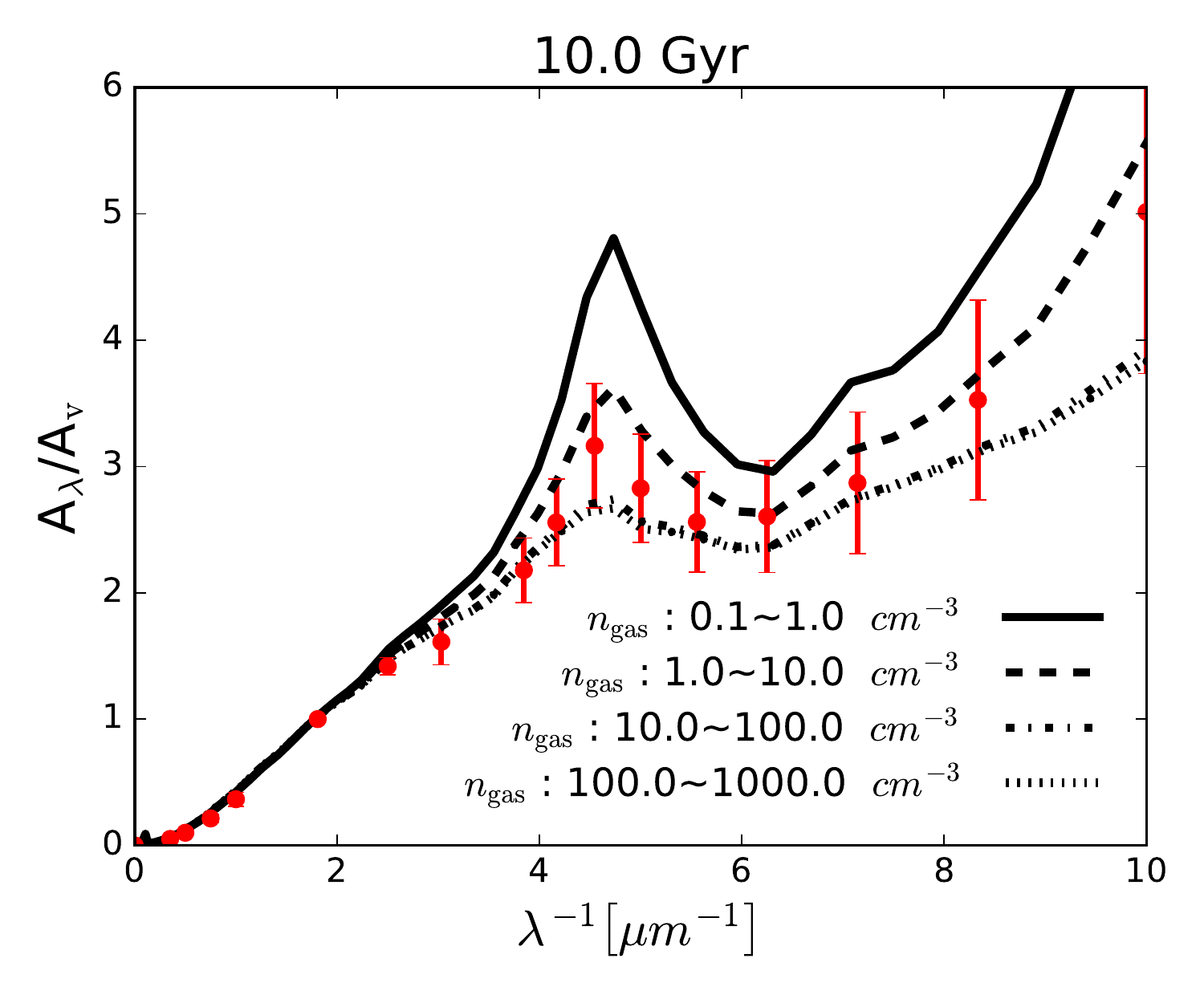}
	\includegraphics[width=0.45\textwidth]{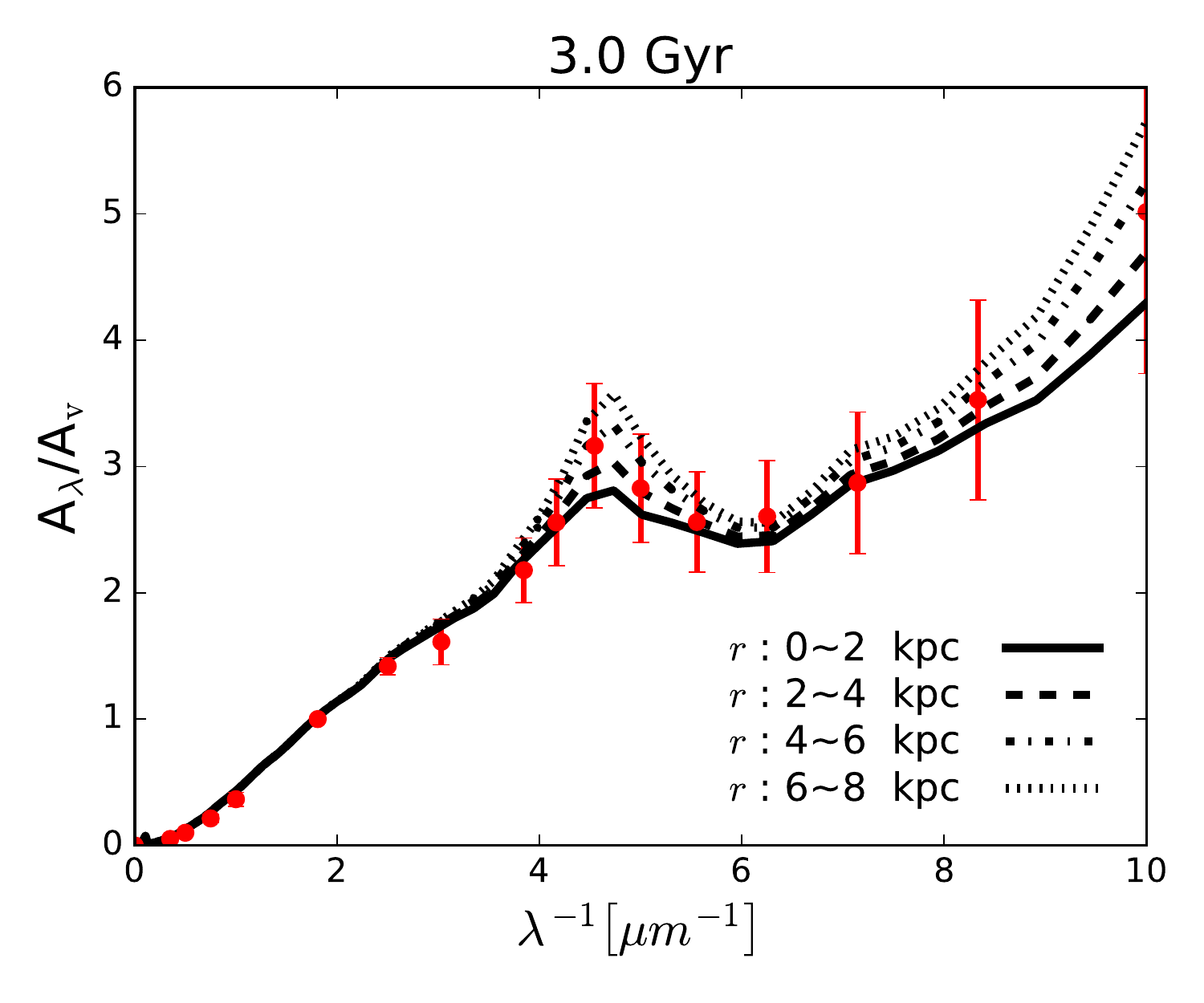}
	\includegraphics[width=0.45\textwidth]{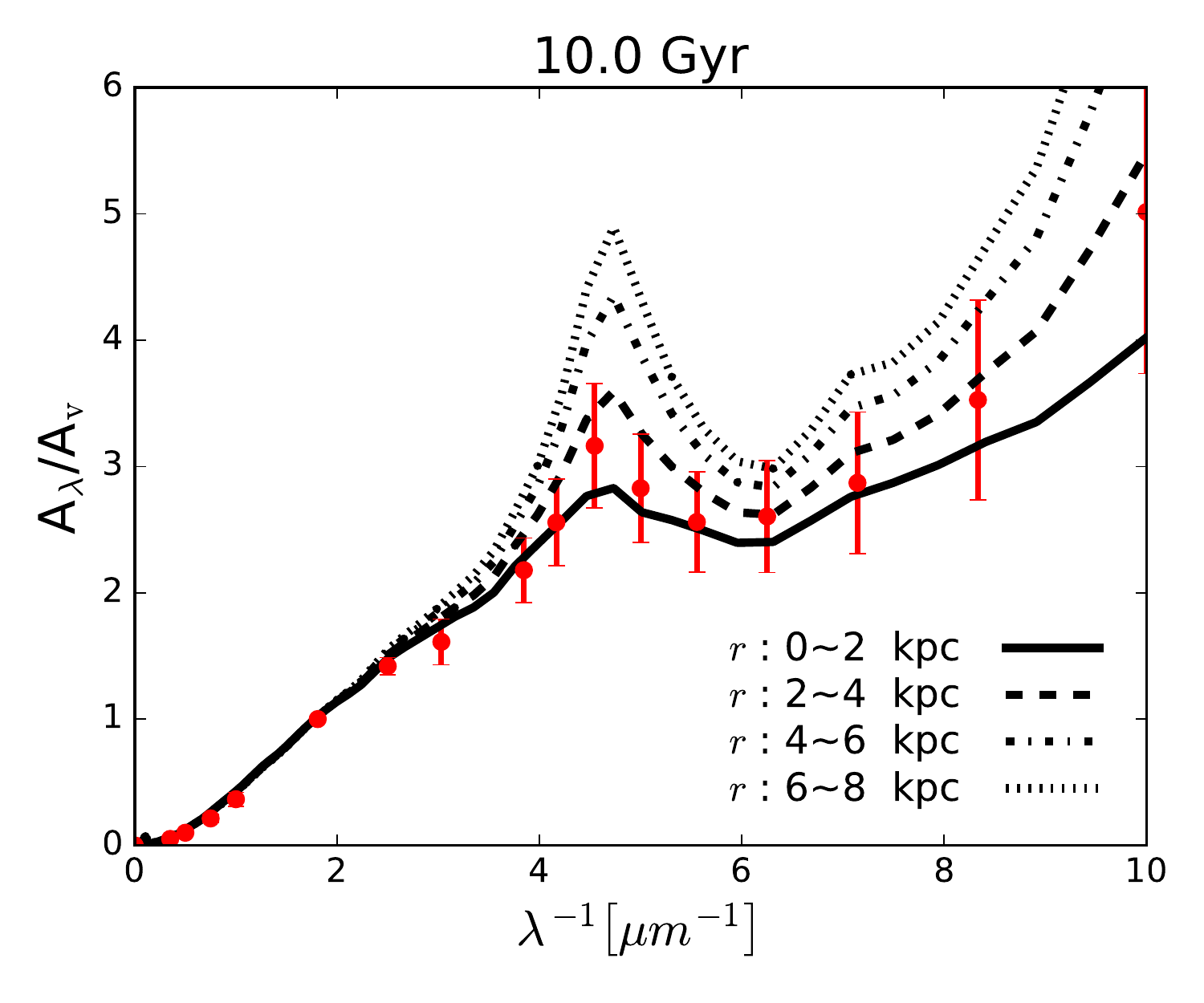}

	\caption{Comparison with observed mean Milky Way extinction curve. 
	{\it Upper panels}: solid, dashed, dash-dotted, 
	and dotted lines represent the extinction curves within gas 
	density ranges, 0.1--1, 1--10, 10--$10^2$ and $10^2$--$10^3$ cm$^{-3}$, 
	respectively.
	{\it Bottom panels}: solid, dashed, dash-dotted, and dotted lines represent the extinction curves within radial ranges, 0--2, 2--4, 4--6 and 6--8 kpc, respectively.
	The left and right columns show the results at ages 3 and 10\,Gyr, respectively.
	The red dots and vertical error bars show the observed mean extinction curve and the 1$\sigma$ dispersion taken from 
	\citet{Fitzpatrick:2007aa} .}
	\label{Fig:MW_extinction}
	\end{center}
\end{figure*}

In Fig. \ref{Fig:MW_extinction}, we compare our results 
at 3 and 10\,Gyr with 
the observed mean Milky Way 
extinction curve and the 1$\sigma$ dispersion of 
the extinction curve in various lines of sight.
The Milky Way extinction curves are determined 
by the stars at distances $\sim$0.1--2 kpc from the Sun
\citep{Fitzpatrick:2007aa}.
Because spatially resolved information on a single line of sight
(or on a scale below the path length)
cannot be obtained, we need to smooth the spatially resolved extinction curves
obtained in our simulation over a length comparable to the path length
in the observations.
%%Because observed extinction curves are not spatially resolved
%%on the direction of the line of sight,
%%we smooth the simulated spatial structure to compare 
%%with observed extinction curves,
%%although our simulation has 80 pc resolution for gravitational forces, 
%%and $\sim30$\,pc resolution for baryonic density contrast.
However, because every observational line of sight 
(for every extinction curve) has different path length,
it is difficult to apply a single smoothing length
for comparison with observational data. Instead,
we estimate a range of extinction curves 
predicted for various densities and radii as possible 
sources of the variation in extinction curves, 
and compare them with the observed curves and their variance.

We select 3 and 10\,Gyr as two appropriate epochs 
because the representative metallicities of these epochs 
are close to that of the Milky Way.
Our results are 
consistent with both the mean value and the dispersion
except that the extinction curves at low density and in the 
outer disc region at 10\,Gyr show a stronger 2175 \AA~ bump
than the observation.
\citet{Cardelli:1989aa} and \citet{Fitzpatrick:2007aa} also show that
the 2175 \AA\ bump strength in the Milky Way extinction curves
has a positive correlation with the FUV slope. 
This correlation can be interpreted as a sequence of
small-to-large grain abundance ratio and is
consistent with our results above.
Reproducing the Milky Way extinction curve implies
that we implemented all the processes that drive dust evolution 
in Milky-Way-like galaxies. We emphasize that
we basically adopted the parameters adopted in Paper I 
without any fine-tuning specific for the extinction curve modelling.

\citet{Schlafly:2016aa} recently showed that 
the Milky Way extinction curves are rather uniform across the entire disc,
but that a significant variation in extinction curves is indeed observed. This
variation occurs on scales much larger than the typical molecular-cloud size. 
As shown in Fig.\ \ref{Fig:extinction_density}, the large dispersion in
$\stol$ at a given density shows that
the grain size distribution (i.e.\ $\stol$)
is not governed uniquely by the density. This is firstly because
the current grain size distribution is determined by
the past environments in which the dust has resided.
The second reason is that the ISM density
structure also changes especially by stellar feedback, which also
disperses dust in a wide area. Therefore, the large-scale variation in
extinction curves shown in the above study indicates the
importance of treating dust processing in a consistent manner with
the variation of ISM density structures.

We mention the following two caveats. First,
most of the Milky Way extinction curves are taken from 
solar neighbourhood stars ($\lesssim 2$ kpc) that only reflects the 
local properties of dust around the Sun. Extinction curves in 
galactic centre and outer disc still remain uncertain because
of lacking observation. 
Although there are some observational studies of extinction curves toward 
the centre of the Milky Way at near-infrared wavelengths 
\citep{Nishiyama:2006aa,Nishiyama:2008aa,Nataf:2013aa}, 
it is hard to compare our results with them because 
our model is not able to predict precise extinctions at infrared wavelengths.

Second, we focus on dust evolution with 
a commonly used dust species model and succeed in reproducing observations.
However, some previous efforts of modelling various dust species and 
grain size distributions suggest that there are some different
dust species that fit the Milky Way extinction curves 
\citep{Zubko:2004aa,Jones:2013aa}. 
Although the trend that small grains produce steep and strong-feature
extinction curves does not change against the change of dust
materials, the effect of varying dust material properties is
worth investigating in the future.

\subsection{$\stol$ in silicate and carbonaceous dust}\label{subsection:c_to_si}

Overall, silicate and carbonaceous dust share the same
dust enrichment mechanisms. The differences between these
two species lie in the 
condensation efficiency ($f_\mathrm{{in,X}}$), 
mass density ($\rho_\mathrm{X}$) and elemental abundance ($Z_\mathrm{X\odot}$; 
these quantities are listed in Table \ref{table:Adopted_quantities}).
Carbonaceous dust has a higher $f_\mathrm{{in,X}}$ than silicate so
that stars produce more carbonaceous dust than silicate in our models.
With a fixed dust mass, carbonaceous dust has higher shattering and
coagulation efficiencies because a lower mass density means a larger
number of dust grains.

At the beginning of galaxy evolution, the important 
processes are stellar dust production and shattering:
Stars produce only large grains so that $\stol$ is determined by
shattering. From the above argument, carbonaceous dust has, on average,
higher $\stol$ than silicate at $t \lesssim 0.3$\,Gyr 
(Figs.\ \ref{Fig:extinction_radii}--\ref{Fig:extinction_metal}). 

Equations (\ref{eq:tau_acc_C}) and (\ref{eq:tau_acc_Si}) show that 
the elemental abundance ($Z_\mathrm{X}$) determines
the efficiency of accretion: because we set the
Milky Way elemental abundance pattern for the metal yields, 
accretion of silicate is more efficient than that of carbonaceous dust 
(Table \ref{table:Adopted_quantities}; recall that Si occupies a 
mass fraction of 0.166 in silicate).
Thus, when accretion dominates the dust abundance (at $0.3 \lesssim t \lesssim 3$\,Gyr), 
silicate has, on average, higher $\stol$ than carbonaceous dust
(Figs.\ \ref{Fig:extinction_radii}--\ref{Fig:extinction_metal}).

Carbonaceous dust has higher shattering and coagulation efficiencies than 
silicate has as mentioned above, 
so in the late evolutionary stage ($t \gtrsim 3$\,Gyr), 
carbonaceous dust shows a clearer decrease
of $\stol$ by coagulation in the dense medium or in metal-rich
regions while it presents higher $\stol$ in the diffuse medium or sub-solar
metallicity regions because of shattering
(Figs.\ \ref{Fig:extinction_radii}--\ref{Fig:extinction_metal}).

\subsection{Dust species abundance ratio}\label{subsec:species_ratio}

The mass ratio of carbonaceous dust to silicate (C/Si) 
is important in producing various 2175 \AA\ bump strengths under 
a similar FUV slope or vice versa.
We adopted the classical graphite--silicate dust species model to 
estimate the extinction curve. In this model, the 2175 \AA\ bump is contributed from 
small graphite grains while the FUV rise is mainly dominated by small silicates.
Although the behaviour of extinction curve is broadly understood by the
small-to-large grain abundance ratio,
we here inspect C/Si to analyse further details
of the extinction curve shapes.

\begin{figure}%figure
	\begin{center}
	\includegraphics[width=0.5\textwidth]{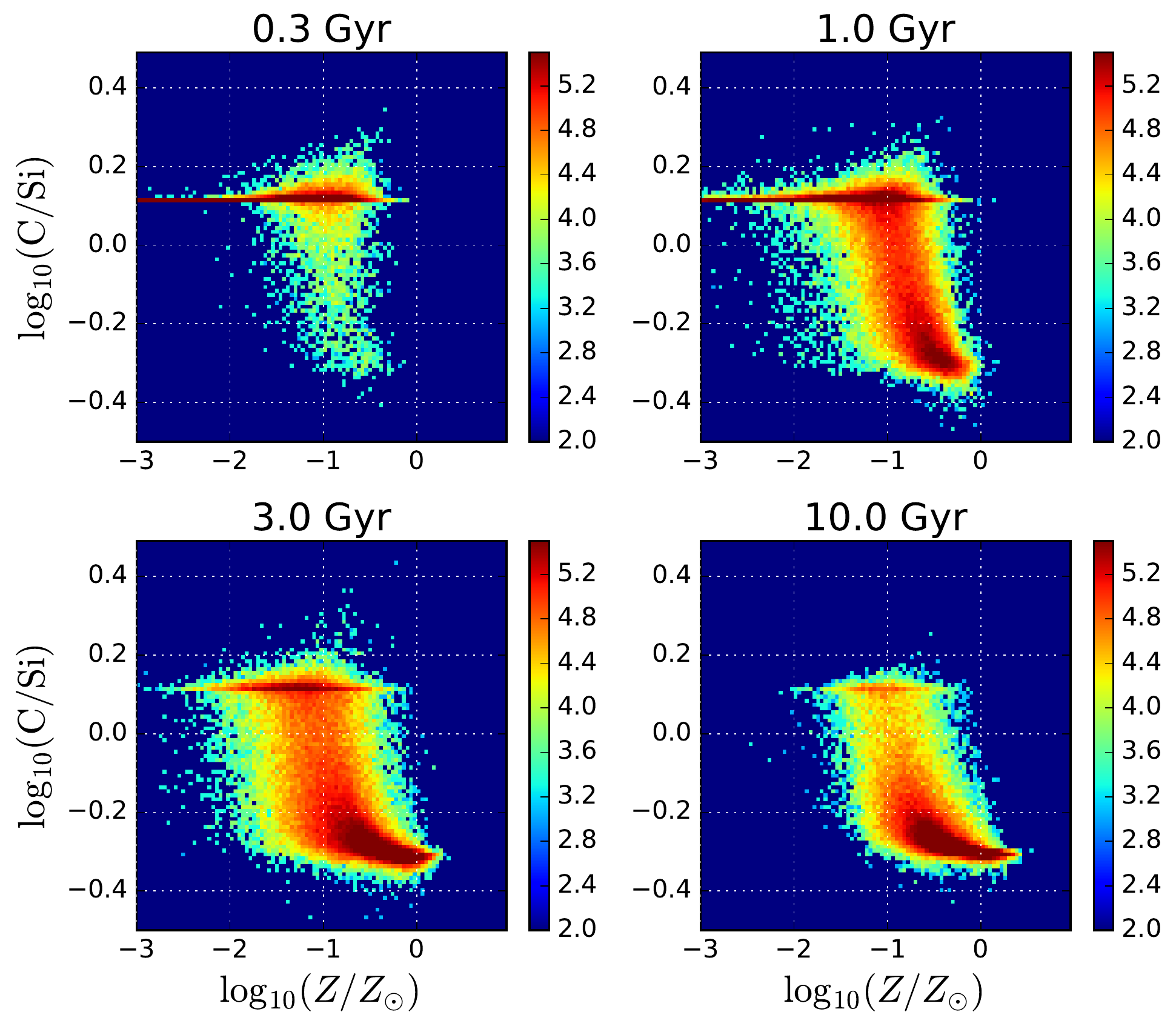}
	\caption{Carbonaceous dust-to-silicate abundance ratio 
	as a function of $Z$ at 0.3, 1, 3 and 10\,Gyr. 
	The meaning of the colours is the same as in Fig. \ref{Fig:extinction_radii}.}
	\label{Fig:c2si}
	\end{center}
\end{figure}

In Fig. \ref{Fig:c2si}, we find that C/Si decreases with increasing age. 
In the early phase of galaxy evolution ($t \lesssim$ 0.3\,Gyr), 
C/Si is determined by the dust condensation efficiency 
in stellar ejecta ($f_{\rm in,X}$)  
multiplied by the available elements composing the dust (C/Si $\sim$ 1.2). 
After accretion becomes efficient, C/Si is 
rather governed by the available gas-phase elements.
Thus, in the late evolution stage ($t \sim$ 10\,Gyr),
C/Si approaches a value ($\sim$0.5) determined by the elemental 
abundance pattern assumed in the model 
(i.e.\ the solar abundance pattern). 
We note that the solar abundance pattern we adopted 
should be achieved as a result of contribution from various
stellar populations including
AGB stars (note that we did not include the contribution from
AGB stars in the metal yield adopted in the simulation). Therefore, at
ages younger than 1\,Gyr, when less stars evolve into the AGB phase
than at the age of the Milky Way ($\sim 10$ Gyr),
adopting the solar metallicity pattern
would overproduce the carbon abundance, which is raised
significantly by AGB stars \citep{Ferrarotti:2006aa}.
Thus, the 2175 \AA\ bump may be overestimated at 
the early evolutionary stage.
The dust species abundance ratio is also important
when we attempt to explain the SMC/LMC extinction curves
in Section \ref{subsec:MCs}, where we mention a possible
variation of C/Si again from a different point of view.

\subsection{Extinction curves inside and outside the disc}\label{feedback}

To examine the impact of stellar and SN feedback on the extinction curve, 
we examine the distribution of various quantities by separating the gas 
spatially inside and outside of galactic disc by the following criteria: 
$r \leq 15$\,kpc and $|z| \leq 2$\,kpc for inside, and vice versa for outside, 
where $|z|$ is the distance perpendicular to the disc plane.

\begin{figure}%figure
	\begin{center}
	\includegraphics[width=0.5\textwidth]{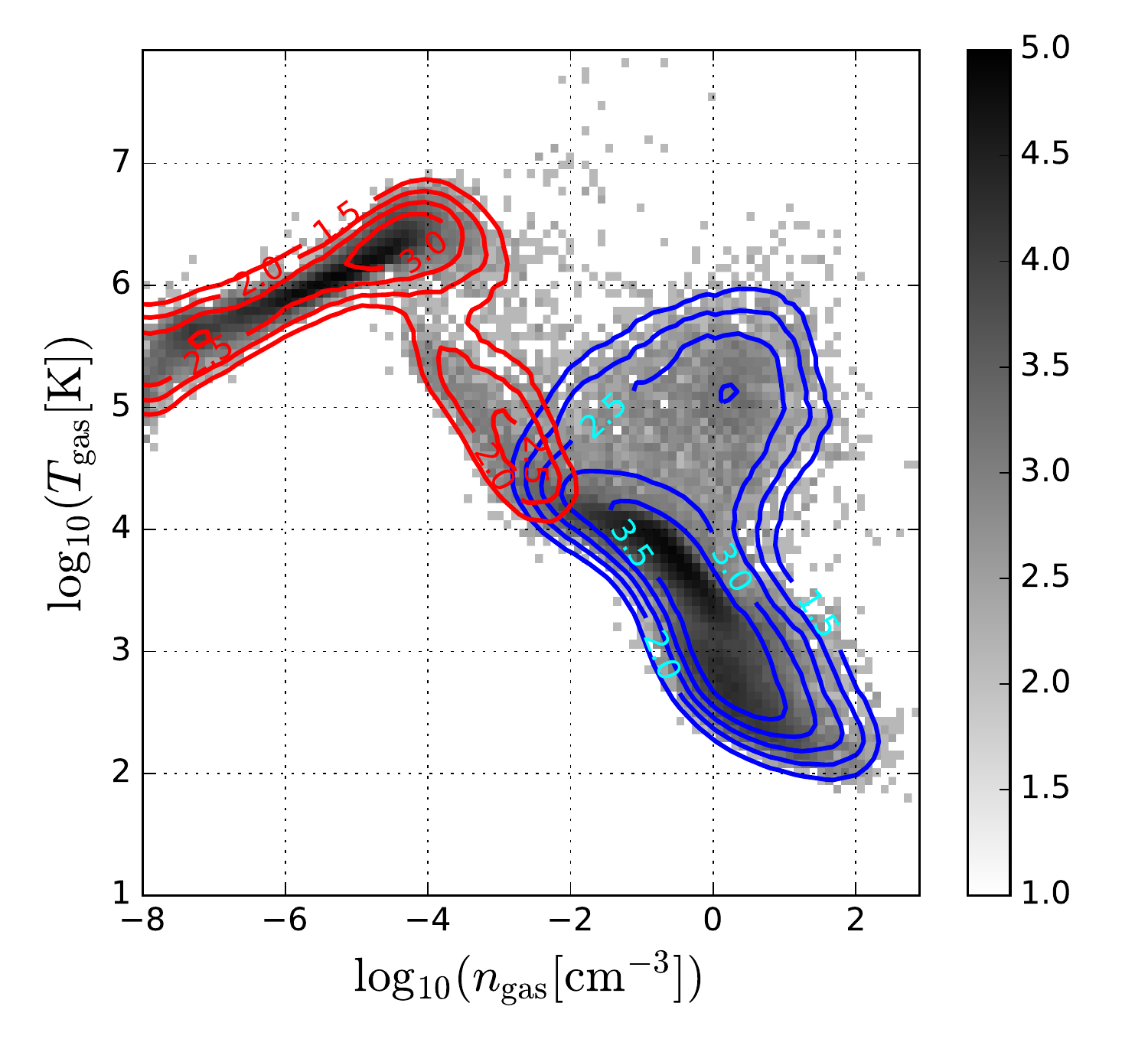}
	\caption{Distribution of gas particles on the temperature-density
	phase diagram at $t = 10$\,Gyr. The grey scale indicates the relative 
	logarithmic surface density on the phase diagram
	for all gas particles; blue and red contours show
	the contribution from particles inside and outside the disc. 
	The contours and grey scale share the same scale.
	We set $r = 15$ and $|z| = 2$ kpc as the disc boundary.}
	\label{Fig:feedback_phase}
	\end{center}
\end{figure}

Before calculating the extinction curves, we produce
a phase diagram at 10\,Gyr (Fig.\,\ref{Fig:feedback_phase}) 
to illustrate how stellar and SN feedback kick gas particles out of the disc and heat them up. 
Particles outside the disc occupy a low density ($n_\mathrm{gas} < 10^{-2}$ cm$^{-3}$) 
and high temperature ($T_\mathrm{gas} > 10^4$\,K) regions, as shown by the red contour. 
Gas inside the disc (blue contours) have higher densities and lower temperature, but there is also a plume of gas at $n_\mathrm{gas} \sim 1$ cm$^{-3}$ which is heated up to 10$^6$\,K in the disc by feedback.  The heated gas will try to escape from the disc and either rain down onto the disc to become cold gas again, or they will expand and adiabatically cool and join the red contour by moving to the left. 

\begin{figure*}%figure
	\begin{center}
	\includegraphics[width=0.33\textwidth]{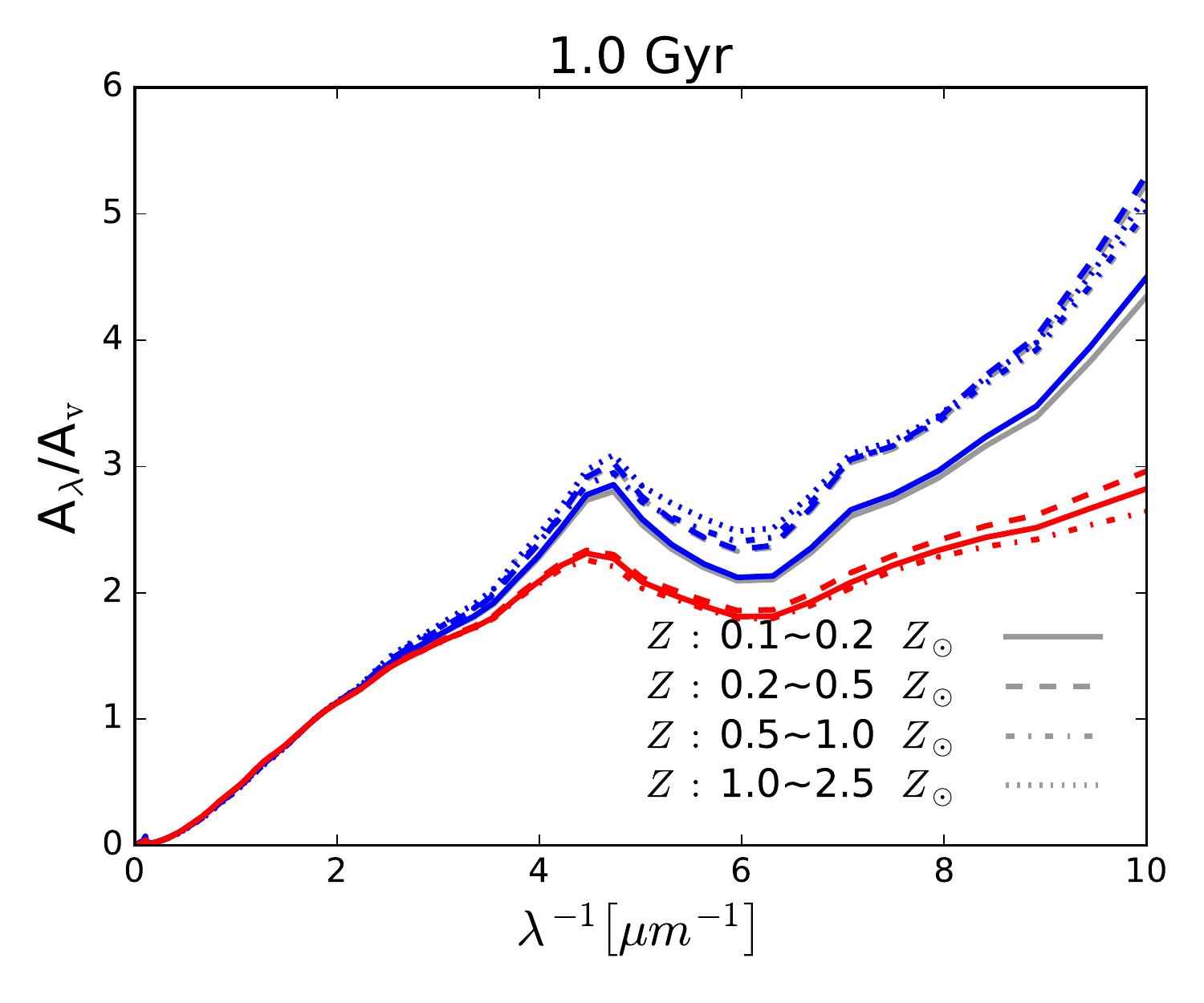}
	\includegraphics[width=0.285\textwidth]{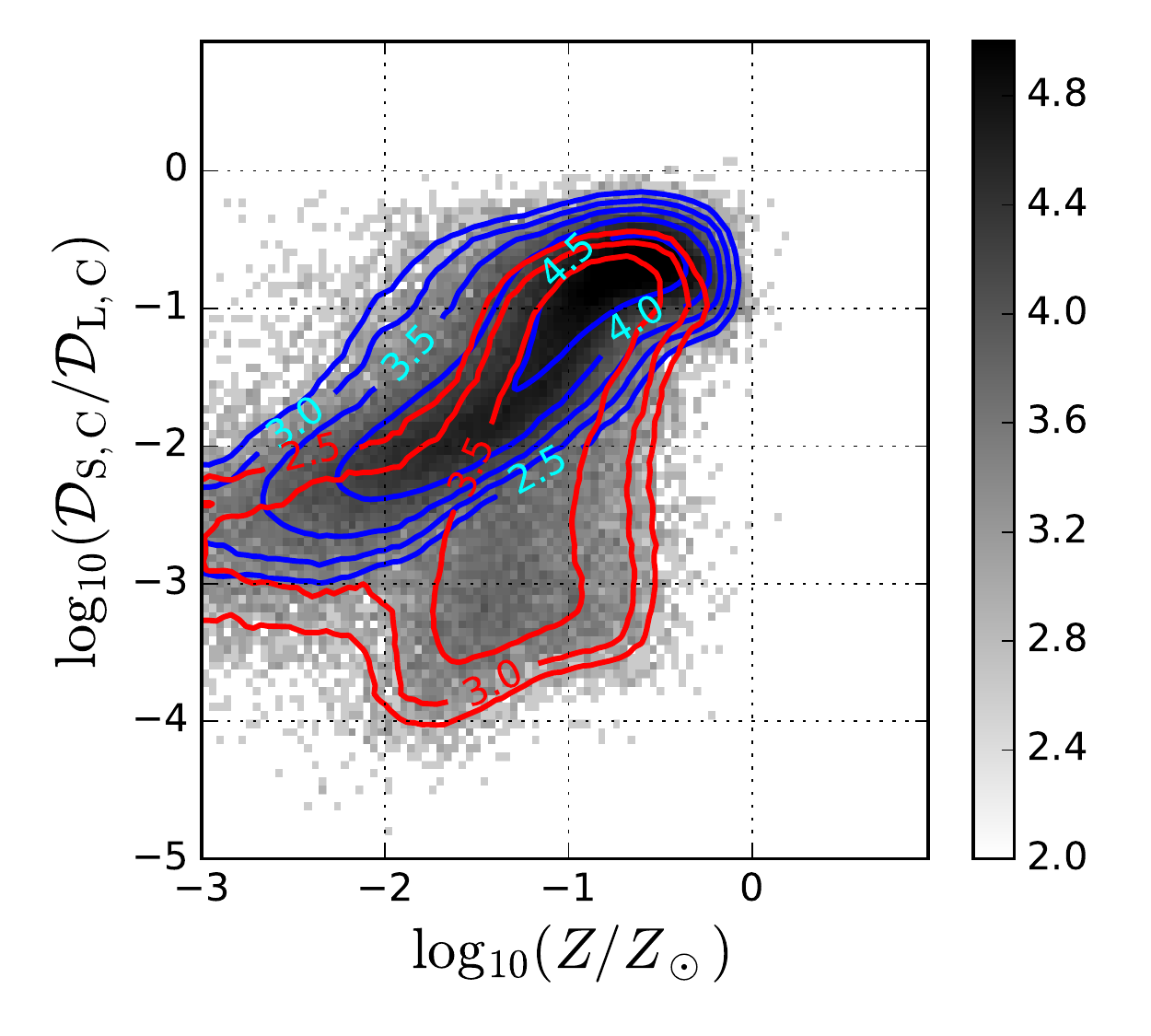}
	\includegraphics[width=0.285\textwidth]{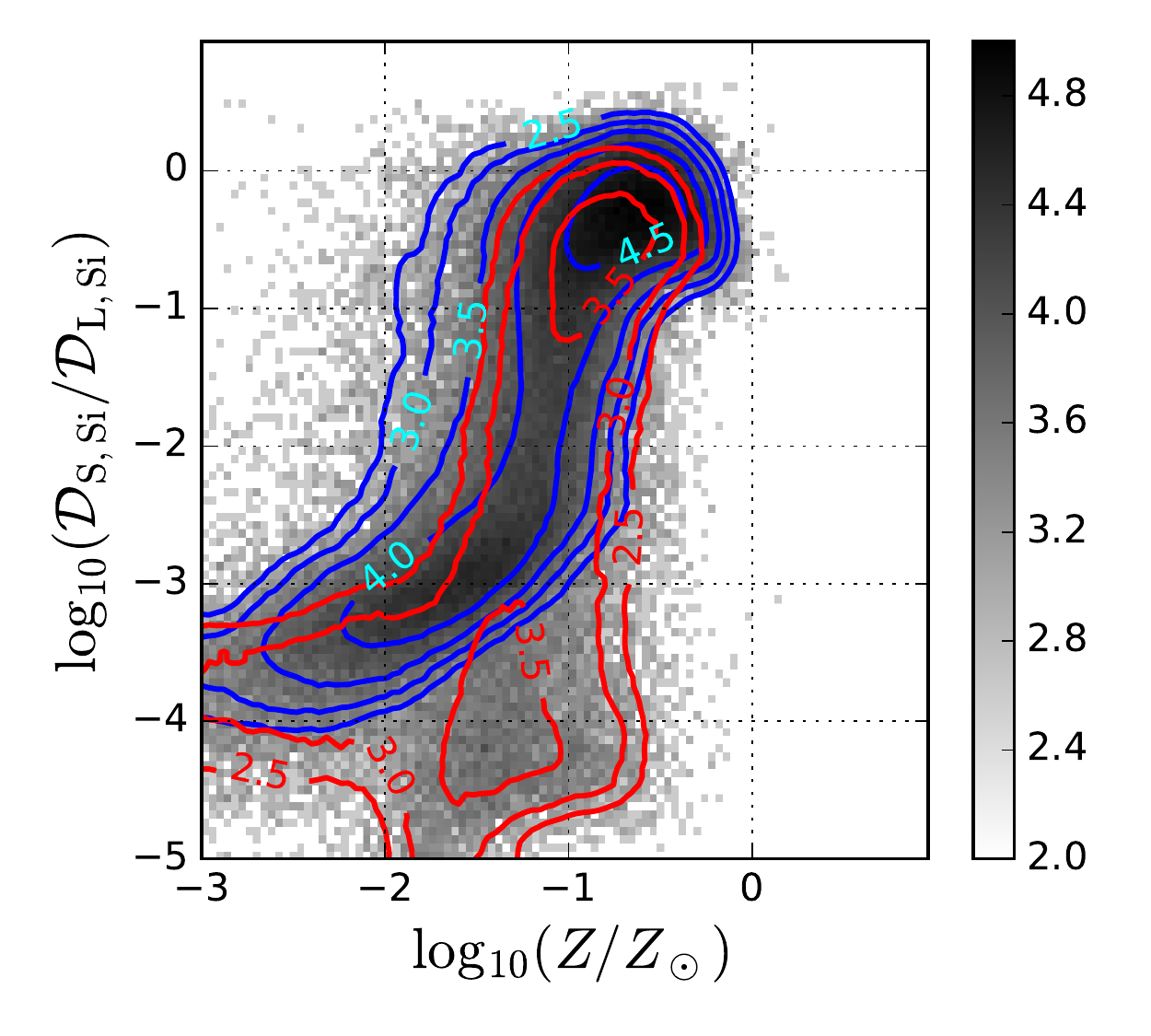}

	\includegraphics[width=0.33\textwidth]{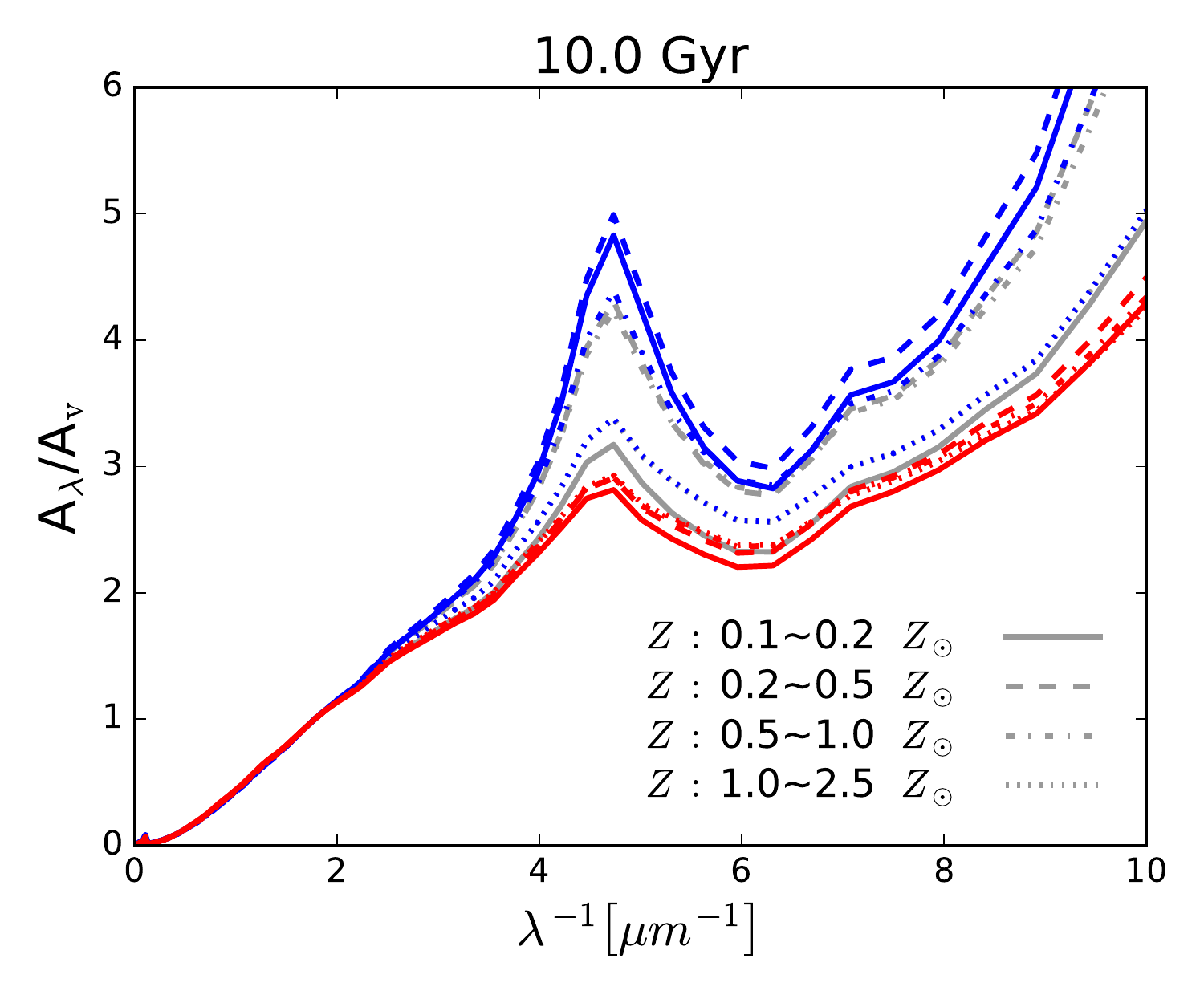}
	\includegraphics[width=0.285\textwidth]{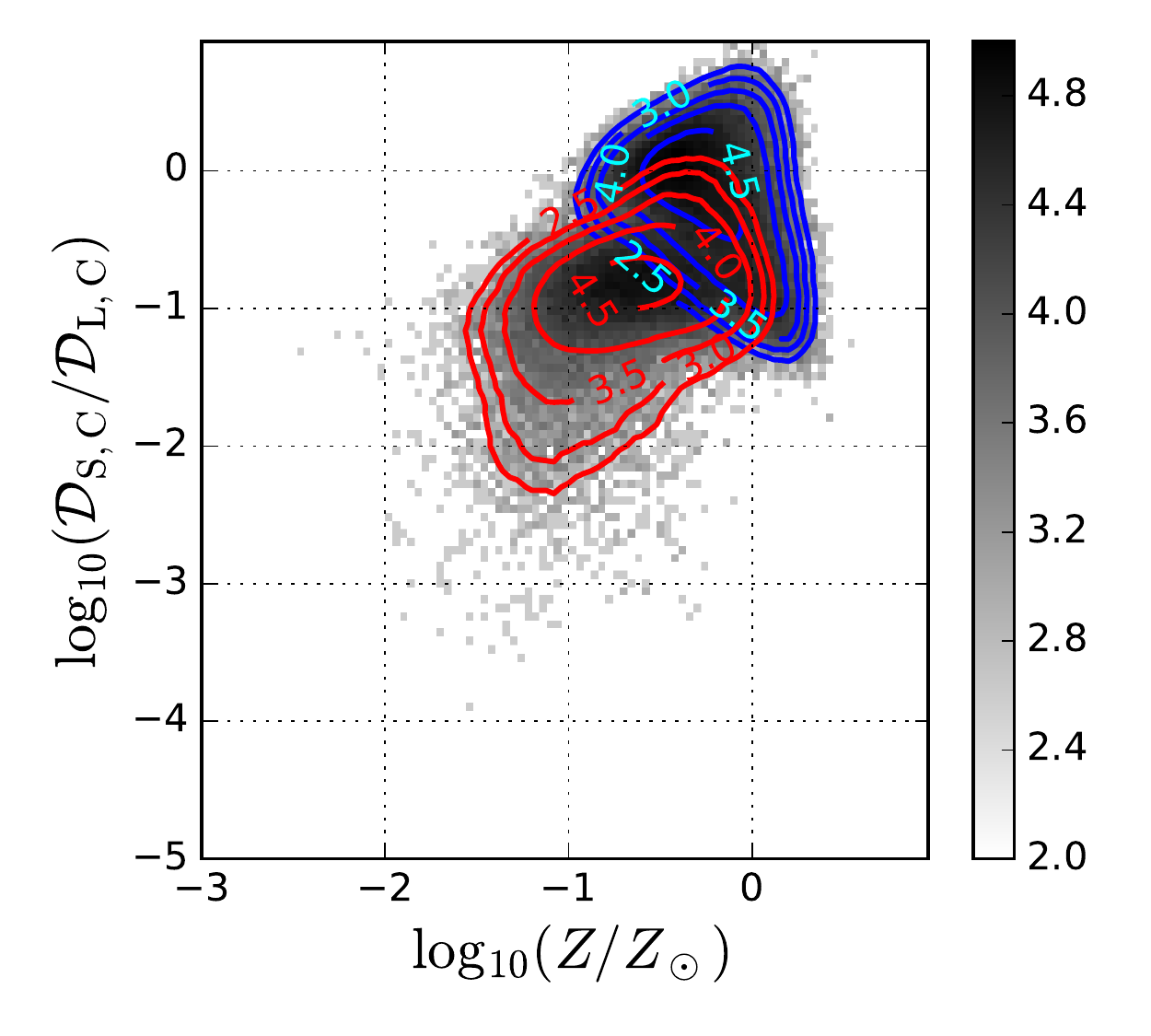}
	\includegraphics[width=0.285\textwidth]{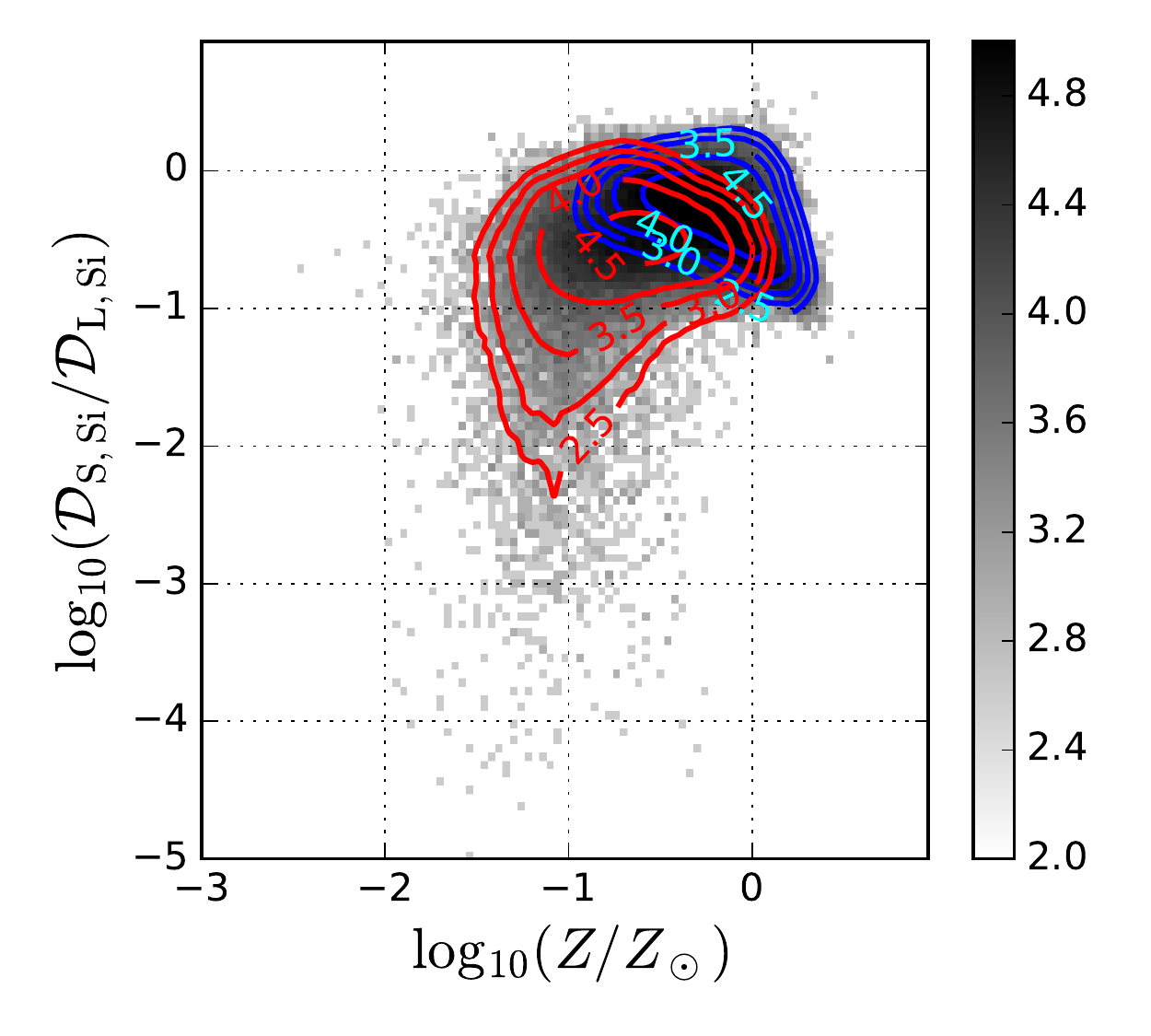}
	\caption{Same as Fig. \ref{Fig:extinction_metal} but for further separating 
	the inside and outside disc components. In the left column, 
	blue and red lines represent the extinction curves inside and outside the disc, respectively; grey lines show extinction curves of all particles. 
	In the right two panels, the grey scale and the contours are the same as in 
	Fig. \ref{Fig:feedback_phase}.}
	\label{Fig:feedback_extinc}
	\end{center}
\end{figure*}

We examine the metallicity dependence of extinction curves again
while this time separating inside and outside the disc components 
at 1 and 10\,Gyr (Fig.\,\ref{Fig:feedback_extinc}). 
In general, the outer disc extinction curves have weaker features 
than the inner disc curves.
At 1\,Gyr, the particles inside the disc dominate the 
total particle number; 
thus, the extinction curves inside the disc are almost 
the same as those of all particles.
Extinction curves of the outer disc show the weak 2175\,\AA\ bump
and a flat FUV slope, and there is no particle outside the disc 
with $Z > 1 Z_\odot$.
At 10\,Gyr, except $Z > 1 Z_\odot$, 
the features of extinction curves inside the disc 
become more prominent than those in the entire galaxy,
because the $\stol$ values of carbonaceous dust concentrate on
a higher value owing to accretion. 

The $\stol$ of silicate does not show a 
clear bimodal distribution at 10\,Gyr for the following reasons:
(i) Silicate has a higher accretion efficiency than carbonaceous dust
(Table\,\ref{table:Adopted_quantities}).
Thus, silicate can reach higher $\stol$ than carbonaceous dust 
before the particle is ejected out of the disc.
(ii) Shattering efficiency is lower for silicate than for carbonaceous dust,
so small silicate grains are not efficiently produced from 
large silicate grains in the disc.
We also notice that $\stol$ outside the disc evolves 
more slowly than that inside the disc. 
Once the particles are kicked out of the disc, 
only shattering occurs because of low density,
and shattering is not efficient in such low density
environment (see Eq.\,\ref{tau_sh}). Thus, the extinction curves outside the disc
show weaker 2175\,\AA\ bumps and flatter FUV slopes
compared to those inside the disc.

In Fig.\,\ref{Fig:extinction_density}, we calculated the extinction curves
in the density range from 0.1 to $10^3$ cm$^{-3}$, which automatically 
chose the particles inside the disc. The feedback effect is not apparent 
in the radial dependence of extinction curve shown in Fig.\,\ref{Fig:extinction_radii}, 
because we considered only up to $r = 8$\,kpc, where the contribution 
from the particles outside the disc is negligible in the resulting extinction curves.

Because we are mainly interested in extinction curves in
the galactic disc, we only considered supernova shocks as sources of
dust destruction, but neglected sputtering in the diffuse hot gas which
exists out of the disc. For more precise treatment of the extinction
curves out of the disc, we need to include thermal sputtering.
We expect that, if we include thermal sputtering, the extinction curves
in the gas out of the disc would be further flattened because smaller grains
are more efficiently destroyed by thermal sputtering
\citep{Hirashita:2015ab}. Thus, the tendency of flat extinction curves
out of the galactic disc would be strengthened if we included thermal
sputtering.

\subsection{SMC/LMC extinction curves}\label{subsec:MCs}

In the above analysis, we have adopted graphite for the carbonaceous
species. Small graphite grains have a strong 2175\,\AA\ bump, while
some extinction curves in extragalactic objects show a much weaker
bump than the Milky Way curve. We here examine a possibility of explaining
the extinction curves in the SMC as a representative of bumpless
extinction curves.

\citet{Nozawa:2015aa} and \citet{Hou:2016aa}
adopted amorphous carbon instead of graphite to explain
bumpless extinction curves in the nearby and distant Universe.
Although the evolutionary link between amorphous carbon and
graphite has not been clarified yet, it is worth examining if
the bumpless SMC extinction curve can be reproduced with amorphous carbon.

\begin{figure}%figure
	\begin{center}
	\includegraphics[width=0.45\textwidth]{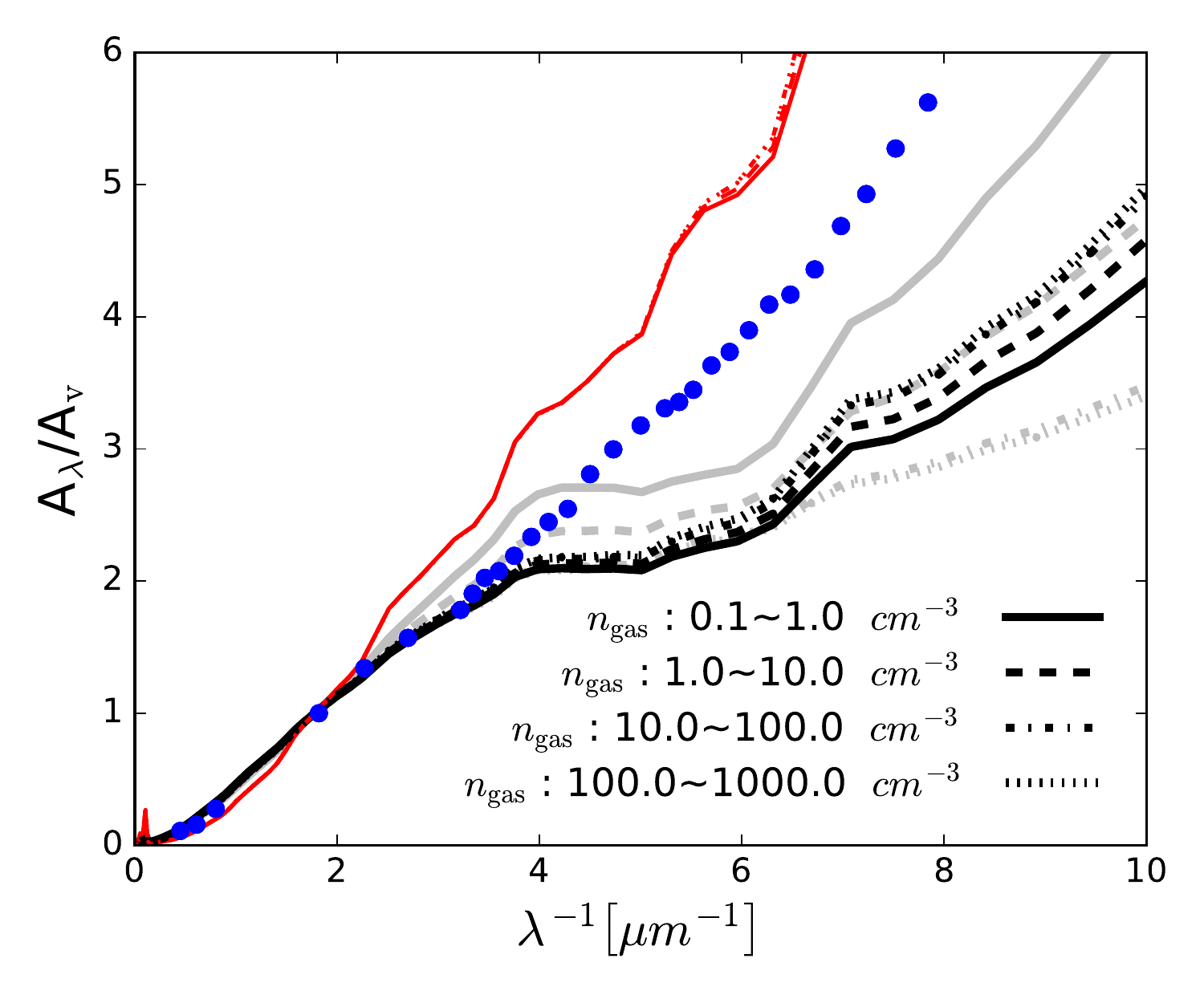}
	\includegraphics[width=0.45\textwidth]{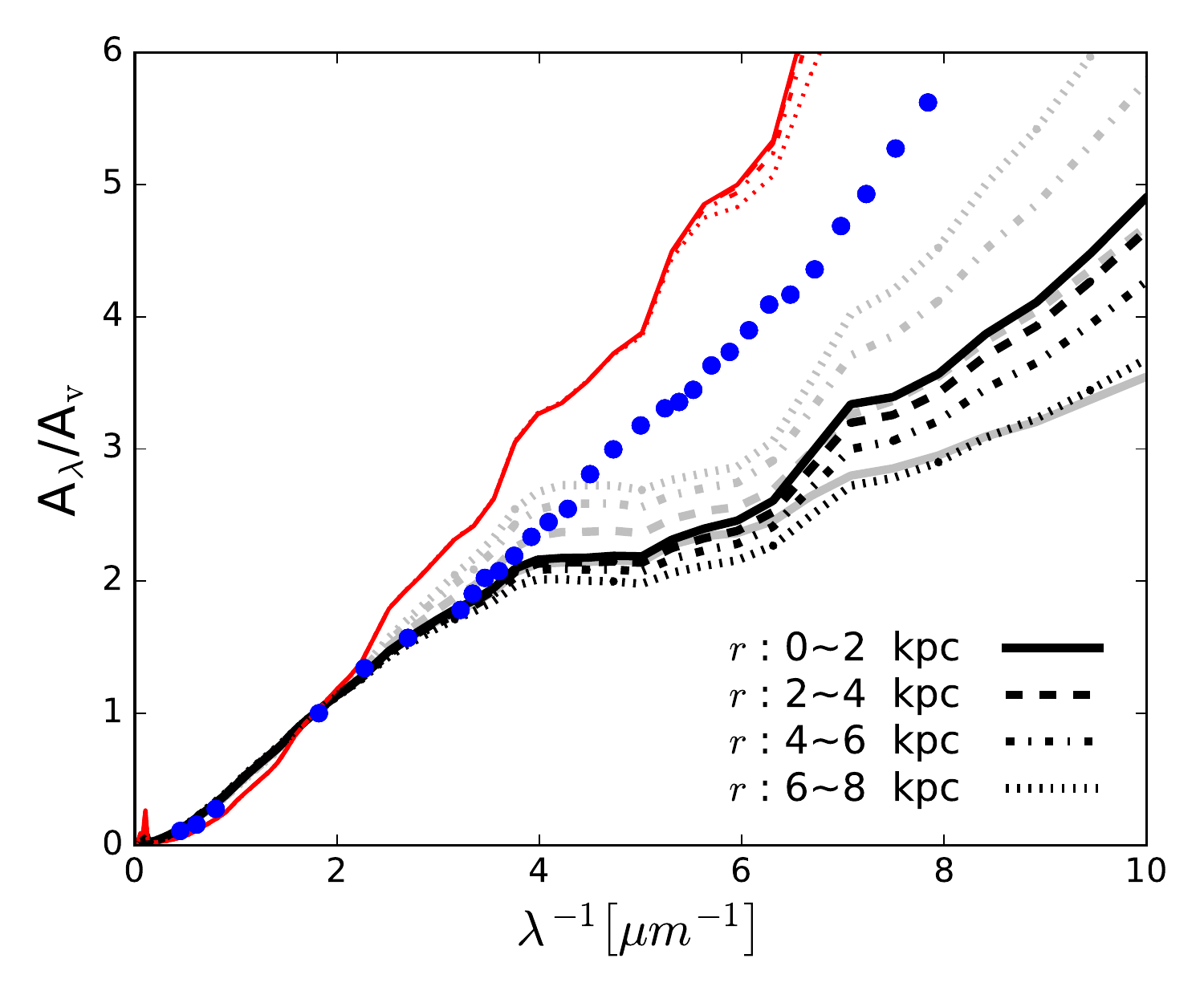}
	\caption{Comparison with the SMC extinction curve. 
	For carbonaceous dust, we used amorphous carbon instead of graphite.
	Upper panel:
	The black solid, dashed, dash-dotted, 
	and dotted lines represent the extinction curves at 1\,Gyr, 
	when the metallicity of the simulated galaxy is similar to that of the SMC
	in gas density ranges, 
	0.1--1, 1--10, 10--$10^2$ and $10^2$--$10^3$ cm$^{-3}$, respectively.
	Bottom panel: 
	The black solid, dashed, dash-dotted, and dotted lines represent 
	the extinction curves at 1\,Gyr in radial ranges, 
	0--2, 2--4, 4--6 and 6--8 kpc, respectively.
	The blue dots in both panels show the observed extinction curve of the SMC
	taken from \citet{Pei:1992aa}.
	Thin red lines show the extinction curves at 1\,Gyr with only considering
	the contribution of silicate. The grey lines show the extinction 
	curves at 10\,Gyr, when C/Si has the lowest value 
	(the line species of the red and grey lines have the same meanings as 
	those of the black lines).}
	\label{Fig:SMC_extinction}
	\end{center}
\end{figure}

There are other possible carbonaceous materials \citep{Jones:2013aa},
but our two-size approach is not suitable for distinguishing detailed
dust properties. Therefore, we just compare graphite and 
amorphous carbon in this paper
to examine the two extremes in terms of the bump strength.
In Fig.~\ref{Fig:SMC_extinction}, we show the extinction curve
at 1\,Gyr, when the typical metallicity is $Z \sim 0.2 Z_\odot$ 
(Fig.\ \ref{Fig:extinction_metal}), 
with adopting amorphous carbon instead of graphite. 
Here, we adopt \citet{Zubko:1996aa} for the
optical constants for amorphous carbon (we adopt their ACAR).
To keep consistency with the simulation, 
we still adopt the material density of graphite ($\rho_\mathrm{C} = 2.24$ g cm$^{-3}$) 
instead of that of amorphous carbon ($\rho_\mathrm{C} = 1.81$ g cm$^{-3}$) 
in \citet{Zubko:2004aa} for carbonaceous dust. 
Because the difference in the material density between 
those two carbonaceous species is not large, 
the following results are hardly affected by the choice of the material density. 
Fig.\ \ref{Fig:SMC_extinction} shows that, although the 2175\,\AA\ bump is eliminated successfully, 
there is a large discrepancy toward FUV wavelengths. 
This discrepancy is resolved only if the relative 
contribution of small silicate grains is increased in our model.
\citet{Weingartner:2001aa} suggested a smaller graphite-to-silicate 
mass ratio than that of the Milky Way to explain the SMC extinction curve, 
and \citet{Pei:1992aa} fitted the SMC extinction curve exclusively with silicate.
However, even the values of C/Si at 10\,Gyr, when C/Si 
is the lowest (Fig. \ref{Fig:c2si}), 
cannot reproduce the SMC extinction curve
(light grey lines in Fig. \ref{Fig:SMC_extinction}). 
This means that our simulation has difficulty 
in reproducing such a steep FUV rise as observed in the SMC extinction curve.

There have been some ideas proposed to reproduce the 
SMC extinction curve.
\citet{Bekki:2015ab} suggested
that small carbonaceous dust grains are selectively
transported outside the SMC by the most recent starburst. 
In their model, decoupling of grain motion from gas motion
is important; however, our simulation assumes a complete dynamical
coupling between dust and gas. This decoupling could be
included in a scheme developed by \citet{Bekki:2015aa},
who proposed an SPH modelling in which dust and gas are treated by
different particle species. Thus, we will leave the possibility of
selective transport of small carbonaceous grains to the future work.
\citet{Hou:2016aa} proposed more efficient SN destruction of 
small carbonaceous grains than that of silicate and showed that 
the extinction curve could become as steep as the SMC curve 
because of a resulting high fraction of small silicate grains.

The above previous ideas of reproducing the SMC extinction curve
are based on selective elimination of small carbonaceous dust.
Thus, we also examine an extreme case by assuming that the extinction curve 
is only contributed from silicate, i.e.\ we eliminate carbonaceous dust
(red thin lines in Fig.~\ref{Fig:SMC_extinction}). 
The results show too steep FUV slopes, which implies that 
a small fraction of carbonaceous dust is still necessary to 
reproduce the SMC extinction curve. The fact that the observed
SMC extinction curve is between the
pure silicate curve and our predictions supports the idea that
a certain fraction of carbonaceous dust is lost by certain mechanisms
such as outflow and SN destruction.

\begin{figure}%figure
	\begin{center}
	\includegraphics[width=0.45\textwidth]{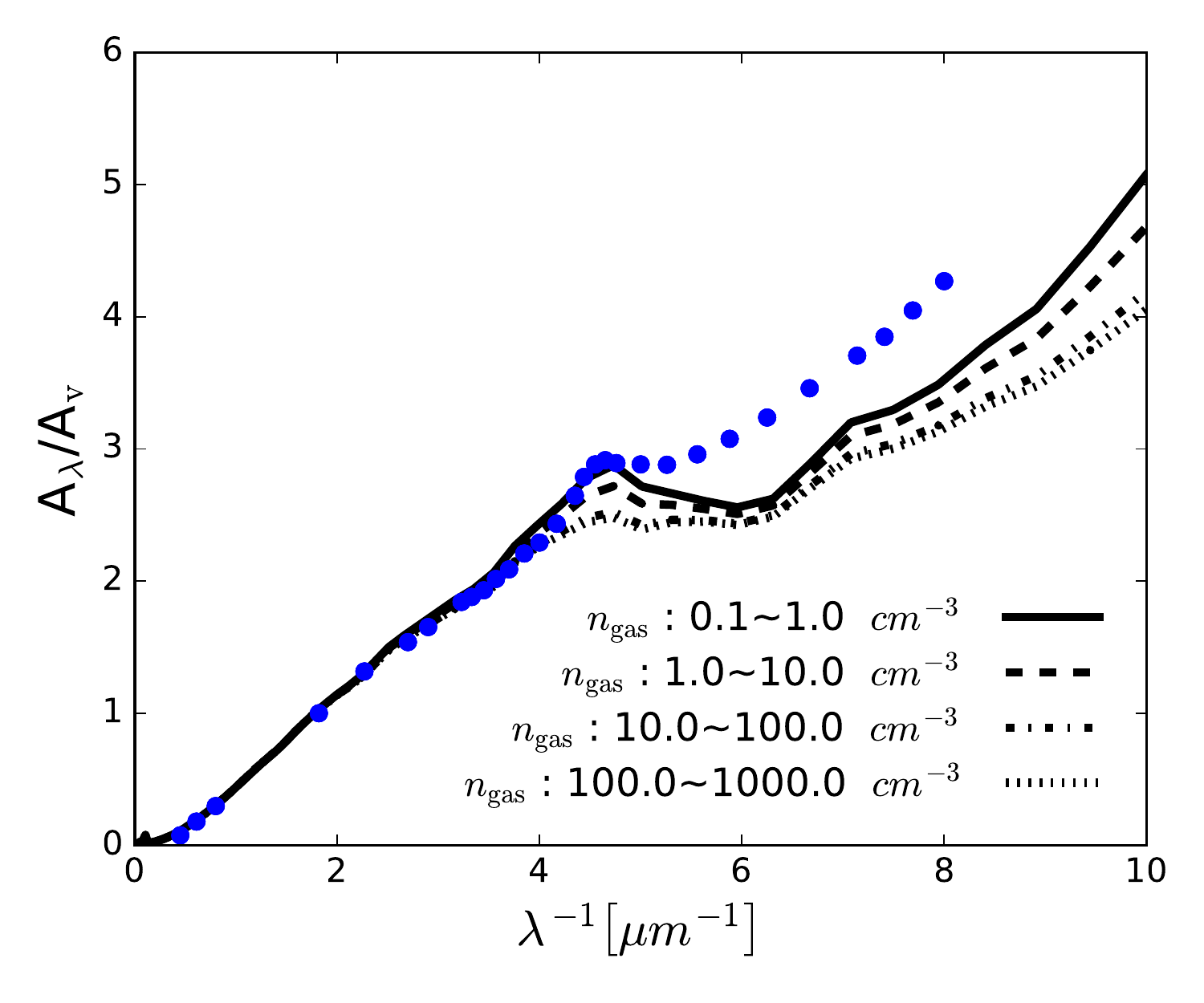}
	\includegraphics[width=0.45\textwidth]{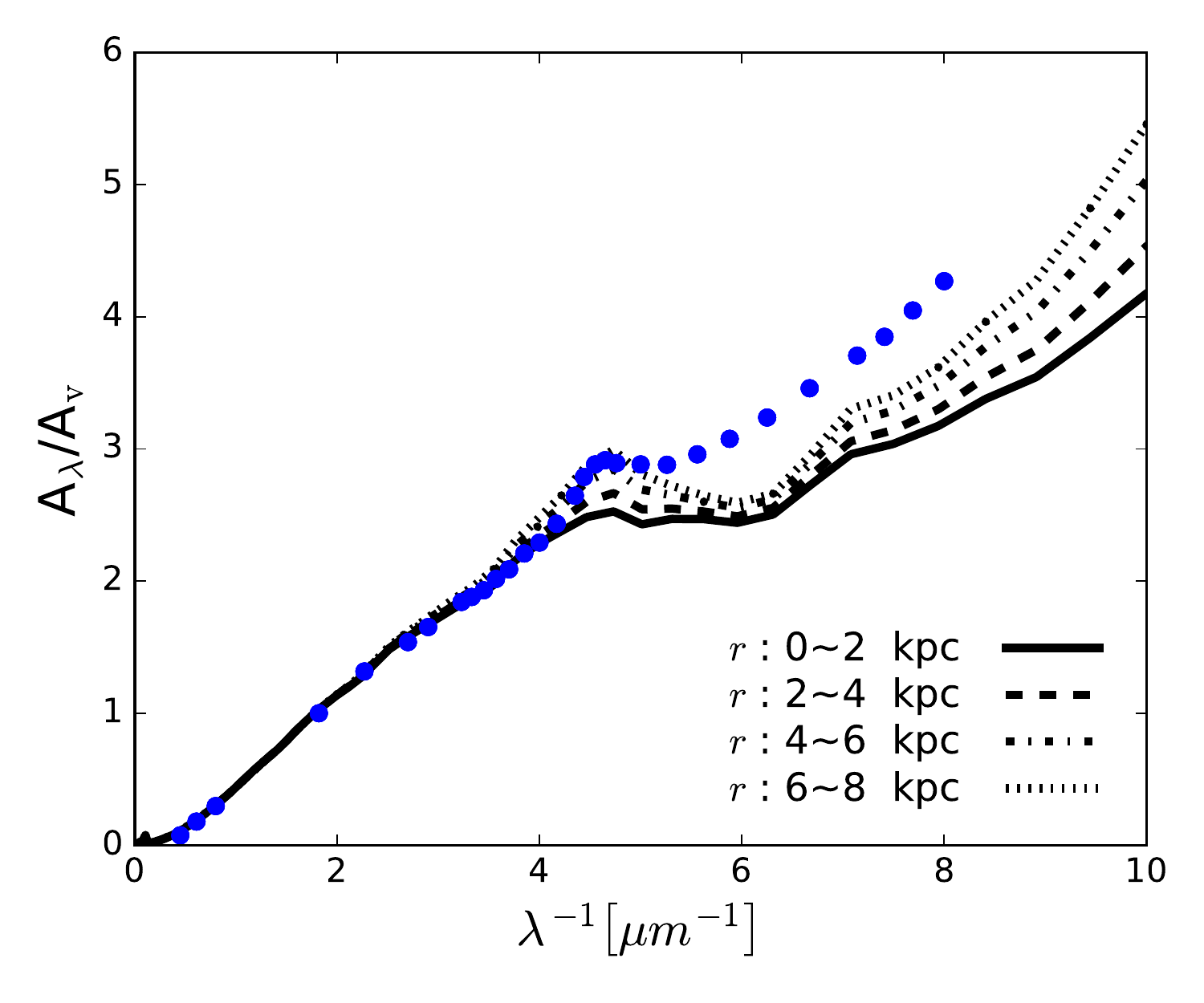}
	\caption{Comparison with the LMC extinction curve
	with the results at 3\,Gyr, when the metallicity of the 
	simulated galaxy is similar to that of the LMC. 
	For carbonaceous dust, we assumed 50 per cent is amorphous carbon 
	and 50 per cent is graphite.
	Upper panel:
	The solid, dashed, dash-dotted, 
	and dotted lines represent the extinction curves in gas 
	density ranges, 0.1--1, 1--10, 10--$10^2$ and $10^2$--$10^3$ cm$^{-3}$, 
	respectively.
	Bottom panel: 
	The solid, dashed, dash-dotted, and dotted lines represent 
	the extinction curves in radial ranges, 
	0--2, 2--4, 4--6 and 6--8 kpc, respectively.
	The blue dots show the observed extinction curve of the LMC
	taken from \citet{Pei:1992aa}.}
	\label{Fig:LMC_extinction}
	\end{center}
\end{figure}

In addition, we also attempt to predict the LMC extinction curve
by changing the graphite-to-amorphous carbon ratio and selecting 
snapshots with a higher metallicity than the SMC.
We adopt the snapshot at 3\,Gyr and assume that 
the graphite-to-amorphous carbon ratio is 1:1. 
The resulting extinction curves are shown in Fig.~\ref{Fig:LMC_extinction}.
In the figure,
the 2175 \AA\ bump strength is roughly reproduced in $n_\mathrm{gas}$ =
0.1--1 cm$^{-3}$ and $r = 6$--8 kpc. Thus, the weak 2175 \AA\ bump in
the LMC compared to the Milky Way can be represented 
by the mixture of graphite and amorphous carbon.
However, a discrepancy remains at FUV wavelengths as was also
the case for the SMC.
This implies that reducing C/Si is also necessary to reproduce the FUV slope
of the LMC extinction curve.

\subsection{Extinction curves in high redshift galaxies}

\begin{figure}%figure
	\begin{center}
	\includegraphics[width=0.45\textwidth]{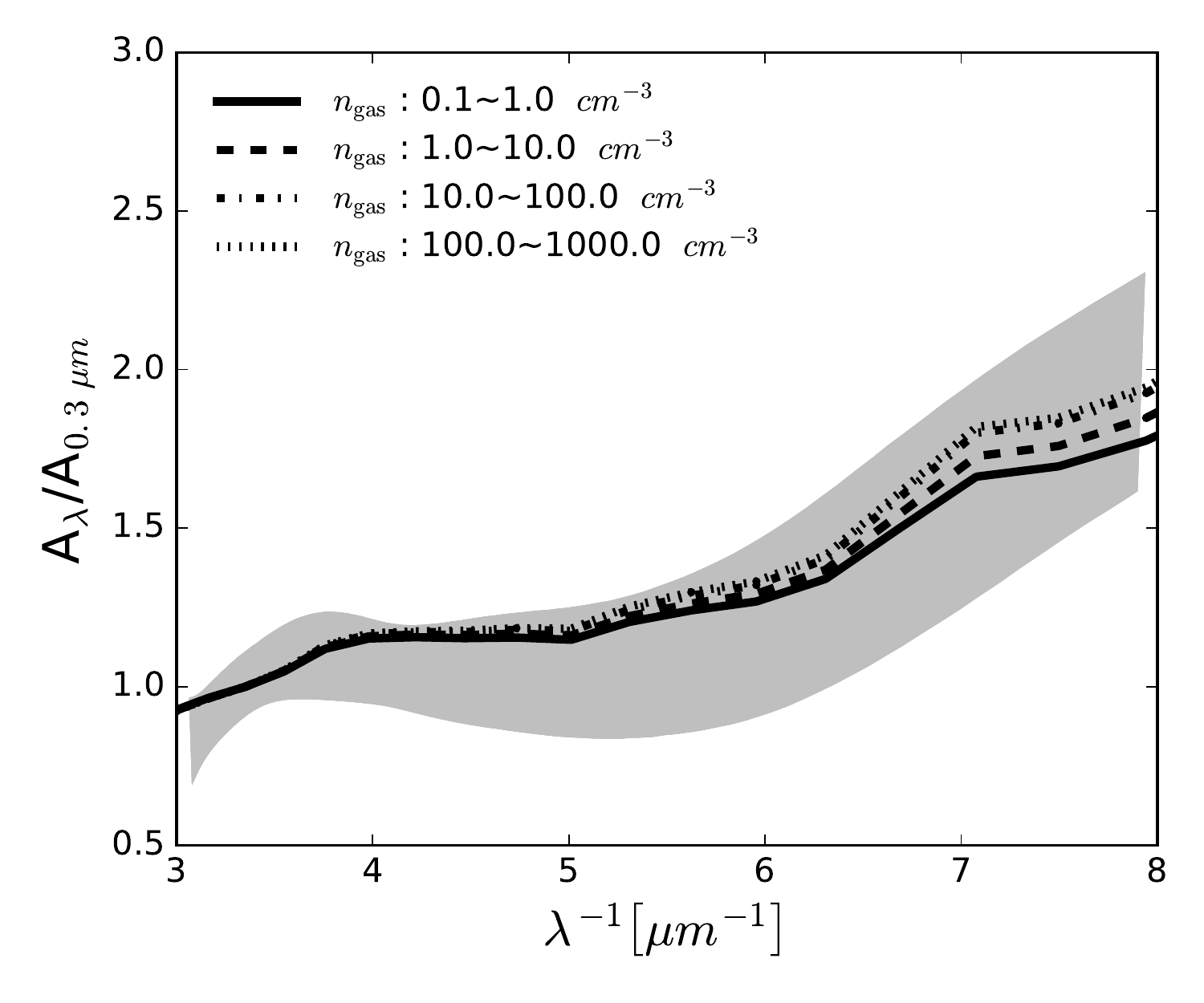}
	\caption{Comparison with the extinction curve of a $z = 6.2$ high redshift quasar. 
	The solid, dashed, dash-dotted, 
	and dotted lines represent the extinction curves at 1\,Gyr in gas 
	density ranges, 0.1--1, 1--10, 10--$10^2$ and $10^2$--$10^3$ cm$^{-3}$, 
	respectively. For carbonaceous dust, we used amorphous carbon instead of graphite.
	The grey shaded region is the observed extinction curve for the $z = 6.2$ quasar
	taken from \citet{Maiolino:2004aa}.
	The width of the shaded region shows the observational error.}
	\label{Fig:quasar_extin}
	\end{center}
\end{figure}

At high redshift, very bright point sources such as 
quasars and gamma-ray bursts provide opportunities to derive
the extinction curves in their host galaxies, taking advantage of their
simple power-law-like SEDs at rest optical and
UV wavelengths \citep{Maiolino:2004aa,Stratta:2007aa,Li:2008ab,
Eliasdottir:2009aa,Gallerani:2010aa,Perley:2010aa,Zafar:2010aa,
Zafar:2011aa,Zafar:2015aa,Schady:2012aa}.
Here we examine whether our simulation could also produce extinction curves
in the distant Universe.
In Fig.\ \ref{Fig:quasar_extin}, 
we show the extinction curves at 1\,Gyr (cosmic age at $z\sim 5$)
using amorphous carbon for the carbonaceous dust species. The
extinction curves are
normalized to the value at rest 0.3 $\micron$ to compare with the 
extinction curve of a quasar, SDSS J1048+4637, at $z = 6.2$
\citep{Maiolino:2004aa} as a representative extinction curve.
As shown by \citet{Gallerani:2010aa}, the extinction curves in
quasars at $z>5$ are similar in the sense that they show much
flatter bumpless extinction curves than the SMC curve 
\citep[but see][for observational uncertainties]{Hjorth:2013aa}.
Our results are in good agreement with the observational data 
and are almost the same as the prediction by \citet{Nozawa:2015aa},
who successfully fitted the same extinction curve 
by using their single-zone dust evolution model but by taking account of 
the full grain size distribution (i.e.\ without the two-size approximation).

Since we have reproduced both the high-redshift extinction curve
and the Milky Way extinction curve,
there may exist an evolutionary link between these two extinction curves (or
extinction curves in various redshifts in general).
\citet{Li:2016aa} examined extinction curves of 
star-forming disc galaxies at $0.5 < z < 2.5$
and found that extinction curves tend to be flatter 
at higher redshift.
However, we note that the extinction curves 
(more precisely attenuation curves) 
derived by them
represent effective extinction as a result of radiative transfer
(absorption and scattering) effects of light emitted from
multiple stellar populations 
\citep[see also][]{Calzetti:1994aa,Inoue:2005aa,Mancini:2016aa}. 
Radiative transfer effects, especially scattering,
skew the shape of the extinction curve in
highly extinguished objects \citep[e.g.][]{Scicluna:2015aa}.
\citet{Kriek:2013aa} analysed SEDs
to derive attenuation curves for galaxies at $0.3<z<2$.
They found that there is a positive correlation between 
2175 \AA~ bump strength and FUV slope. 
This correlation can be interpreted as a sequence of 
small-to-large grain abundance ratio.
They also showed that galaxies with higher sSFR 
tend to have flatter attenuation curve. 
In this work, the radiative transfer calculation 
is not included yet, but simulation results tend
to predict flat extinction curves at the early evolutionary
stage (or in the phase of high sSFR; see Sec 5.1 of Paper I),
which is similar to the above observational tendency of 
attenuation curves with sSFR.

\section{conclusion}

In this work, we investigated the evolution of extinction curve
using the SPH simulation in which dust evolution is implemented.
As discussed in Paper I, our simulation adopts the two-size model of dust by 
\citet{Hirashita:2015aa}, and solves the formation and destruction of dust 
as a function of time together with gas dynamics.
Additionally, dust species are separated into carbonaceous dust and silicate, 
and the two species are assumed to evolve independently. 
They share the same dust formation 
and destruction mechanisms but have different condensation efficiencies, 
material properties and elemental abundances.

We compare our results with observed 
galaxy extinction curves in the local Universe.
To estimate the extinction curves, 
we adopt the optical constants of graphite and silicate based on 
\citet{Draine:1984aa} and \citet{Weingartner:2001aa}.
The simulation enables us to examine the dependence of extinction 
curves on the position, gas density, and metallicity in the galaxy
as well as the time evolution.

Our results show that
extinction curves are flat at the earliest evolutionary
stage $t \lesssim$  0.3\,Gyr, because stellar sources 
dominate the dust production at  
that epoch (stars only produce large grains in our model). 
Dust growth by accretion increases the small-to-large 
grain abundance ratio $\stol$ at
$t \gtrsim$ 0.3\,Gyr, making the 2175\,\AA\ bump 
and FUV rise more prominent.
Coagulation becomes efficient after
small grains become abundant, especially in the central region
where the gas is dense and metal-rich. Because of coagulation,
the 2175\,\AA\ bump and FUV rise are suppressed 
at $t \gtrsim$ 3\,Gyr in dense regions.
At $t \sim$ 10\,Gyr,
shattering dominates the dust evolution in low-density, low-metallicity, or
outer regions of the galactic disc. 
Because shattering continuously converts large 
grains to small grains, the 2175\,\AA\ bump and FUV rise are
strong in those regions.

To further constrain the grain size distribution, 
we make a comparison with observed extinction curves and 
find that the predicted extinction curves at $t \gtrsim 3$\,Gyr are
consistent with the mean Milky Way extinction curve
and the dispersion of extinction curves toward  
different lines of sight. This implies that 
all dust processes that drive dust evolution in Milky-Way-like galaxies
are implemented successfully in our simulation.

We also examine the difference in $\stol$ 
between carbonaceous dust and silicate.
The two dust species have different
condensation efficiencies, mass densities and elemental abundances.
At the stellar production dominated epoch ($t \lesssim 0.3$\,Gyr), 
carbonaceous dust has higher $\stol$ than silicate because
the shattering efficiency of carbonaceous dust is higher than that of silicate.
In the period in which accretion dominates the dust abundance 
($0.3 \lesssim t \lesssim 3$\,Gyr), 
$\stol$ of silicate is, on average, higher than that of carbonaceous dust
because under the solar elemental abundance pattern we assumed, silicate has more gas-phase elements available for accretion than carbonaceous dust.
After 3\,Gyr, carbonaceous dust presents higher $\stol$ in 
the diffuse medium or sub-solar metallicity regions and 
lower $\stol$ in the dense medium or solar metallicity regions
than silicate since the efficiencies of shattering and coagulation of carbonaceous dust are higher than those of silicate.
The abundance ratio of carbonaceous dust to silicate, C/Si, 
is also affected by different efficiencies of enrichment mechanisms. 
Carbonaceous dust is more abundant than silicate in the 
stellar production dominated epoch ($t \lesssim 0.3$\,Gyr)
because carbonaceous dust has higher 
dust condensation efficiency in stellar ejecta than silicate has; 
after the accretion dominated epoch ($t \gtrsim 1$\,Gyr), 
silicate becomes more abundant than carbonaceous dust
because of the higher abundance of available gas-phase elements for accretion.

There are two sequences in the metallicity dependence of $\stol$ 
corresponding to the inside and outside disc components. 
The outside disc component is transported there 
mainly by stellar and SN feedback.
The extinction curves of the outside disc component show shallower FUV slopes 
and weaker 2175\,\AA\ bumps than those of the inside disc component 
because $\stol$ evolves more slowly outside the disc than inside.
This indicates that the extinction curves in the circum-galactic medium 
are flatter than those in galactic discs.

We also attempt to reproduce the SMC extinction curve 
by adopting amorphous carbon instead of graphite.
Although 2175\,\AA\ bump is successfully eliminated, 
there is a large discrepancy at FUV wavelengths.
The steep FUV rise in the observed SMC extinction curve is reproduced
only by a higher silicate fraction than calculated in this paper,
which indicates that carbonaceous dust is selectively lost
by, e.g., outflow and SN destruction. 
In addition, reproducing the LMC extinction curve
by changing graphite-amorphous carbon ratio
is not fully successful because
we underpredict the FUV rise also for the LMC.

Finally, we also reproduced
a representative extinction curve of high-redshift
quasars using our simulation results at 1\,Gyr.
The observational trend that extinction curves are flatter for 
higher specific star formation rates is also interpreted by our model, 
in the sense that our model also predicts the change 
from flat to steep extinction curves 
as the specific star formation rate drops.

\section*{Acknowledgements}

We thank Y.-H. Chu, T.-H Chiueh, and W.-H. Wang for useful
comments.
HH is supported by the Ministry of Science and Technology
grant MOST 105-2112-M-001-027-MY3.
KN, SA and IS acknowledges the support from JSPS KAKENHI Grant Number 26247022.  Numerical simulations were in part carried out on the XC30 at the Centre for Computational Astrophysics, National Astronomical Observatory of Japan.

%%%%%%%%%%%%%%%%%%%%%%%%%%%%%%%%%%%%%%%%%%%%%%%%%%



\bibliographystyle{mnras}
\bibliography{kchou}
% \bibliography{/Users/loveluthien/bitbucket/kchou_paper/kchou_osaka}

%%%%%%%%%%%%%%%%%%%%%%%%%%%%%%%%%%%%%%%%%%%%%%%%%%

% Don't change these lines
\bsp	% typesetting comment
\label{lastpage}
\end{document}